\documentclass[twocolumn]{aastex63}

\usepackage{amsmath}
\usepackage{amssymb}
\usepackage{fontawesome}
\usepackage{hyperref}
\usepackage{float} 
\usepackage{longtable}
\let\tablenum\relax
\usepackage{siunitx}

\newcommand{\tess}{\emph{TESS}}

\newcommand{\gaia}{\emph{Gaia}}

\newcommand{\jwst}{\emph{JWST}}
\newcommand{\kepler}{\emph{Kepler}}
\newcommand{\angd}{\ang[angle-symbol-over-decimal]}

\shorttitle{TOI 560}
\shortauthors{El Mufti et al.}

\begin{document}

\title{TOI 560 : Two Transiting Planets Orbiting a K Dwarf Validated with iSHELL, PFS and HIRES RVs}

\correspondingauthor{Mohammed El Mufti}
\email{mabdall@gmu.edu}

\author[0000-0001-8364-2903]{Mohammed El Mufti}
\affiliation{Department of Physics and Astronomy, George Mason University, 4400 University Drive, Fairfax, VA 22030, USA}
\affiliation{University of Khartoum, Faculty of Science, Department of Physics, P.O.BOX 321, Khartoum 11111, Sudan}

\author[0000-0002-8864-1667]{Peter P. Plavchan}
\affiliation{Department of Physics and Astronomy, George Mason University, 4400 University Drive, Fairfax, VA 22030, USA}

\author[0000-0002-0531-1073]{Howard Isaacson}
\affiliation{{Department of Astronomy,  University of California Berkeley, Berkeley CA 94720, USA}}
\affiliation{Centre for Astrophysics, University of Southern Queensland, Toowoomba, QLD, Australia}

\author[0000-0002-2078-6536]{Bryson L. Cale}
\affiliation{IPAC, 770 S Wilson Avenue, Pasadena, CA 91106, USA}
\affiliation{Jet Propulsion Laboratory, California Institute of Technology, 4800 Oak Grove Drive, Pasadena, CA 91109, USA}

\author[0000-0002-2457-7889]{Dax L. Feliz}
\affiliation{Department of Physics and Astronomy, Vanderbilt University, 2201 West End Avenue, Nashville, TN 37235, USA}

\author[0000-0003-4701-8497]{Michael A. Reefe}
\affiliation{Department of Physics and Astronomy, George Mason University, 4400 University Drive, Fairfax, VA 22030, USA}

\author[0000-0000-0000-0000]{Coel Hellier}
\affiliation{Astrophysics Group, Keele University, Staffordshire, ST5 5BG, United Kingdom}

\author[0000-0000-0000-0000]{Keivan Stassun}
\affiliation{Department of Physics and Astronomy, Vanderbilt University, 2201 West End Avenue, Nashville, TN 37235, USA}

\author[0000-0003-3773-5142]{Jason Eastman}
\affiliation{Center for Astrophysics, Harvard University, 60 Garden St, Cambridge, MA 02138, USA}

\author{Alex Polanski}
\affiliation{Department of Physics \& Astronomy, The University of Kansas, 1251 Wescoe Hall Dr, Lawrence, KS 66045}

\author{Ian J. M. Crossfield}
\affiliation{Department of Physics \& Astronomy, University of Kansas, 1082 Malott, 1251 Wescoe Hall Dr., Lawrence, KS 66045, USA}

\author[0000-0002-5258-6846]{Eric Gaidos}
\affiliation{Department of Earth Sciences, University of Hawai`i at M\"{a}noa, Honolulu, HI 96822}

\author{Veselin Kostov}
\affiliation{NASA Goddard Space Flight Center, 8800 Greenbelt Rd, Greenbelt, MD 20771, USA}

\author[0000-0002-7424-9891]{Justin M. Wittrock}
\affiliation{Department of Physics and Astronomy, George Mason University, 4400 University Drive, Fairfax, VA 22030, USA}


\author{Joel Villase{\~ n}or}
\affiliation{Department of Physics and Kavli Institute for Astrophysics and Space Research, Massachusetts Institute of Technology, Cambridge, MA 02139, USA}

\author[0000-0001-5347-7062]{Joshua~E.~Schlieder}
\affiliation{NASA Goddard Space Flight Center, 8800 Greenbelt Rd, Greenbelt, MD 20771, USA}

\author[0000-0002-0514-5538]{Luke~G.~Bouma}
\affiliation{Department of Astrophysical Sciences, Princeton University, 4 Ivy Lane, Princeton, NJ 08544, USA}


\author[0000-0003-2781-3207]{Kevin I. Collins}
\affiliation{Department of Physics and Astronomy, George Mason University, 4400 University Drive, Fairfax, VA 22030, USA}

\author[0000-0003-2872-9883]{Farzaneh Zohrabi}
\affiliation{Department of Physics and Astronomy, Louisiana State University, 202 Nicholson Hall, Baton Rouge, LA 70803, USA}

\author[0000-0001-7058-4134]{Rena A. Lee}
\affiliation{Department of Earth Sciences, University of Hawai`i at M\"{a}noa, Honolulu, HI 96822}

\author{Ahmad Sohani}
\affiliation{Mississippi State University, Department of Physics and Astronomy, 355 Lee Boulevard, Mississippi State, MS 39762} 

\author{John Berberian}
\affiliation{Department of Physics and Astronomy, George Mason University, 4400 University Drive, Fairfax, VA 22030, USA}
\affiliation{Woodson High School, 9525 Main St, Fairfax, VA 22031, USA}

\author[0000-0002-4501-564X]{David Vermilion}
\affiliation{Department of Physics and Astronomy, George Mason University, 4400 University Drive, Fairfax, VA 22030, USA}
\affiliation{NASA Goddard Space Flight Center, 8800 Greenbelt Rd, Greenbelt, MD 20771, USA}

\author{Patrick Newman}
\affiliation{Department of Physics and Astronomy, George Mason University, 4400 University Drive, Fairfax, VA 22030, USA}

\author[0000-0001-9596-8820]{Claire Geneser}
\affiliation{Mississippi State University, Department of Physics and Astronomy, 355 Lee Boulevard, Mississippi State, MS 39762}

\author{Angelle Tanner}
\affiliation{Mississippi State University, Department of Physics and Astronomy, 355 Lee Boulevard, Mississippi State, MS 39762}

\author[0000-0002-7030-9519]{Natalie M. Batalha}
\affiliation{Department of Astronomy and Astrophysics, University of California, Santa Cruz, CA 95060, USA}

\author[0000-0001-8189-0233]{Courtney Dressing}
\affiliation{501 Campbell Hall, University of California at Berkeley, Berkeley, CA 94720, USA}

\author[0000-0003-3504-5316]{Benjamin Fulton}
\affiliation{NASA Exoplanet Science Institute/Caltech-IPAC, MC 314-6, 1200 E. California Blvd., Pasadena, CA 91125, USA}

\author[0000-0001-8638-0320]{Andrew W. Howard}
\affiliation{Department of Astronomy, California Institute of Technology, Pasadena, CA 91125, USA}

\author[0000-0001-8832-4488]{Daniel Huber}
\affiliation{Institute for Astronomy, University of Hawai`i, 2680 Woodlawn Drive, Honolulu, HI 96822, USA}

\author[0000-0002-7084-0529]{Stephen R. Kane}
\affiliation{Department of Earth and Planetary Sciences, University of California, Riverside, CA 92521, USA}

\author[0000-0003-0967-2893]{Erik A. Petigura}
\affiliation{Department of Physics \& Astronomy, University of California Los Angeles, Los Angeles, CA 90095, USA}

\author[0000-0003-0149-9678]{Paul Robertson}
\affiliation{Department of Physics \& Astronomy, University of California Irvine, Irvine, CA 92697, USA}

\author[0000-0001-8127-5775]{Arpita Roy}
\affiliation{Space Telescope Science Institute, 3700 San Martin Drive, Baltimore, MD 21218, USA}
\affiliation{Department of Physics and Astronomy, Johns Hopkins University, 3400 N Charles St, Baltimore, MD 21218, USA}

\author[0000-0002-3725-3058]{Lauren M. Weiss}
\affiliation{University of Notre Dame, Notre Dame, IN 46556 USA}

\author[0000-0003-0012-9093]{Aida Behmard}
\altaffiliation{NSF Graduate Research Fellow}
\affiliation{Division of Geological and Planetary Science, California Institute of Technology, Pasadena, CA 91125, USA}

\author[0000-0001-7708-2364]{Corey Beard}
\affiliation{Department of Physics \& Astronomy, University of California Irvine, Irvine, CA 92697, USA}

\author[0000-0003-1125-2564]{Ashley Chontos}
\altaffiliation{NSF Graduate Research Fellow}
\affiliation{Institute for Astronomy, University of Hawai`i, 2680 Woodlawn Drive, Honolulu, HI 96822, USA}

\author[0000-0002-8958-0683]{Fei Dai}
\affiliation{California Institute of Technology, Division of Geological and Planetary Sciences, 1200 E California Blvd, Pasadena, CA, 91125, USA}

\author[0000-0002-4297-5506]{Paul A.\ Dalba}
\altaffiliation{NSF Astronomy and Astrophysics Postdoctoral Fellow}
\affiliation{Department of Astronomy and Astrophysics, University of California, Santa Cruz, CA 95064, USA}
\affiliation{Department of Earth and Planetary Sciences, University of California, Riverside, CA 92521, USA}

\author[0000-0002-3551-279X]{Tara Fetherolf}
\altaffiliation{UC Chancellor's Fellow}
\affiliation{Department of Earth and Planetary Sciences, University of California, Riverside, CA 92521, USA}

\author[0000-0002-8965-3969]{Steven Giacalone}
\affil{Department of Astronomy, University of California Berkeley, Berkeley, CA 94720, USA}

\author{Michelle L. Hill}
\affiliation{Department of Earth and Planetary Sciences, University of California, Riverside, CA 92521, USA}

\author{Lea A.\ Hirsch}
\affiliation{Kavli Institute for Particle Astrophysics and Cosmology, Stanford University, Stanford, CA 94305, USA}

\author[0000-0002-5034-9476]{Rae Holcomb}
\affiliation{Department of Physics \& Astronomy, University of California Irvine, Irvine, CA 92697, USA}

\author[0000-0001-8342-7736]{Jack Lubin}
\affiliation{Department of Physics \& Astronomy, University of California Irvine, Irvine, CA 92697, USA}

\author{Andrew Mayo}
\affil{Department of Astronomy, University of California Berkeley, Berkeley, CA 94720, USA}

\author[0000-0003-4603-556X]{Teo Mo\v{c}nik}
\affiliation{Gemini Observatory/NSF's NOIRLab, 670 N. A'ohoku Place, Hilo, HI 96720, USA}

\author[0000-0001-8898-8284]{Joseph M. Akana Murphy}
\altaffiliation{NSF Graduate Research Fellow}
\affiliation{Department of Astronomy and Astrophysics, University of California, Santa Cruz, CA 95064, USA}

\author{Lee J.\ Rosenthal}
\affiliation{Department of Astronomy, California Institute of Technology, Pasadena, CA 91125, USA}

\author[0000-0003-3856-3143]{Ryan A. Rubenzahl}
\altaffiliation{NSF Graduate Research Fellow}
\affiliation{Department of Astronomy, California Institute of Technology, Pasadena, CA 91125, USA}

\author[0000-0003-3623-7280]{Nicholas Scarsdale}
\affiliation{Department of Astronomy and Astrophysics, University of California, Santa Cruz, CA 95060, USA}

\author[0000-0003-2163-1437]{Christopher Stockdale}
\affiliation{Hazelwood Observatory, Australia}
\author[0000-0001-6588-9574]{Karen Collins}
\affiliation{Center for Astrophysics ${\rm \mid}$ Harvard {\rm \&} Smithsonian, 60 Garden Street, Cambridge, MA 02138, USA}
\author[0000-0001-5383-9393]{Ryan Cloutier}
\altaffiliation{Banting Fellow}
\affiliation{Center for Astrophysics ${\rm \mid}$ Harvard {\rm \&} Smithsonian, 60 Garden Street, Cambridge, MA 02138, USA}
\author{Howard Relles}
\affiliation{Center for Astrophysics ${\rm \mid}$ Harvard {\rm \&} Smithsonian, 60 Garden Street, Cambridge, MA 02138, USA} 

\author[0000-0001-5603-6895]{Thiam-Guan Tan}
\affiliation{Perth Exoplanet Survey Telescope, Perth, Western Australia}  
\affiliation{Curtin Institute of Radio Astronomy, Curtin University, Bentley, Western Australia 6102} 

\author{Nicholas J Scott}
\affiliation{NASA Ames Research Center, Moffett Field, CA 94035, USA} 
\author{Zach Hartman}
\affiliation{Gemini observatory, Maunakea Access Rd, Hilo, HI 96720, USA}
\author{Elisabeth Matthews }
\affiliation{Observatoire de l’Université de Genève, Chemin Pegasi 51,1290 Versoix, Switzerland}
\author[0000-0002-5741-3047]{David R. Ciardi} 
\affiliation{ Caltech/IPAC-NASA Exoplanet Science Institute, 770 S. Wilson Ave, Pasadena, CA 91106, USA}
\author{Erica Gonzales}
\affiliation{Department of Astronomy and Astrophysics, University of California, Santa Cruz, CA 95064, USA}
\author[0000-0001-7233-7508]{Rachel~A.~Matson}
\affiliation{U.S. Naval Observatory, Washington, D.C. 20392, USA}
\author{Charles Beichman}
\affiliation{Jet Propulsion Laboratory, California Institute of Technology, 4800 Oak Grove Drive, Pasadena, CA 91109, USA}
\affiliation{Caltech/IPAC-NASA Exoplanet Science Institute, 770 S. Wilson Ave, Pasadena, CA 91106, USA}
\author[0000-0001-6637-5401]{Allyson~Bieryla} 
\affiliation{Center for Astrophysics ${\rm \mid}$ Harvard {\rm \&} Smithsonian, 60 Garden Street, Cambridge, MA 02138, USA}

\author[0000-0001-9800-6248]{E. Furlan}
\affiliation{NASA Exoplanet Science Institute, Caltech/IPAC, Mail Code 100-22, 1200 E. California Blvd., Pasadena, CA 91125, USA}

\author[0000-0003-2519-6161]{Crystal~L.~Gnilka}
\affiliation{NASA Ames Research Center, Moffett Field, CA 94035, USA}

\author[0000-0002-2532-2853]{Steve~B.~Howell}
\affiliation{NASA Ames Research Center, Moffett Field, CA 94035, USA}

\author{Carl Ziegler}
\affiliation{Department of Physics, Engineering and Astronomy, Stephen F. Austin State University, 1936 North St, Nacogdoches, TX 75962, USA}

\author{C\'{e}sar Brice\~{n}o}
\affiliation{Cerro Tololo Inter-American Observatory, Casilla 603, La Serena, Chile}

\author{Nicholas Law}
\affiliation{Department of Physics and Astronomy, The University of North Carolina at Chapel Hill, Chapel Hill, NC 27599-3255, USA}

\author[0000-0003-3654-1602]{Andrew W. Mann}
\affiliation{Department of Physics and Astronomy, The University of North Carolina at Chapel Hill, Chapel Hill, NC 27599-3255, USA}
\author[0000-0003-2935-7196]{Markus Rabus}
\affiliation{Departamento de Matem\'atica y F\'isica Aplicadas, Facultad de Ingenier\'ia, Universidad Cat\'olica de la Sant\'isima Concepci\'on, Alonso de Rivera 2850, Concepci\'on, Chile}
\author[0000-0002-5099-8185]{Marshall C. Johnson}
\affiliation{Department of Astronomy, The Ohio State University, 4055 McPherson Laboratory, 140 West 18$^{\mathrm{th}}$ Ave., Columbus, OH 43210 USA}

\author{Jessie Christiansen}
\affiliation{ Infrared Processing and Analysis Center, Caltech, Pasadena CA 91125, USA}
\author[0000-0003-0514-1147]{Laura Kreidberg}
\affiliation{Max-Planck-Institut f\"ur Astronomie, K\"onigstuhl 17, D-69117 Heidelberg, Germany}
\author{David Anthony Berardo}
\affiliation{MIT Kavli Institute for Astrophysics and Space Research, 77 Massachusetts Avenue, Cambridge, MA 02139, USA}
\author{Drake Deming}
\affiliation{Department of Astronomy, University of Maryland, College Park, MD 20742-2421 USA}
\author{Varoujan Gorjian}
\affiliation{Jet Propulsion Laboratory, California Institute of Technology, 4800 Oak Grove Dr., MS 306-392, Pasadena, CA 91109, USA}
\author{Farisa Y. Morales}
\affiliation{Jet Propulsion Laboratory, California Institute of Technology, 4800 Oak Grove Drive, Pasadena, CA 91109, USA}
\author{Björn Benneke}
\affiliation{Department of Physics and Institute for Research on Exoplanets, Universit´e de Montr´eal, Montreal, QC, Canada}
\author{Diana Dragomir}
\affiliation{Department of Physics and Astronomy, University of New Mexico, 210 Yale Blvd NE, Albuquerque, NM 87106, USA}

\author[0000-0001-9957-9304]{Robert A. Wittenmyer}
\affiliation{University of Southern Queensland, Centre for Astrophysics, West Street, Toowoomba, QLD 4350 Australia}
\author{Sarah Ballard}
\affiliation{Department of Astronomy, University of Florida, 211 Bryant Space Science Center, Gainesville, FL, 32611, USA}
\author{Brendan P. Bowler}
\affiliation{Department of Astronomy, The University of Texas at Austin, Austin,TX 78712, USA}
\author[0000-0002-1160-7970]{Jonathan Horner}
\affiliation{University of Southern Queensland, Centre for Astrophysics, West Street, Toowoomba, QLD 4350 Australia}

\author{John Kielkopf}
\affiliation{Department of Physics and Astronomy, University of Louisville, Louisville, KY 40292, USA}
\author{Huigen Liu}
\affiliation{School of Astronomy and Space Science, Key Laboratory of Modern Astronomy and Astrophysics in Ministry of Education, Nanjing University, Nanjing 210046, Jiangsu, China}
\author{Avi Shporer}
\affiliation{Department of Physics and Kavli Institute for Astrophysics and Space Research, Massachusetts Institute of Technology, Cambridge, MA 02139, USA}
\author{C.G. Tinney}
\affiliation{Exoplanetary Science at UNSW, School of Physics, UNSW Sydney, NSW 2052, Australia}
\author{Hui Zhang}
\affiliation{Shanghai Astronomical Observatory, Chinese Academy of Sciences, Shanghai 200030, China}
\author[0000-0001-7294-5386]{Duncan J. Wright}
\affiliation{University of Southern Queensland, Centre for Astrophysics, West Street, Toowoomba, QLD 4350 Australia}
\author[0000-0003-3216-0626]{Brett C. Addison}
\affiliation{University of Southern Queensland, Centre for Astrophysics, West Street, Toowoomba, QLD 4350 Australia}
\author{Matthew W. Mengel}
\affiliation{University of Southern Queensland, Centre for Astrophysics, West Street, Toowoomba, QLD 4350 Australia}
\author{Jack Okumura}
\altaffiliation{University of Southern Queensland, Centre for Astrophysics, West Street, Toowoomba, QLD 4350 Australia}

\begin{abstract} 

We validate the presence of a two-planet system orbiting the 0.15--1.4 Gyr K4 dwarf TOI 560 (HD 73583). The system consists of an inner moderately eccentric transiting  mini-Neptune (TOI 560 b, $P = 6.3980661^{+0.0000095}_{-0.0000097}$ days, $e=0.294^{+0.13}_{-0.062}$, $M= 0.94^{+0.31}_{-0.23}M_{Nep}$) initially discovered in the Sector 8 \tess\ mission observations, and a transiting mini-Neptune (TOI 560 c, $P = 18.8805^{+0.0024}_{-0.0011}$ days, $M= 1.32^{+0.29}_{-0.32}M_{Nep}$) discovered in the Sector 34 observations, in a rare near-1:3 orbital resonance. We utilize photometric data from \tess\, \textit{Spitzer}, and ground-based follow-up observations to confirm the ephemerides and period of the transiting planets, vet false positive scenarios, and detect the photo-eccentric effect for TOI 560 b. We obtain follow-up spectroscopy and corresponding precise radial velocities (RVs) with the iSHELL spectrograph at the NASA Infrared Telescope Facility and the HIRES Spectrograph at Keck Observatory to validate the planetary nature of these signals, which we combine with published PFS RVs from Magellan Observatory. We detect the masses of both planets at $> 3-\sigma$ significance. We apply a Gaussian process (GP) model to the \tess\ light curves to place priors on a chromatic radial velocity GP model to constrain the stellar activity of the TOI 560 host star, and confirm a strong wavelength dependence for the stellar activity demonstrating the ability of NIR RVs in mitigating stellar activity for young K dwarfs. TOI 560 is a nearby moderately young multi-planet system with two planets suitable for atmospheric characterization with James Webb Space Telescope (JWST) and other upcoming missions. In particular, it will undergo six transit pairs separated by $<$6 hours before June 2027.
\end{abstract}

\keywords{Infrared: stars, methods: data analysis, stars: individual (TOI 560), techniques: radial velocities}

\section{Introduction}   
\label{sect:intro} 

The \kepler\ mission, after its launch in 2009, discovered over 4000 transiting exoplanet candidates \citep{Borucki2011,Howard2012,2021AJ....161...36B}. However, many of these planets orbited relatively faint 14th-16th visual magnitude stars that were difficult to follow-up, validate and confirm \citep{2015arXiv150301770P,2014ApJ...783L...6W,2016ApJ...825...19W,2015ApJ...801...41R,2014ApJS..210...20M}. The NASA Transiting Exoplanet Survey Satellite (\tess) mission launched in 2018 to detect and identify relatively nearby and brighter transiting exoplanets, particularly those orbiting smaller and cooler M dwarf stars which are the most abundant spectral type, and are known to host compact multi-planet systems \citep{Ricker2015,Dressing_2015,Howard2012}. To date, the \tess\ mission has expanded the pool of nearby, bright transiting exoplanet candidates considerably, with over 4,000 candidates identified. Many are amenable to ground-based follow-up and characterization \citep{Ricker2015}.

\kepler\ has shown that compact multi-planet Neptune and terrestrial planets are more commonly found to orbit M dwarf stars \citep{Dressing_2015,Howard2012}.  Additionally, the relatively small size of M dwarfs in comparison to transiting exoplanets leads to larger, easier-to-detect transit depths, and M dwarfs are the most common spectral types in the Milky Way \citep{Chabrier_2003, Henry_2006}. The relatively nearby M dwarf exoplanet discoveries from \tess\ are therefore some of the most suitable targets for exoplanet atmospheric characterization with ground-based and future space-based facilities such as JWST \citep{Kempton_2018}. Before jumping to atmospheric characterization, however, the \tess\ candidates need further supporting observations to validate and confirm that they are not false-positives. \tess\ pixels are \SI{21}{\arcsecond} in size on the sky, a relatively large value compared to typical ground-based seeing-limited angular resolution of $\sim$1'', a design choice for the \tess\ mission to optimize its cameras' fields of view \citep{Ricker2015}. As a result, the light from nearby main sequence stars can be blended with fainter visual eclipsing binaries, and produce false-positives, especially when only a singly transiting planet is identified in a 27-day \tess\ sector of observations at lower ecliptic latitudes away from the ecliptic poles \citep{Ricker2015,Cale2021,vanderburg2019,rodriguez2020,hobson2021,addison2021,osborn2021,dreizler2020,brahm2020,nowak2020,teske2020,sha2021,gan2021,bluhm2020}.  

Ground-based panchromatic light curves, high-resolution imaging with high contrasts, and spectroscopic follow-up can search for close visual companions to the candidate host star to validate that the target is not a false positive from a nearby, background, blended, or grazing eclipsing binary, and to ensure that the planetary radius is unbiased by additional flux sources within the \tess\ aperture. Additional light curve analyses such as EDI-Vetter Unplugged \citep{Zink_2020} can check for neighboring flux contamination from nearby stars in the sky, examine the abundance of outliers from the transit model, and consider any signal variations between even and odd transits and the presence of any secondary eclipses. The Discovery and Vetting of Exoplanets \citep[DAVE,][]{Kostov_2019} tool evaluates \tess\ light curve-based diagnostics such as photocenter motion, odd-even transit-depth consistency, searches for secondary eclipses, and other sources of likely false-positive scenarios.

Constraining and/or measuring the masses of transiting planets with precise radial velocities (RVs) derived from high-resolution spectroscopy further enables validation and confirmation of exoplanet candidates, while also providing constraints on the planet bulk densities \citep{2007ApJ...669.1279S,2010ApJ...712..974R,2019PNAS..116.9723Z}. However, RV analysis can be complicated from stellar photospheric activity from young and/or active host stars that can produce RV signals comparable in amplitude of the Keplerian signals of interest. Stellar surface inhomogeneities (e.g., cool spots, hot plages) driven by the dynamic stellar magnetic field rotate in and out of view, leading to photometric variations over time. The presence of such active regions breaks the symmetry between the approaching and receding limbs of the star, introducing apparent RV variations over time as well \citep{Desort2007}. These active regions further affect the integrated convective blue-shift over the stellar disk, and will therefore manifest as an additional net red- or blue-shift \citep{Meunir2013,Dumusque2014}. Various techniques have been introduced to lift the degeneracy between activity- and planetary-induced signals in RV datasets such as line-by-line analyses \citep{Dumusque2018,Wise2018,Cretignier2020} and Gaussian process (GP) modeling \citep[e.g.][]{Haywood2014,Grunblatt2015}, but such measurements remain challenging due to the sparse cadence of typical RV datasets compared to the activity timescales. 

This validation and confirmation process can be greatly aided when multiple transiting planets orbiting the same star are detected, as it is much more difficult to contrive a false-positive scenario that mimics two or more transiting planets at different orbital periods \citep{Lissauer_2012,2014ApJ...784...45R}. The probability of randomly finding two background eclipsing binaries in the same \tess\ pixel is vanishingly small ($\ll$1\%).  However, \kepler\ did find one single example -- KOI-284 -- out of $\sim$190,000 target stars, consisting of two stars in a binary, one hosting a single transiting planet and the other hosting two transiting planets in the same \kepler\ postage stamp. This ``false positive'' scenario was uncovered with dynamical stability analysis due to the very similar orbital periods for two of the planets \citep{Lissauer_2012}. Furthermore, multi-planet systems can potentially have ``double transits'', making possible the optimal use of limited telescope resources for atmospheric characterization such as the James Webb Space Telescope \citep[\jwst\ , e.g.][]{Bean2018,Lissauer_2011}, and also permitting differential exoplanet characterization \citep[e.g., ][]{Ciardi2013,Weiss2018,Weiss2020}. Multi-planet systems are also of interest for studying dynamical interactions between planets, and inferring population-level information on their formation and evolution \citep[e.g.,][]{Lissauer2007,Zhu2012,Anglada-Escude2013,Mills&Mazeh2017,Morales2019}. 

The discovery of a young multi-planet system can be especially valuable for improving our understanding of exoplanet formation and evolution, both the demographics of the exoplanet architectures, and the atmospheric evolution as a function of stellar age, orbital period, and host star spectral type \citep{Fulton2017,Marchwinski2015,Newton2019,Ilin2022,Klein2022,Flagg2022,Feinstein2022,Cohen2022,Alvarado2022,Benatti2021,Carolan2020,Hirano2020}. We know that the orbital properties of planets change over time as shown by the existence of hot Jupiters; we also see evidence for orbital distance dependent mass loss via photoevaporation which may be responsible for producing the planet-radius gap and may impact our understanding of the occurrence rate of terrestrial planets at larger orbital separations \citep{Klein2020,Mann2020,Plavchan2020,Pascucci2019}. Therefore, we would benefit from examining the orbital architecture and atmospheric properties of multi-planetary systems over a wide range of evolutionary phases.  Multi-planet systems in particular enable a differential comparison within the system between their atmospheric properties, and/or escape from, because any differences will be isolated to differences in the planet mass and orbital separation / irradiation at a common age and spectral type for the host star. For example, the Kepler-51 system ($\sim$530 Myr) of three planets contains not only the least dense planets ever discovered but they also lie in a close resonant chain of 1:2:3 \citep{Nava2020}. The existence of this system as well as others around more mature, compact multi-transiting systems such as Kepler-89 \citep{Zechmeister2020} hints at a formation mechanism which includes convergent disk migration and resonance capture \citep{David2019,Walkowicz2013}. An even younger multi-planet system, V1298 Tau, also hints at being a 1:2:3 resonant chain based on transit data \citep{dreizler2020}. TTV observations of AU Mic have led to the tentative hypothesis of a third candidate planet, d, which may complete a resonant chain this time with a 4:6:9 configuration \citep{Wittrock2022}. Both the v1298 and AU Mic planetary systems show unexpectedly higher densities for some of the planets in the system at very young ages, in contrast to the Kepler-51 system \citep{Maggio,Tejada,Zicher,Cale2021}. Already, there are complexities in the architecture of systems over time.

In this work, we identify a two-planet transiting system orbiting the $\sim$0.5 Gyr host star TOI 560 (HD 73583 \& GAIA EDR3 5746824674801810816; Table \ref{tab:star}), which we validate with ground-based photometry, high-resolution imaging, and optical and near-infrared RVs. The first planet candidate was identified in Sector 8 observations of the \tess\ mission, and we identify in this work the second candidate with the release of the Sector 34 \tess\ light curve. The host star TOI 560 is active, and we present a chromatic RV analysis to characterize the stellar activity jointly with $> 3-\sigma$ detections of the exoplanet dynamical masses.

This paper is organized as follows: In Section \ref{sect:observations} we present an overview of our observations. In Section \ref{sect:analysis} we present our analysis of the \textit{TESS} light curve, one ground-based transit, high-contrast imaging, and RVs from the iSHELL, PFS and HIRES spectrometers to validate the planetary nature of the transit signals. In Section \ref{sect:resu} we present the results of our analysis of the \tess\ light curve and RVs.  In Section \ref{sect:discussion} we consider and simulate the dynamical stability of the TOI 560 system, discuss the chromatic RV analysis of the stellar activity, and perfom a search for additional RV companions. In Section \ref{sect:conclusions} we present our conclusions and future work.

\begin{table}
    \centering
    \begin{tabular}{ccc}
        \hline
        Parameter & Value & Reference \\
        \hline
        & \textit{Identifiers:} & \\
        TIC & 101011575 & S19 \\
        TOI & 560 & G21 \\
        HIP & 42401 & S07 \\
        2MASS & J08384526-1315240 & S06 \\ 
        Gaia DR2 \& EDR3 & 5746824674801810816 & Gaia \\
         \hline 
        & \textit{Coordinates and Distance:} & \\
        $\alpha$ & 08:38:45.260 & S19 \\
        $\delta$ & -13:15:24.09 & S19 \\
        Distance [pc] & $31.5666 \pm 0.03205$ & S19 \\
        Parallax [mas] & $31.657 \pm 0.0152$ &  Gaia \\
        $\mu_{\alpha}$ cos $\delta$ [mas y $r^{-1}$] & -63.8583 $\pm$ 0.0505  & Gaia  \\
        $\mu_{\sigma}$ [mas y r$^{-1}$] & 38.3741 $\pm$ 0.0406 & Gaia\\ 
        Absolute RV [km $s^{-1}$] & 21.52 & this work$^{^a}$\\
        \hline
        & \textit{Physical Properties:} & \\
         Age [Gyr] & 0.15 - 1.4 & this work${^b}$ \\  
        M$_{*}$ (M$_{\odot}$)& $0.702^{+0.026}_{-0.025}$ & this work${^c}$ \\
        R$_{*}$(M$_{\odot}$)& $0.677\pm 0.017$ & this work${^c}$ \\
        T$_{eff}$ [K]& $4582^{+64}_{-62}$ & this work${^c}$ \\ 
        log g [cgs]& $4.623^{+0.025}_{-0.024}$ & this work${^c}$ \\ 
        Spectral Type & K4 & S05\\
        $v sin i$ [km s$^{-1}$] & \textless 3& this work${^d}$\\ 
        P$_{rot}$ [days] & 12$\pm$0.1 & this work${^e}$\\ 
        $\rho$ [g cm$^{-3}$] & $3.17 \pm 0.23$ & S19\\ 
        Luminosity [L$_\odot$] & $0.1802 \pm 0.0058$ & S19\\ 
        \hline
        & \textit{Magnitudes:} & \\
        TESS [mag] & $8.59 \pm 0.01$ & S19 \\
        B [mag] & $10.74 \pm 0.07$ & S19 \\
        V [mag] & $9.67 \pm  0.03$ & S19 \\
        Gaia G [mag]& $9.270 \pm 0.005$ & G18 \\
        Gaia BP [mag] & $9.905 \pm 0.001$ & G18 \\
        Gaia RP [mag] & $8.546 \pm 0.002$ & G18 \\
        J [mag]& $7.65 \pm 0.03$ & S06 \\
        H [mag]& $7.09 \pm 0.05$ & S06 \\
        K [mag]& $6.95 \pm 0.02$ & S06 \\
        WISE 1 [mag] & $6.850 \pm 0.037$ & W10 \\
        WISE 2 [mag] & $6.963 \pm 0.021$ & W10 \\
        WISE 3 [mag] & $6.921 \pm 0.017$ & W10 \\
        WISE 4 [mag] & $6.723 \pm 0.084$ & W10 \\
        \hline
    \end{tabular}
    \caption{Stellar parameters of TOI 560. 
    References: 
    G18: \citep{Gaia_Collaboration_2018}, G21: \citep{Guerrero_2021},
    S19: \citep{Stassun_2019}, S07: \citep{vanLeeuwen2007}, S06: \citep{Skrutskie_2006}, S05: \citep{Scholz_2005}, W10: \citep{Wright_2010} \\ ${^a}$ the average of the two observations in Table \ref{tab:starr-1}.\\ ${^b}$ estimated in Section \ref{WASP_period}. \\${^c}$ from our ExoFASTv2 isochrone-based jointed transit light curve analysis in Table \ref{table:long1}; consistent results are obtained in our reconnaissance spectroscopy and broadband SED modeling.\\${^d}$ calculated from the rotation period and stellar radius, assuming the stellar rotation axis is rotating in or near the plane of the sky.\\${^e}$ Section \ref{WASP_period} and Figure \ref{fig:WASP}.}
    
    \label{tab:star}
\end{table}
\section{Observations}  
\label{sect:observations}

In this Section, we present an overview of all observational data used throughout this analysis. \textit{TESS} photometric light curve data and ground-based photometry are presented in ${\S}$\ref{photometry}, a brief description of reconnaissance spectroscopy is summarized in ${\S}$\ref{recon}, high resolution imaging observations are presented in ${\S}$\ref{high}, and RV observations are detailed in ${\S}$\ref{radial}.   

\subsection{Photometric Light Curves} 
\label{photometry} 

Herein, we present space and ground-based light curves of the TOI 560 system, respectively, in the following two subsections, the latter of which are summarized in Table \ref{tab:photom}.

\begin{table*}
    \centering
    \begin{tabular}{llrrrrrrrrl}
        \hline
        UT Date & Site & Aperture & Filter & $N_{exp}$ & $t_{exp}$ & Duration & Pixel Scale & FWHM & Aperture & Transit\\
        (YYYYMMDD) &  & (m) &  & & (s) & (min) & ($''$/px) & ($''$) & ($''$) & Coverage \\
        \hline
        20190427 & LCO-SSO & 1 & B & 213 & 30 & 151 & 0.389 & 5.16 & 7.4,12.4 & Egress \\  
        20190427 & LCO-SSO & 1 & zs & 84 & 30 & 79 & 0.389 & 4.19 & 7.4,12.4 & Egress \\
        20190503 & LCO-SSO & 1 & zs  & 59 & 30 & 55 & 0.389 & 2.42 & 6.6,10.9 & Ingress \\
        20191207 & NGST & 4x0.2 & NGTS & 4840 & 12 & 260 & 4.97 & 20 & 40 & full \\
        20191207 & LCO-CTIO & 1 & B & 211 & 15 & 120 & 0.389 & 1.93 & 7.4,12.4 & Ingress \\  
        20191213 & LCO-SSO & 1 & B & 342 & 15 & 256 & 0.389 & 3.56 & 6.24 & Full \\
        20200114 & PEST & 0.3 & Rc & 482 & 30 & 357 & 1.23 & 4.93 & 7.4 & Full \\
        20200127 & LCO-HAL & 0.4 & zs & 363 & 30 & 319 & 0.57 & 10.7 & 11.4 & Full \\
        20200127 & LCO-SSO & 0.4 & zs & 288 & 30 & 247 & 0.57 & 13.3 & 12.5 & Full \\
        20200202 & LCO-SAAO & 1 & zs & 295 & 30 & 308 & 0.389 & 8.8 & 10 & Full \\
        20200331 & LCO-SSO & 1 & B & 281 & 30 & 297 & 0.389 & 5.25 & 6.6 & Full \\
        \hline
\end{tabular}
    \caption{Summary of available information from ExoFOP-TESS for the ground-based light curve observations of TOI 560 b ; Site Abbreviations: LCO-CTIO (Cerro Tololo, Chile), LCO-SAAO (South Africa), LCO-SSO (Siding Springs, Australia), LCO-HAL (Haleakala, Hawaii, USA), NGST (Paranal, Chile)}
    \label{tab:photom}
\end{table*}

\subsubsection{\tess\ Observations}

TOI 560 (TIC 101011575; HD 75383; GAIA EDR3 5746824674801810816) was observed first in \textit{TESS} Sector 8 from UT February 2 2019 to UT February 28 2019, then again in Sector 34 during the \textit{TESS} extended mission from UT January 13 2021 to UT February 9 2021.  Stellar properties are listed in Table \ref{tab:star}.  TOI 560 is in the Hydra constellation and is relatively nearby  (31.6 pc) and bright (V=9.67 mag) making it an ideal candidate for study by \textit{TESS}. The data collection pipeline was developed by the \textit{TESS} Science Processing Operations Center \citep[SPOC,][]{Jenkins_2016} and uses a wavelet-based matched filter \citep{2002ApJ...575..493J,2010SPIE.7740E..0DJ,2020TPSkdph}. One transit signal was detected during the Sector 8 observations, and fitted with a limb-darkened transit model \citep{Li:DVmodelFit2019} with various vetting and diagnostic tests \citep{Twicken:DVdiagnostics2018} and labeled TOI 560 b  \citep{Guerrero_2021}, which we hereafter refer to as TOI 560 b. This prompted us to explore the target further with radial velocity measurements (${\S}$\ref{radial}) to verify the \textit{TESS} candidate and perhaps uncover additional companions in the system.  With the release of the Sector 34 light curve, we independently identified by eye a second transiting candidate in the TOI 560 system, which was vetted and adopted as a TOI by the \tess\ mission as TOI 560 c, which hereafter we refer to as TOI 560 c.

\begin{figure*}
    \centering
    \includegraphics[width=.45\textwidth]
    {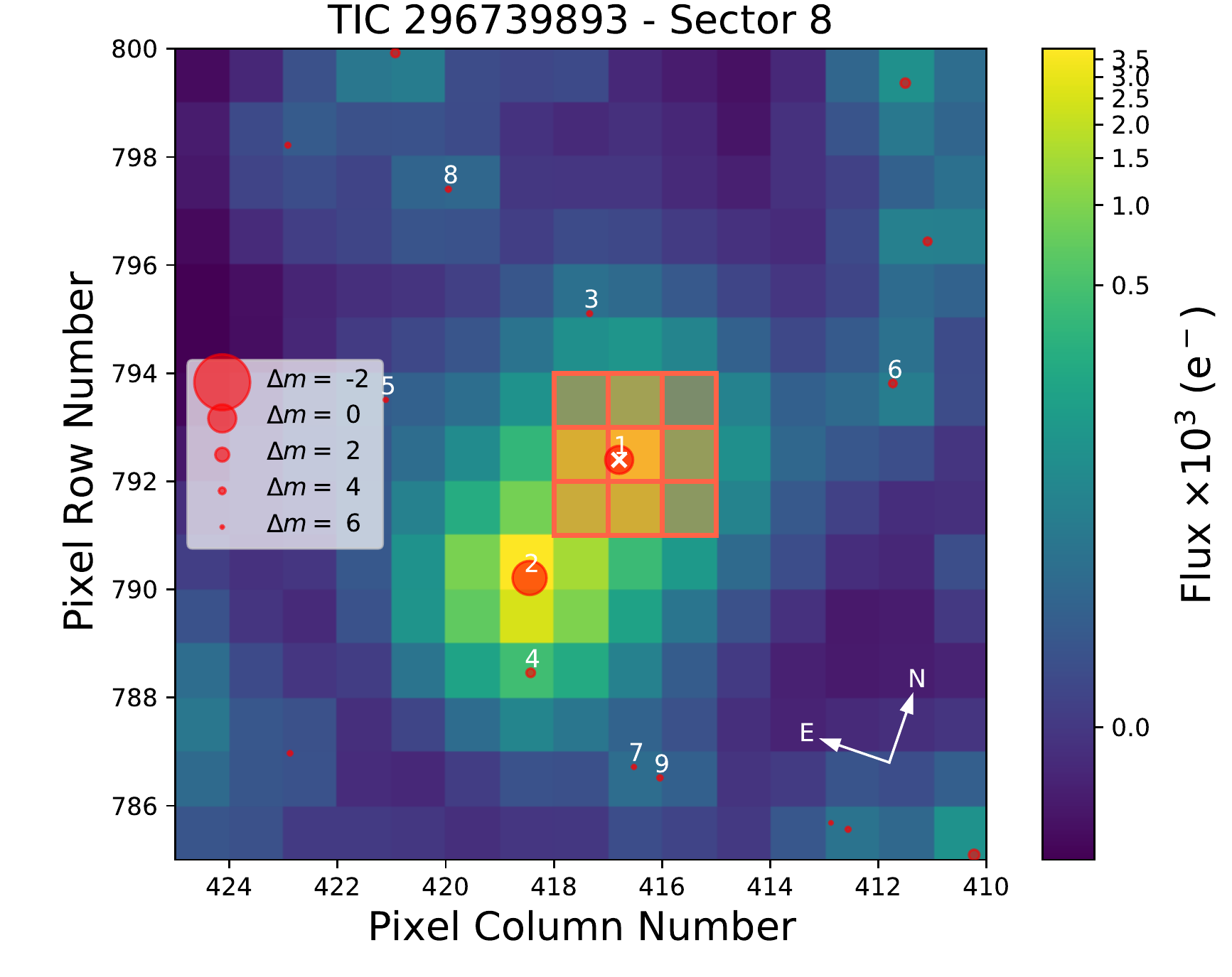}
    \includegraphics[width=.45\textwidth]
    {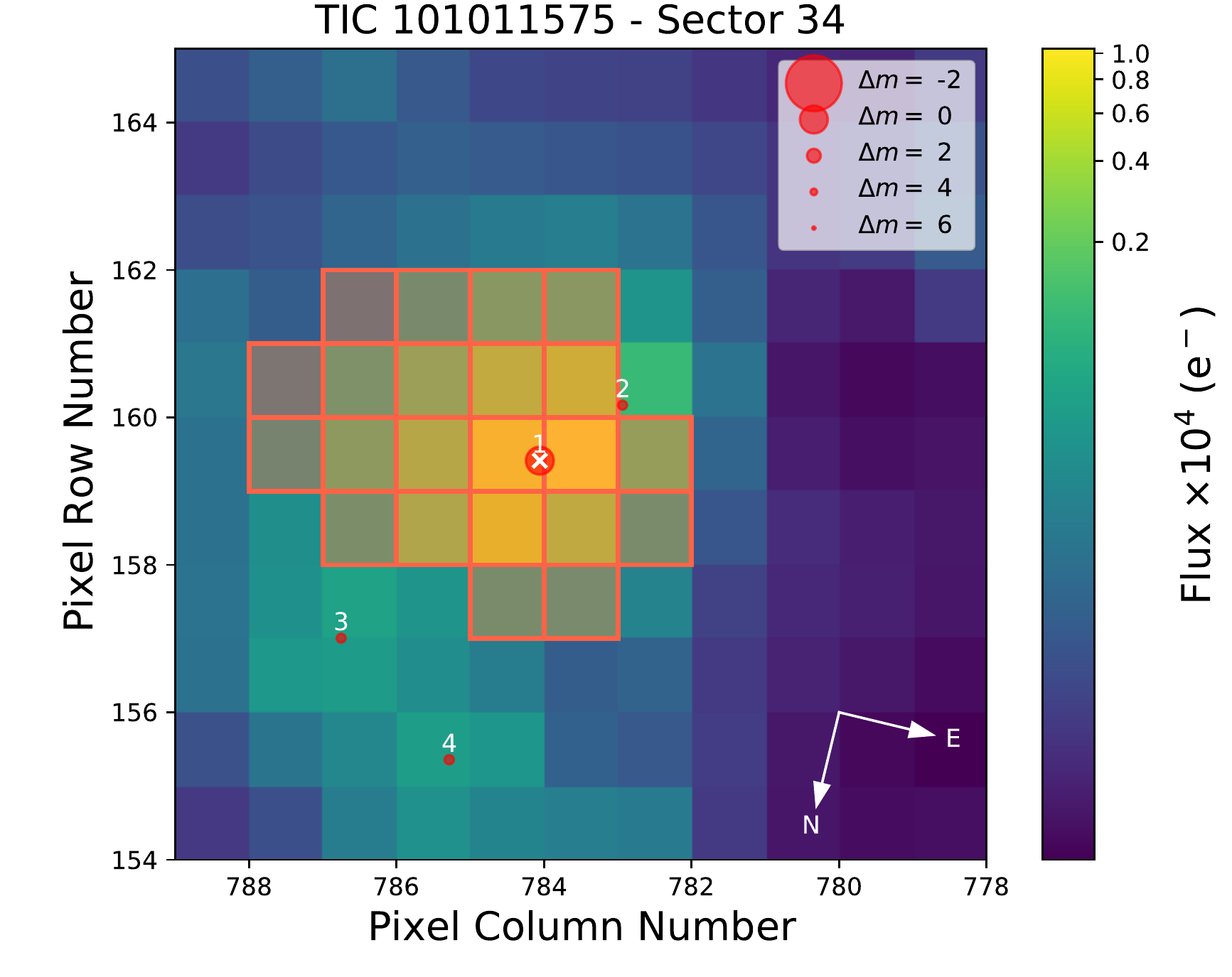}
    \caption{TESS target pixel file (TPF) data from sector 8 (left) and sector 34 (right) for TOI 560, created with \texttt{tpfplotter} \citep{Aller_2020}.  The pixels shown outlined in orange were the ones used to extract the light curve, while point sources from the Gaia DR2 catalog are labeled in red, with sizes in accordance to their relative magnitude from the target star.}
    \label{fig:tpfs1}
\end{figure*}

In this work, we specifically analyze the detrended Presearch Data Conditioning Simple Aperture Photometry (PDC-SAP) light curves for Sectors 8 and 34 \citep{Smith_2012, Stumpe_2012, Stumpe_2014} obtained from the Mikulski Archive for Space Telescopes (MAST)\footnote{\href{https://mast.stsci.edu/portal/Mashup/Clients/Mast/Portal.html}{https://mast.stsci.edu/portal/Mashup/Clients/Mast/Portal.html}}.  After gathering this data, we normalize the detrended flux in $e^- s^{-1}$ so that the median for each sector is unity. In Figure \ref{fig:tpfs1}, we show the \tess\ target pixel files (TPF) around the target star in Sectors 8 and 34, where orange outlines show the pixels used to create the TESS light curve.  There are other point sources (notated by circular red points) that are located within the TESS apertures, so these needed to be subtracted from the SAP flux when creating the PDC-SAP flux.

\subsubsection{\textit{Spitzer} Light Curve}  

TOI 560 b was observed with \textit{Spitzer} on 2019 August 20, using Director's Discretionary Time \citep{Crossfield2018}. A single transit was observed using the 4.5~$\mu$m channel \citep[IRAC2,][]{Fazio2004} in subarray mode with an integration time of 0.36 seconds. The transit observation spanned 6 hr 15 min totaling 823 frames with short observations taken before and after transit to check for bad pixels. Peak-Up mode was used to place the star as close as possible to the well-characterized ``sweet spot'' of the detector.

To extract photometry from the \textit{Spitzer} observations, we use the Photometry for Orbits Eclipses and Transits (\texttt{POET} \footnote{\url{https://github.com/kevin218/POET}}) package \citep{Cubillos2013,May2020}. In summary, \texttt{POET} creates a bad pixel mask and discards bad pixels based on the \textit{Spitzer} Basic Calibrated Data (BCD). Outlier pixels are also discarded using sigma-rejection. Then, the center of the point spread function (PSF) is determined using a 2-D Gaussian fitting technique. After the center of the PSF is found, the lightcurve is extracted using aperture photometry in combination with a BiLinearly-Interpolated Subpixel Sensitivity (BLISS) map described in \cite{Stevenson2012}. The resulting data are then simultaneously fit with a model that accounts for both the lightcurve itself and a temporal ramp-like trend attributed to ``charge trapping''. The posterior distribution is sampled using an MCMC algorithm with chains initialized at the best fit values. 

Aperture photometry was performed with various aperture sizes (ranging from 2 - 6 pixels in increments of 1 pixel). Visual inspection of the raw data indicated the temporal ramp model was likely either linear or constant. The optimal aperture size was found to be 3 pixels as this size returned the lowest standard deviation of the normalized residuals (SDNR) for both ramp models. We then tested bin sizes of 0.1, 0.03, 0.01, and 0.003 pixels square for the BLISS map finding that a bin size of 0.01 minimized the SDNR. Both the linear ramp and constant models gave similar SDNR for these tests so we chose the constant model, the simplest of the two. 

\subsubsection{Las Cumbres Observatory Global network of Telescopes (LCOGT)}  

In 2019 and 2020, seven partial or full transit observations of TOI 560 b were collected with five observatory sites from the Las Cumbres Observatory Global network of Telescopes \citep[LCOGT][]{Brown_2013}. These observations are summarized in Table \ref{tab:photom}. Data were collected in either the B or Sloan $z'$ filters with exposure times of 15 or 30 seconds to check for transit depth chromaticity such as would be produced by a false positive eclipsing binary scenario. For one transit (UT 2019/04/27), data were collected simultaneously in both filters.  For a second transit (UT 2020/01/27), data was collected simultaenously in the same filter from two sites. The other five transits were observed in a single band at a single telescope/site. Data were analyzed using AIJ, and the data reduction was done using the BANZAI LCOGT facility pipeline \citep{McCully2018}. 

\subsubsection{The Perth Exoplanet Survey Telescope (PEST)}  
The Perth Exoplanet Survey Telescope (PEST) near Perth, Australia is a 0.3m telescope equipped with a 1530 × 1020 SBIG ST-8XME camera with an image scale of $\angd{;;1.2}\ /{\rm pixel}$, resulting in a 31$''$ × 21$''$ field of view. Observations of a transit of TOI 560 b  were obtained on UT 2020 January 14  in the Rc filter with an exposure time of 30 seconds.  A custom pipeline based on C-Munipack was used to calibrate the images and extract differential photometry. 

\subsubsection{NGTS / Paranal}

On UT 2019 December 07 the TOI 560 system was observed by the Next Generation Transit Survey \citep[NGTS;][]{Wheatly2018ngts} using 4 telescopes \citep[see][]{bryant2020multicam}, consist of 0.2\,m diameter robotic telescopes, located at ESO's Paranal Observatory, Chile. A custom NGTS filter in wavelength range 520-890 nm was used. A transit was detected and confirmed for planet b. The NGTS data were reduced using a custom aperture photometry pipeline version 2, which performs source extraction and photometry using the \textsc{SEP} Python library \citep{bertin96sextractor, Barbary2016} and is detailed in \citep{bryant2020multicam}. The pipeline uses \gaia\ DR2 \citep{Barbary2016, Gaia_Collaboration_2018} to automatically identify comparison stars which are similar in brightness, colour, and CCD position to TOI 560.

\subsubsection{WASP Light Curves} 

The SuperWASP team monitored TOI 560 for four seasons from 2009-2012 inclusive as part of a larger all-sky survey \citep{2006PASP..118.1407P}. Data were reduced and light curves generated using the standard SuperWASP analyses \citep{2011PASP..123..547M}.

\subsection{Recon Spectroscopy}
\label{recon}
The purpose of reconnaissance spectroscopy is to provide spectroscopic parameters that will more precisely constrain the properties of planet host stars, to detect false positives caused by spectroscopic binaries as manifested by large (e.g. $>$1 km/s) RV changes and/or line-doubling in multi-epoch spectroscopy, and to identify stars unsuitable for precise RV measurements, such as rapid rotators.

\subsubsection{NRES} 
Reconnaissance spectroscopy for TOI 560 was obtained with the Las Cumbres Observatory (LCOGT) Network of Robotic Echelle Spectrographs (NRES). NRES is a fiber-fed echelle spectrograph at several sites mounted on the respective LCOGT 1-m telescopes with wavelength range 380-860 nm and spectral resolution $R \sim53000$. Observations were executed at the Sutherland Observatory, South Africa, and the McDonald Observatory, USA, on the nights 2019 November 04, 2019 October 29, and 2019 May 12. We took dark, bias, flat, and arc lamp calibration images at the beginning of each night. The target was centered on one fiber, and the second fiber was used to observe a ThAr calibration lamp simultaneously. On each night, we took 3 consecutive exposures with 480 seconds on each individual spectrum. SNR for each night were 33, 30 and 33 respectively. We obtained the wavelength-calibrated spectra using the NRES commissioning \texttt{IDL} pipeline and obtained improved and stacked spectra for each night using the Stage2 IDL pipeline with a total integration time of 1440s and a resulting SNR of $\sim$35. 

\subsubsection{TRES}
The Tillinghast Reflector Echelle Spectrograph \citep[TRES;][]{Furesz2008} team obtained two reconnaissance (median SNR=33.25) spectra of TOI 560 near opposite quadratures of TOI 560 b (phases 0.25 and 0.72 on the initial Sector 8 released ephemeris) at 2019 April 16 04:37 and 2019 April 19 03:49 UT. TRES has a resolving power R $\sim$ 44,000 and the wavelength range of the spectrograph is 390-910 nm. The Stellar Parameter Classification (SPC) tool specifically uses a $\sim$310 Angstrom region of the spectrum ($\sim$5050-5360 Angstroms) to derive stellar parameters (${\S}$\ref{sect:recon}). Spectra are processed following methods described in \cite{Buchhave2010}. 

\subsection{High Resolution Imaging}
\label{high} 

High-resolution imaging is used to obtain images of faint companions located near bright target stars. Typical targets are stars with companions including exoplanets and circumstellar structures and disks of gas and dust. In seeing-limited imaging, a faint companion could be swamped and lost in the noise of scattered light in the point spread function (PSF) wings of the target star. High contrast imaging methods -- speckle imaging and adaptive optics -- were designed to mitigate the impact of target star scattered light at the position of the companion, in order to make the companion detectable against the residual PSF noise \cite[e.g.][]{2005ApJ...629..592G,2014ApJ...792...97M,2007ApJ...660..770L,2006ApJ...641..556M}. 

High-contrast imaging of \tess\ candidates is crucial to ensure that the target is not a false positive from a background eclipsing binary, and to ensure that the planetary radius is unbiased by additional, unidentified, flux sources within the aperture. Spatially close stellar companions can create a false- positive transit signal if, for example, the fainter star is an eclipsing binary (EB). However, even more troublesome is ``third-light'' flux contamination from a close companion (bound or line of sight) which can lead to underestimated derived planetary radii  if not accounted for in the transit model \citep[e.g.,][]{Ciardi_2015} and even cause total non-detection of small planets residing within the same exoplanetary system \citep{Lester2021}. Given that close bound companion stars exist in nearly one-half of FGK type stars \citep{Matson2018} high-resolution imaging provides crucial information toward our understanding of exoplanetary formation, dynamics and evolution \citep{Howell2021}.  Herein we present high-contrast imaging of TOI 560 obtained with Gemini with NIRI and Zorro and the SOAR telescope respectively.

\subsubsection{Gemini North / NIRI}
We searched for close visual companions to the target using high resolution imaging with both AO and speckle imaging at Gemini. The AO and speckle images are highly complementary: the speckle images reach higher resolutions in the optical, while the AO images reach deeper sensitivities beyond a few hundred mas in the near-IR, and are therefore more sensitive to more widely separated low-mass stars. We collected high-resolution AO images of TOI 560 with Gemini/NIRI \citep{hodapp2003} on 2019 May 26. Given that the star is bright in the K band, we collected images with individual exposure time 0.9s with the Br$\gamma$ filter (2.166$\mu$m), to avoid saturating the detector. Our sequence consisted of nine such images, with the telescope dithered in a grid pattern between each frame, and we also collected daytime flats. The science images themselves can be used to construct a sky background frame, by median combining the dithered images.

\subsubsection{Gemini South / Zorro} 

TOI 560 was observed twice on 2020 March 16 and 2019 May 22 UT using the Zorro speckle instrument on the Gemini South 8-m telescope\footnote{https://www.gemini.edu/sciops/instruments/alopeke-zorro/}. Six sets of 1000 by 0.06 sec exposures were collected for TOI 560 and processed with Fourier analysis in our standard reduction pipeline (see \citealp{Howell2011}).
 
\subsubsection{SOAR telescope}
We searched for stellar companions to TOI 560 with speckle imaging on the 4.1-m Southern Astrophysical Research (SOAR) telescope \citep{Tokovinin2018} on 18 May 2019 UT, observing in Cousins I-band, a similar visible bandpass as TESS. This observation was sensitive to a 7.5-magnitude fainter star at an angular separation of 1$''$ from the target. More details of the observation, data reduction and analysis are available in \cite{Ziegler2020}.

\subsection{Radial Velocities}  
\label{radial}

Herein we present ground-based precise radial velocity observations collected with high-resolution echelle spectrographs PFS, HIRES, and MINERVA-Australis at visible wavelengths, and iSHELL in the NIR in the following subsections.

\subsubsection{iSHELL RVs}  

We have gathered a total of 204 observations of TOI 560 over 30 nights using the iSHELL instrument at NASA InfraRed Telescope Facility (IRTF) atop Maunkea, Hawaii, USA from UT 26 January 2020 to UT 29 May 2021. We observe with iSHELL in KGAS mode covering the wavelengths of 2.18--2.47 $\mu$m. Our exposure times were always set at 300 seconds, and were repeated anywhere from 6--14 times consecutively per night to attempt obtaining a signal-to-noise ratio (SNR) per spectral pixel of $\sim$120, though our actual results varied from 46-152 due to variable seeing and atmospheric transparency conditions. A methane isotopologue ($^{13}\mathrm{CH}_4$) gas cell is used in the instrument \citep{Plavchan2013,Cale2019} to constrain the line-spread function (LSF) and provide a common reference for the optical path wavelength. Along with each set of exposures within a night, we also collect a set of 5x15-second flat-field images with the gas cell removed for data reduction purposes, and particularly to mitigate flexure-dependent and time-variable fringing present in the spectra.
In \cite{Cale2019}, the RV pipeline is adapted from the CSHELL RV code described in \cite{Gao2016}, and the CSHELL code was re-written in a Python script \textit{pychell}\footnote{\href{ Documentation: https://pychell.readthedocs.io/en/latest/}{ Documentation: https://pychell.readthedocs.io/en/latest/}} to adapt to iSHELL’s larger spectral grasp with multiple orders.

When the light passes through a gas cell and an echelle spectrograph, the spectrum consists of multi-order spectra of the target with the gas spectrum super-imposed, which can be used to subtract off instrument systematic shifts in the wavelength solution and changes in the line spread function \citep{Butler}. We extract our spectra following the general procedures outlined in \citet{Cale2019}. For each of the 29 orders in the regime that we observe between 2.18--2.47 $\mu$m, we do optimal spectral extraction (weighted summation) so that in each column eventually we get a single value, which gives us a one-dimensional spectrum for each order. 

The extracted spectra from \textit{pychell} were forward-modeled and analyzed using the methods outlined in \cite{Cale2019}, and the updated methods described in \citet[submitted,][]{Reefe2021}, briefly described herein. We forward model the extracted spectra independently for each order to account for the stellar spectrum, gas cell, telluric absorption, the line spread function, and the spectral continuum to construct a complete spectral model. We find that \textit{pychell} performs better when providing a synthetic stellar template for the stellar model to start from, which is based on properties of the host-star.

We assume a solar metallicity and gravity, and we explore synthetic models $\pm$500 K from an initial temperature of 4700 K, based upon the EXO-FOP-TESS values, in increments of 100 K to identify the optimal synthetic stellar template that produces the lowest rms flux residuals in forward modeling our extracted spectra. We use the Spanish Virtual Observatory's (SVO) BT-Settl models accessible from their web server\footnote{\href{http://svo2.cab.inta-csic.es/theory/newov2/index.php?models=bt-settl}{http://svo2.cab.inta-csic.es/theory/newov2/index.php?models=bt-settl}}, which we further refine by Doppler broadening the spectrum to the rotational velocity of the star (3.1 km s$^{-1}$), as found by TRES extracted spectra (${\S}$\ref{sect:recon}).  We find that the $T_{eff}=4900 K$ synthetic model minimizes the flux residuals of our observed iSHELL spectra, even though hotter than our adopted stellar temperature, and we use this stellar template as our initial stellar spectrum model guess. Barycenter velocities are also generated as an input via the \texttt{barycorrpy} library \citep{Kanodia_2018}.After each order is forward-modeled, we generate one RV measurement per-order and per-exposure.  We optimally co-add RVs across orders and exposures within a night using the same procedures as in \cite{Cale2019}.

We have filtered out three individual spectra from UT 7 March 2020 and three more individual spectra from UT 8 March 2020, due to modeled RVs that were in disagreement with other spectra from the same night by $>$1 km s$^{-1}$.  We suspect this is due to an initially poor seeing on the nights of observation of $\angd{;;1.3}$ that was later improved to $\angd{;;1.0}$, giving us an overall SNR of only 54 and 46 respectively for the nights. Finally, we discard nights with RV uncertainties $>$20 m/s.

\subsubsection{PFS RVs}
We obtained 14 RVs of TOI 560 from the Planet Finder Spectrograph (PFS) on the 6.5m Magellan II telescope at Las Campanas Observatory in Chile. These spectra were reduced and RVs were extracted and published in the Magellan-TESS Survey I paper \citep{Teske2021}. PFS Spectrograph has resolution of R=120,000 and wavelength range of $391-734$ nm.The observations were carried out between UT 2019 April 18 and May 24.    

\subsubsection{HIRES RVs} 
We include 14 Keck-HIRES \citep{Vogt1994} observations of TOI 560 in our analyses. Keck-HIRES is located atop Maunkea, Hawaii, USA with spectral resolution R= 67,000. The majority of these observations took place between UT 2019 October 20 and 2020 January 21, with the final night contemporaneous with the start of iSHELL RVs. Exposure times range from 204--500 seconds, yielding a median SNR $\approx$ 234 at 550 nm per spectral pixel. HIRES spectra are processed and RVs computed via methods described in \citet{Howard2010}. 

\subsubsection{{\textsc{Minerva}}-Australis RVs}
{\textsc{Minerva}}-Australis collected two nights of observations of TOI 560 \citep{addison2019,TOI257}. {\sc {\textsc{Minerva}}}-Australis consists of an array of four independently operated 0.7\,m CDK700 telescopes situated at the Mount Kent Observatory in Queensland, Australia \citep{addison2019}.  Each telescope simultaneously feeds stellar light via fiber optic cables to a single KiwiSpec R4-100 high-resolution ($R=80,000$) spectrograph \citep{2012SPIE.8446E..88B} with wavelength coverage from 480 to 620\,nm. We obtained a total of 10 individual spectra of TOI 560 on UT 2019 May 23 and UT 2019 May 29 using three and two of the four {\textsc{Minerva}}-Australis telescopes respectively -- twice per night per telescope -- and exposure times of 30-60 minutes.  Wavelength calibration is achieved using a simultaneous Th-Ar calibration fibre.  Radial velocities were derived by cross-correlation, where the template being matched is the mean spectrum. Given that only two nightly RV data points were obtained, we do not include the {\textsc{Minerva}}-Australis RVs in the remainder of the analysis presented herein.

\section{Analysis}
\label{sect:analysis}

In this Section, we present our analysis of the TOI 560 system. In Section \ref{sect:shc} we present the stellar host characterization. In  Section \ref{lcs} we present the light curve analysis, including the discovery of the transits of TOI 506 c, false-positive vetting metrics, and characterization of the out-of-transit stellar activity in both the \textit{TESS} and SuperWASP light curves. Finally, in Section \ref{rvanalysis} we present the analysis of the TOI 560 RV data aided with a stellar activity model constrained by the light curve analysis.  

\subsection{Stellar Host Characterization} 
\label{sect:shc}

Our analysis and understanding of exoplanets is directly dependent on our understanding of their host stars \citep[e.g.][]{Ballard2014}. We measure planetary masses and radii in terms of their host stars’ masses and radii, we derive elemental abundance ratios from those in the atmosphere of their host star, and we infer surface temperatures and conditions informed by their host stars’ luminosities, ages, and levels of activity. Host star properties are precisely determined through spectroscopic analysis; therefore, herein we analyze in turn the reconnaissance spectra of TOI 560, fitting bulk stellar properties from the broadband spectral energy distribution (SED), and solidify our analysis with high contrast imaging to exclude faint flux-contaminating companions.

\subsubsection{Reconnaissance Spectroscopy} 
\label{sect:recon}
The reconnaissance spectroscopy observations reveal a typical late K dwarf. Stellar parameters from the NRES spectra were obtained using SpecMatch \citep{Vanderburgp2019}, and summarized in Table \ref{tab:starr-2}. For the TRES spectra, there is a small velocity difference that is consistent with the photometric ephemeris, given that the data were collected approximately at opposite quadratures.  In Table \ref{tab:starr-1}, the results for the stellar parameters for the two TRES observations are presented, as well as values from the Stellar Parameter Classification (SPC) tool \citep{2012Natur.486..375B,2014Natur.509..593B}. Given the absence of any large bulk RV changes or spectroscopic binarity, we do not co-add the multi-epoch spectroscopy to improve the cumulative SNR (and consequently reduce the stellar parameter uncertainties), due to current systematic limitations in these approaches \citep[e.g.,][]{duck2022}. We adopt floor errors of $50 K$ in $T_{eff}$, 0.10 in $log(g)$, 0.08 in $[m/H]$, and 0.5 km/s in $V_{rot}$. Note that $V_{rot}$ does not include a correction for the contribution of macroturbulence, and the velocities in Table \ref{tab:starr-1} are on the TRES native system and do not correct for the gravitational redshift. The TRES and NRES analyses yield consistent stellar characterization, with the SPC analysis favoring a slightly higher but not statistically significant rotational velocity and metallicity.

\subsubsection{Bulk Stellar Properties} 

We undertake two independent analyses of the broadband spectral energy distribution (SED) of the star, and we describe each in turn. The first approach is particularly useful for assessing the evidence for any near-UV excess from consideration of the broadband photometry alone.  Our second approach is used for our adopted final stellar parameters and is based upon holistic modeling of the SED jointly with the transit light curves and model isochrones.

\begin{figure}[!ht]
    \centering
    \includegraphics[width=0.75\linewidth,trim=75 75 85 85,clip,angle=90]{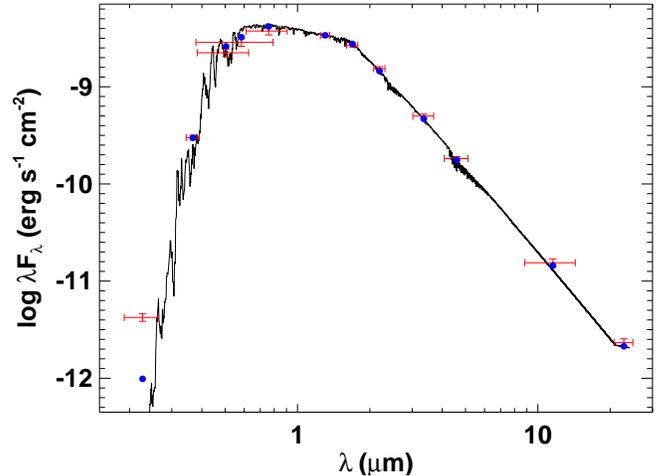}
\caption{Spectral energy distribution of TOI 560 from the first of our two independent SED analyses; this analysis is particularly suited for the identification of the near-UV excess. Red symbols represent the observed photometric measurements, where the horizontal bars represent the effective width of the passband. Blue symbols are the model fluxes from the best-fit NextGen atmosphere model (black).  \label{fig:sed}}
    \label{fig:my_label}
\end{figure}

First, we performed a single-parameter fit of the SED of the star together with the {\it Gaia\/} EDR3 parallax \citep[with no systematic offset applied; see, e.g.,][]{StassunTorres:2021}, in order to determine an empirical measurement of the stellar radius and following the procedures described in \citet{Stassun:2016,Stassun:2017,Stassun:2018}.  We use the $U B V$ magnitudes from the compilation of \citet{Mermilliod:2006}, the $JHK_S$ magnitudes from {\it 2MASS}, the W1--W4 magnitudes from {\it WISE} \citep{Wright_2010}, the $G_{\rm BP} G_{\rm RP}$ magnitudes from {\it Gaia}, and the NUV magnitude from {\it GALEX}. Together, the available photometry spans the full stellar SED over the wavelength range 0.2--22~$\mu$m (see Figure~\ref{fig:sed}).  We used a NextGen stellar atmosphere model, fixing the model effective temperature ($T_{\rm eff}$), metallicity ([Fe/H]), and surface gravity ($\log g$) to the values derived from the TRES SPC analysis in ${\S}$\ref{sect:recon} and Table \ref{tab:starr-1}. The remaining free parameter is the extinction $A_V$, which we fixed at zero given the proximity of the TOI 560 system to Earth. The resulting fit (Figure~\ref{fig:sed}) has a reduced $\chi^2$ of 1.7, excluding the {\it GALEX} NUV flux which indicates a moderate level of activity (see below). Integrating the (unreddened) model SED gives the bolometric flux at Earth, $F_{\rm bol} = 6.01 \pm 0.14 \times 10^{-9}$ erg~s$^{-1}$~cm$^{-2}$. Taking the $F_{\rm bol}$ and $T_{\rm eff}$ together with the {\it Gaia\/} parallax, gives the stellar radius, $R_\star = 0.656 \pm 0.016$~R$_\odot$ (consistent with the NRES analysis). In addition, we can estimate the stellar mass from the empirical relations of \citet{Torres:2010}, giving $M_\star = 0.69 \pm 0.04$~M$_\odot$, which is consistent with the less precise but empirical estimate of $M_\star = 0.75 \pm 0.08$~M$_\odot$ obtained directly from $R_\star$ together with the spectroscopic $\log g$. 

Second, we determine characteristics of the host star, such as effective temperature, gravity, metallicity, etc., by performing a joint amoeba fit followed by  Markov-Chain Monte Carlo (MCMC) posterior sampling of both stellar properties and planet properties of TOI 560 b\&c assuming a two-planet, single-star scenario (from the \tess\ transit data) simultaneously with \texttt{EXOFASTv2}. \texttt{EXOFASTv2} is an (\texttt{IDL}) framework for MCMC simulations of exoplanet transits and radial velocities created by \citet{Eastman_2013}. Details on the planet transit analysis are in the next section ${\S}$\ref{lcs}, but here we present the stellar modeling analysis steps. We start the MCMC with as few assumptions as possible – namely, we place no priors on the spectral type. To ensure we cover an adequate
portion of parameter space to find a believable result, we employ parallel tempering with eight parallel threads, following \cite{Eastman2019}. We place priors on V-band extinction, parallax, and metallicity summarized in Table \ref{tab:exofasts} along with the priors used in the transit analysis. We simultaneously fit with Mesa Isochrones and Stellar Tracks (MIST) and a spectral energy distribution (SED) function. We then use the posteriors of this run as priors for a second iteration MCMC analysis to examine the stability of the solution and adopt our final stellar parameters. 

The results of our \texttt{EXOFASTv2} stellar characterization are shown in 
and Table \ref{table:long1}. The analysis yields a typical K dwarf consistent with the reconnaisance spectroscopy and SED analysis presented in \ref{sect:shc}. We provide two separate estimates for $R_{*}$ and $T_{\rm eff}$ corresponding to the MIST results and the SED results, respectively. SED results are marked with the subscript SED while MIST results have no subscript. We adopt the MIST values for the remainder of the analysis, but both sets of values are consistent with one another and with the analysis carried out with the SED and reconnaissance spectroscopy. Finally, we note that the age estimate from \texttt{EXOFASTv2} shown in Table \ref{table:long1} is drawn from a very flat posterior distribution and is not a reliable estimate; e.g. the stellar age is unconstrained from this particular analysis. Instead, we adopt the final rotation period of 12.2$\pm$0.1 days as described in ${\S}$\ref{WASP_period}.

\begin{table*}
    \begin{center}
    \caption{Median values and 68\% confidence interval for the stellar host, TOI 560, from our \texttt{ExoFASTv2} analysis. See Table 3 in \cite{Eastman2019} for a detailed description of all parameters.}
    \label{table:long1}
    \begin{tabular}{|c|c|c|c|c|} 
        \hline
        Stellar Parameters & Units  & Values & & \\ 
        \hline
        \hline
        M$_{*}$&Mass (M$_{\odot}$) & $0.702^{+0.026}_{-0.025}$ & & \\
        R$_{*}$& Radius (R$_{\odot}$)& $0.677 \pm 0.017$ & & \\
        R$_{*,SED}$& Radius\footnote{This value ignores the systematic error and is for reference only} (R$_{\odot}$)& $0.6819^{+0.0099}_{-0.0094}$ & & \\
        L$_{*}$& Luminosity (L$_{\odot}$)& $0.1823^{+0.0062}_{-0.0060}$ & & \\
        F$_{Bol}$& Bolometric Flux ((erg/s)/cm$^{2}$)& $5.85 \times 10^{-9}\pm 2 \times 10^{-10}$ & & \\
        $\rho_{*}$& Density (g/cm$^{3}$)& $3.19^{+0.26}_{-0.24}$ & & \\
        $\log $g& Surface gravity (cm/s$^{2}$)& $4.623 ^{+0.025}_{- 0.024}$ & & \\
        T$_{eff}$& Effective Temperature (K)& $4582^{+64}_{-62}$& & \\
        T$_{eff,SED}$& Effective Temperature\footnote{This value ignores the systematic error and is for reference only} (K)& $4568 \pm 45$& & \\
        $[Fe/H]$& Metalicity (dex)& $-0.055^{+0.044}_{-0.052}$& & \\
        $[Fe/H_{0}]$& Initial Metallicity\footnote{The metallicity of the star at birth} & $-0.051^{+0.057}_{-0.062}$& & \\
        Age& Age\footnote{This posterior is relatively flat and unconstrained, and does not take into account the stellar rotation period analysis estimated in ${\S}$3.1.3 and /or Figure \ref{fig:WASP}} (Gyr)& $6.7^{+4.7}_{-4.5}$& & \\
        EEP& Equal Evolutionary Phase\footnote{Corresponds to static points in a star’s evolutionary history. See $\S$2 in \cite{Dotter2016}.} & $329^{+13}_{-30}$& & \\
        A$_{v}$& V-band extinction (mag)& $0.076^{0.046}_{0.051}$& & \\
        $\sigma_{SED}$& SED photometry error scaling & $1.68^{+0.77}_{-44}$& & \\
        $\Bar{\omega}$& Parallax (mas)& $31.683 \pm 0.032$& & \\
        d& Distance (pc)& $31.562 \pm 0.032$& & \\
        \hline
        Wavelength Parameters & Units  & B & R & z' \\ 
        \hline
        \hline
        $u_{1}$& linear limb-darkening coeff& $0.50^{+0.037}_{-0.32}$ & $0.68^{+0.037}_{-0.039}$& $0.58 ^{+0.44}_{-0.38}$\\
        $u_{2}$& quadratic limb-darkening coeff&  $0.27^{+0.038}_{-0.044}$ & $0.02^{+0.045}_{-0.038}$& $0.18^{+0.043}_{-0.046}$\\
        $A_{D}$& Dilution from neighboring stars&  -- & -- &-- \\
        \hline
        Telescope Parameters & Units  & HIRES & PFS & iSHELL\\ 
        \hline
        \hline
        $\gamma_{rel}$& Relative RV Offset (m/s)& $0.1^{+3.5}_{-3.6}$& $-10.8 ^{+2.2}_{-2.3}$ & $0.5\pm 3.1$\\
        $\sigma_{J}$& RV Jitter (m/s)&$12.7^{+3.4}_{-2.4}$ & $8.0^{+2.5}_{-1.7}$ & $10.9^{+3.1}_{-2.5}$\\
        $\sigma^{2}_{J}$& RV Jitter Variance& $160^{+97}_{-55}$ & $64^{+47}_{-24}$ & $119^{+79}_{-48}$\\
        \hline
    \end{tabular}
    \end{center}
\end{table*}  

Finally, TOI 560 has a \textit{GAIA} \texttt{RUWE} value of 1.030, which is consistent with a single star \citep{2021ApJ...907L..33S}.  The \texttt{RUWE} statistic is an indication of the goodness of fit of the astrometric solution, and values $>$1.4 have been shown to be associated with unresolved binaries. A search of EDR3 also shows no stars with similar parallaxes within 75', so TOI 560 likely does not possess any distant, resolved binary companions either.

\subsubsection{Age Metrics -- Is TOI 560 a young system?} 
\label{WASP_period}

\begin{figure}[!ht]
    \centering
    \includegraphics[width=0.75\linewidth]{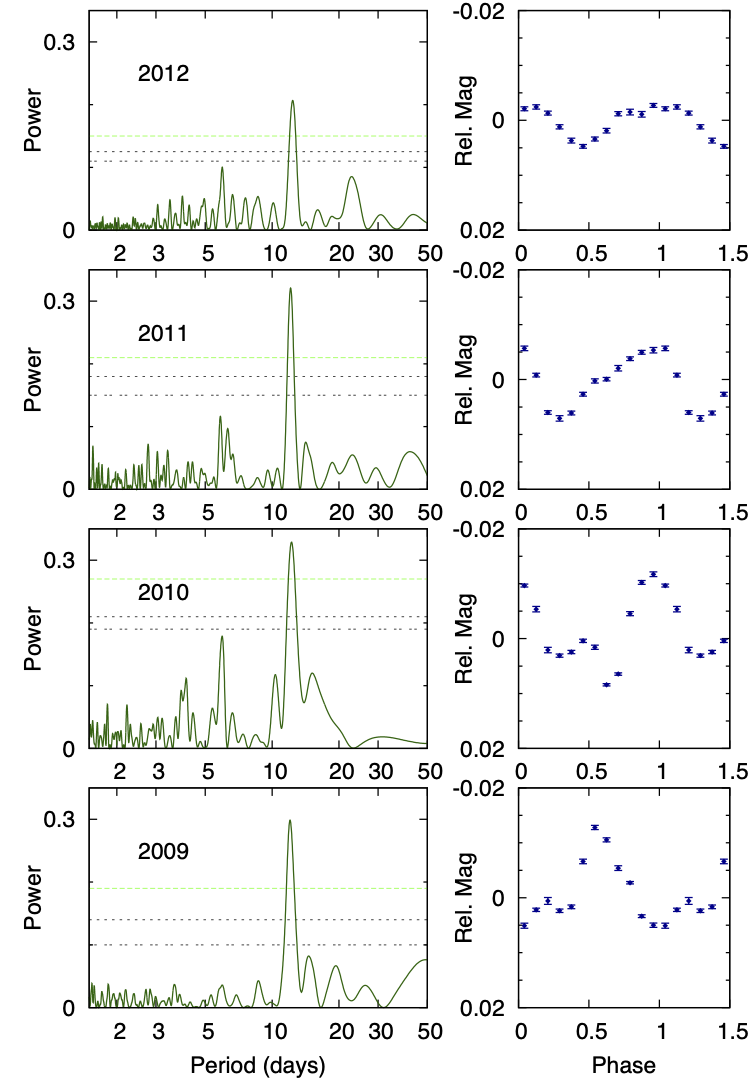} 
\caption{Left panels: Lomb-Scargle periodograms of the light curves of TOI 560 from the SuperWASP suvey in 2009-2012 from bottom to top, plotted in green as a function of period on the horizontal axis and power on the vertical axis. Right panels: Phased seasonal light curves of the SuperWASP TOI 560 seasonal light curves in the same reverse annual order.  The light curves are binned in phase in 0.05 increments in blue with the co-added uncertainties shown, with rotational phase on the horizontal axis and relative and normalized apparent magnitude on the vertical axis.  Spot evolution is readily apparent from year to year, including some double-humped features from spots on the northern and southern hemispheres of the star.}
    \label{fig:WASP}
\end{figure}

Lomb-Scargle periodograms of the SuperWASP seasonal light curves show a clear and consistent rotation period in each season for TOI 560 of 12 days, with evolution of the light curve features between seasons (Figure \ref{fig:WASP}.  The periodogram peak is statistically significant in all seasons (p-value$<$1\%) and the average periodogram peak period from all four seasons yields a rotation period of 12.2$\pm$0.1 days. We next use the star's NUV excess (Fig.~\ref{fig:sed}) to estimate the star's rotation and age via empirical rotation-activity-age relations. The observed NUV excess implies a chromospheric activity of $\log R'_{\rm HK} = -4.5 \pm 0.1$ via the empirical relations of \citet{Findeisen:2011}, which is fully consistent with the measured value of $-4.45\pm0.05$ from \citet{Gomes:2021} and -4.60 from our HIRES spectra. The HIRES and TRES spectra show core Ca II HK flux in emission, consistent with this moderate level of activity. This value of $\log R'_{\rm HK}$ in turn implies a stellar rotation period of $P_{\rm rot} = 14 \pm 4$~d via the empirical relations of \citet{Mamajek:2008}.  This rotation period estimate is also consistent with that estimated from the spectroscopic $v\sin i$ and $R_\star$, which gives $P_{\rm rot}/\sin i = 11\pm3$~d. Thus the observed photometric modulation, NUV excess, and the observed rotational velocity all yield consistent rotation periods. This establishes the rotation period of the TOI 560 host star.

\begin{figure}[!ht]
    \centering
    \includegraphics[width=\linewidth]{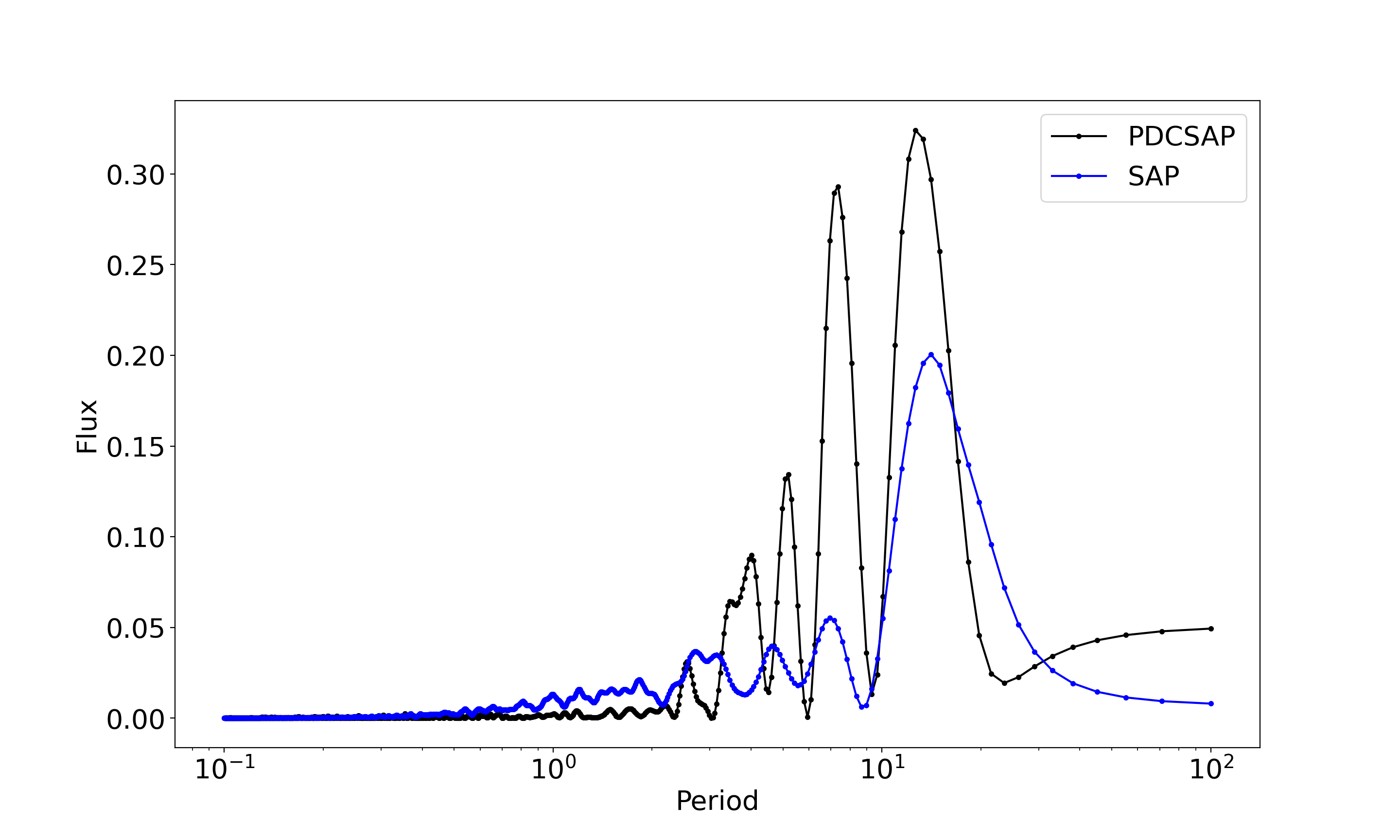}
    \includegraphics[width=\linewidth]{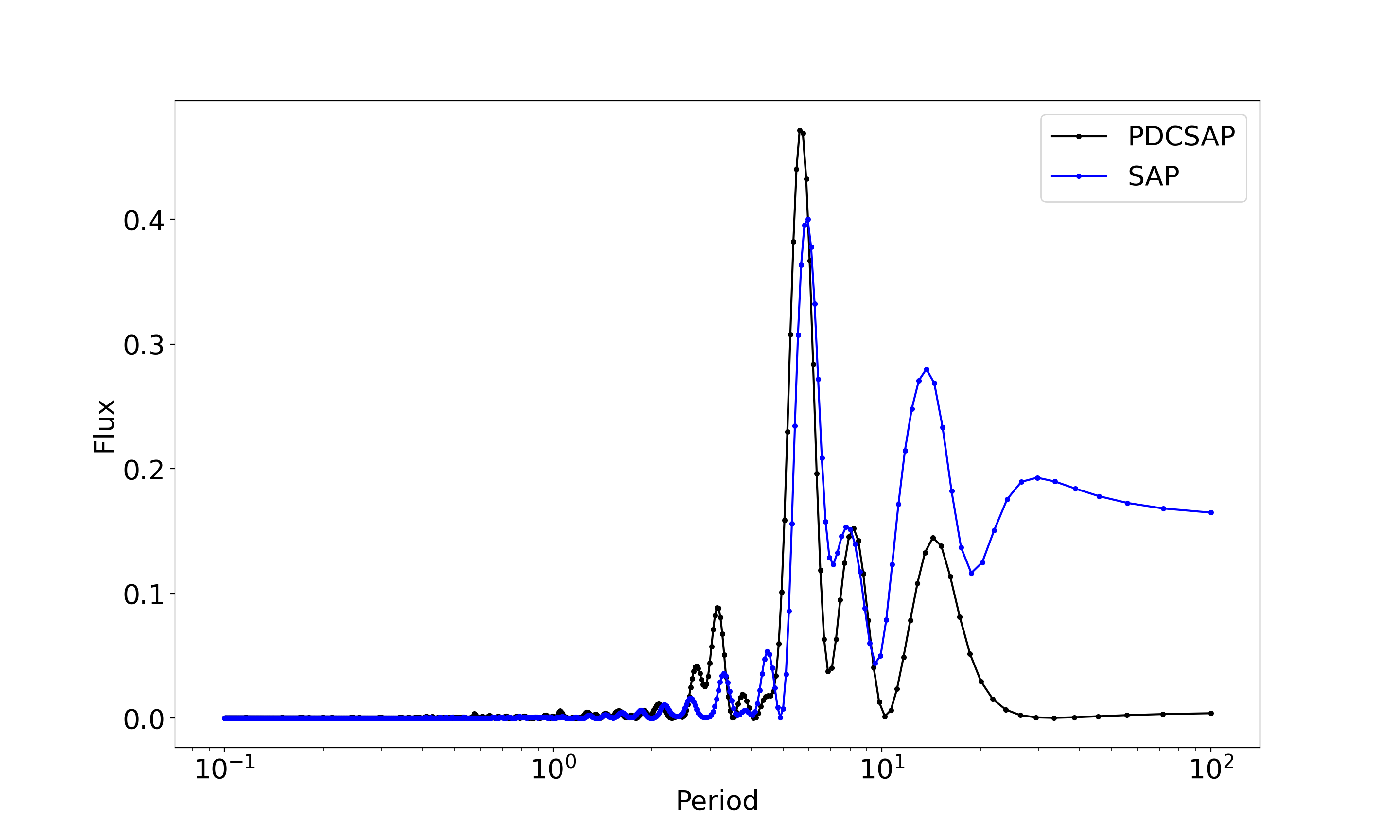}
\caption{\tess\ PDC-SAP and SAP sectors 8 (top) and 34 (bottom) rotation periods.}
    \label{fig:rot-periods}
\end{figure}

However, this 12-day rotation period is roughly a factor of 2 slower than that reported by \citet{Canto:2020}, who report a ``dubious'' rotation period 7.17d from their independent \tess\ light curve analysis. We can also estimate rotation periods for TOI 560 by performing Lomb-Scargle (LS) periodograms (\citealp{Lomb1976}; \citealp{Scargle1982}) of the \textit{TESS} Sector 8 and 34 light curves of TOI 560. In Figure \ref{fig:rot-periods}, the \tess\ light curves contain significant peaks at 12.7 days for PDCSAP and 14.1 days for SAP, both for the Sector 8 respectively, that are not obvious integer fraction multiples of one another nor of 7.17d; we do find a secondary peak at 7.29 days for the Sector 8 light curve, consistent with \citet{Canto:2020}. However, the dominant period in the Sector 8 light curve at 12.7 days is reasonably consistent with the WASP, NUV, and rotational velocity inferred rotation periods. A more detailed FF' modeling of the \tess\ light curves in ${\S}$\ref{sect:challenge} with a Gaussian process and uniform rotation period prior from 2--20 days also favors a rotation period of 7.13$^{+0.31}_{-0.13}$ days, also consistent with \citet{Canto:2020}. However, given that the time segments of the \tess\ light curve in-between data downlinks are themselves $<$12 days in time baseline, we find that the \tess\ light curves themselves are insensitive to a 12-day rotation period and are thus not reliable; the 7.1 and 5.6 day power in the \tess\ light curves may be from a combination of the first harmonic of the stellar rotation period at $P_{rot}/2$, potentially the spot evolution timescale, and the pre-search data conditioning of the \tess\ PDC-SAP light curves. Thus, we adopt the rotation period of 12.2$\pm0.1$ days from the SuperWASP light curves in the remainder of our analysis. 

\begin{figure}[!ht]
    \centering
    \includegraphics[width=0.75\linewidth]{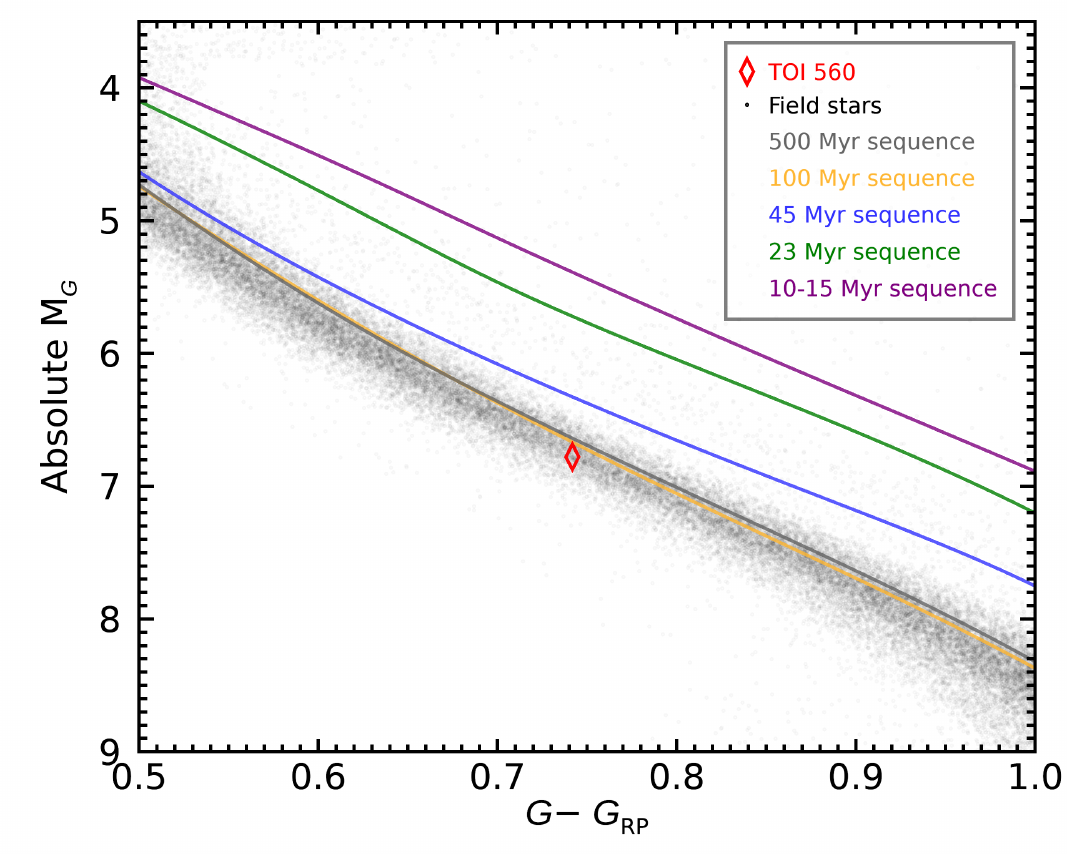}
\caption{Gaia~DR3 color-magnitude diagram for TOI~560 (red diamond) compared to nearby field stars (black dots) and empirical age-dated sequences built from the members of nearby coeval associations (thick colored lines; see \citet{Gagn2021}). The position of TOI~560 is consistent with an age $\approx$\,100\,Myr or older, and ages of 45\,Myr and younger are clearly ruled out.}
    \label{fig:banyan1}
\end{figure}

\begin{figure}[!ht]
    \centering
    \includegraphics[width=0.75\linewidth]{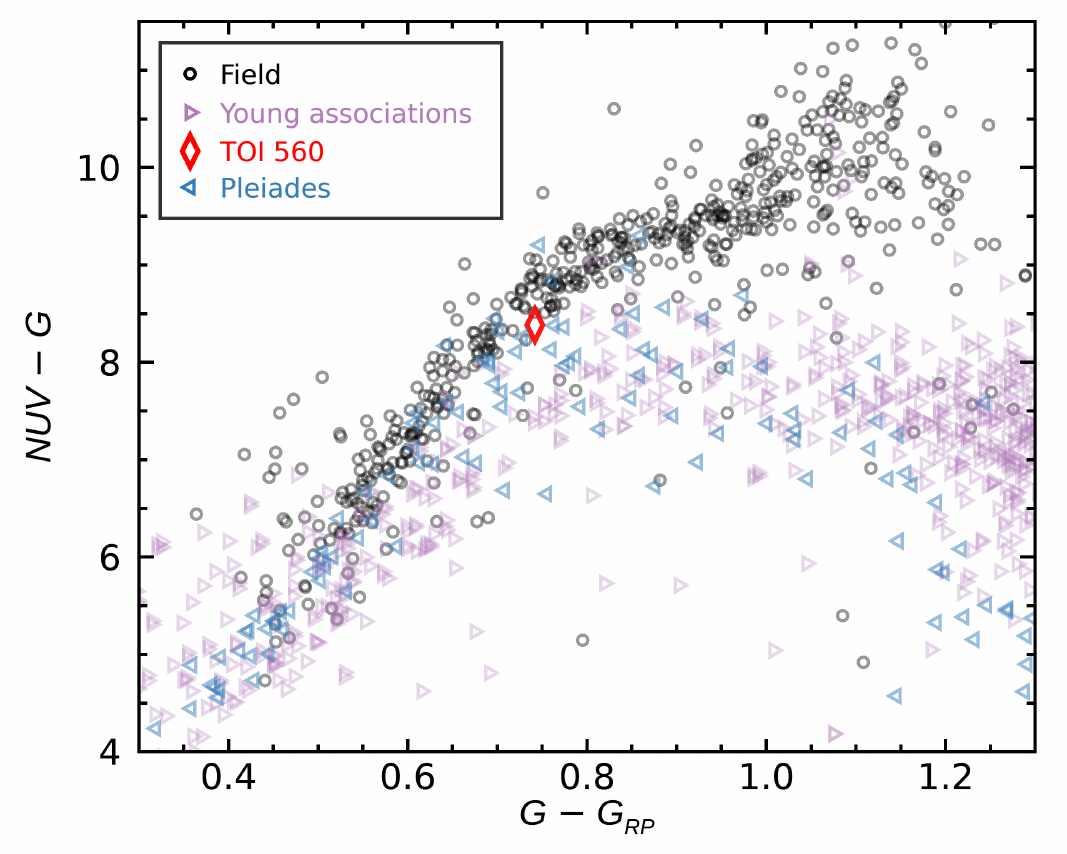}
\caption{GALEX-Gaia color-color diagram for TOI~560 (red diamed) compared to nearby field stars detected in GALEX (black circles), members of various nearby, young associations with ages in the range 5--100\,Myr (right-pointing purple triangles; see \citet{Gagn2018}), and members of the Pleiades (left-pointing blue triangles; $112 \pm 5$\,Myr; Dahm et al. 2015, \citet{Gagne2021}). TOI~560 falls on the UV-bright end of the field distribution, but on the UV-faint end of the Pleiades distribution, compared to other stars of a similar Gaia $G-G_{\rm RP}$ color.}
    \label{fig:banyan2}
\end{figure}

For this rotation period, using the age-rotation relation from \citet{Mamajek:2008} based upon the power-law model of \citet{2007ApJ...669.1167B}, we derive a stellar age estimate of $\tau_\star = 591^{+130}_{-97}$~Myr. 
However, \citet{2020ApJ...904..140C} have recently shown that for K dwarfs like TOI 560, there is an age-rotation ``pile-up'' at rotation periods of $\sim$10--15 days where the spin-down of K dwarfs stalls between ages $\sim$0.6--1.4 Gyr, before resuming for ages $>$1.4 Gyr. We also verified whether TOI~560 may be a plausible member of a nearby moving group or association with the BANYAN~$\Sigma$ \citep{2018ApJ...856...23G} tool which compares the $XYZ$ Galactic coordinates and $UVW$ space velocities of known stellar associations and assigns a membership probability based on Bayes' theorem with the option to marginalize over unknown quantities such as heliocentric radial velocities. We find a negligible membership probability to all 27 young associations included in BANYAN~$\Sigma$. We also compared the $XYZ$ and $UVW$ of TOI~560 to those of 1\,000 moving groups and open clusters identified in the literature so far within 500\,pc of the Sun, and we find no plausible association to which TOI~560 may belong to. Inspecting the position of TOI~560 in a Gaia~eDR3 \citep{Gaia_Collaboration_2018} color-magnitude diagram (Figure~\ref{fig:banyan1}) reveals a picture consistent with an age of at least 100\,Myr. TOI~560 has been detected in GALEX~DR5\citep{Bianchi2011}, and its position in a GALEX-Gaia color-color diagram (Figure~\ref{fig:banyan2}) shows that it located on the UV-bright end of field stars and on the UV-faint end of the Pleiades members with similar Gaia eDR3 $G-G_{\rm RP}$ color (which tracks spectral types), thus indicating that TOI~560 is likely older than the Pleiades, but plausibly on the younger end of the age distribution of field stars, and consistent with the rotation-based age estimate. Additionally, there is no lithium detected, consistent with this star not being $<$100 Myr \citep{delgado}.  We thus conclude from the rotation period of TOI 560 and the K dwarf rotational evolution pile-up that TOI 560 possesses and age between 150 Myr and 1.4 Gyr.

\subsubsection{High Contrast Imaging}  

\begin{figure*}
\centering
    \includegraphics[width=.45\textwidth]{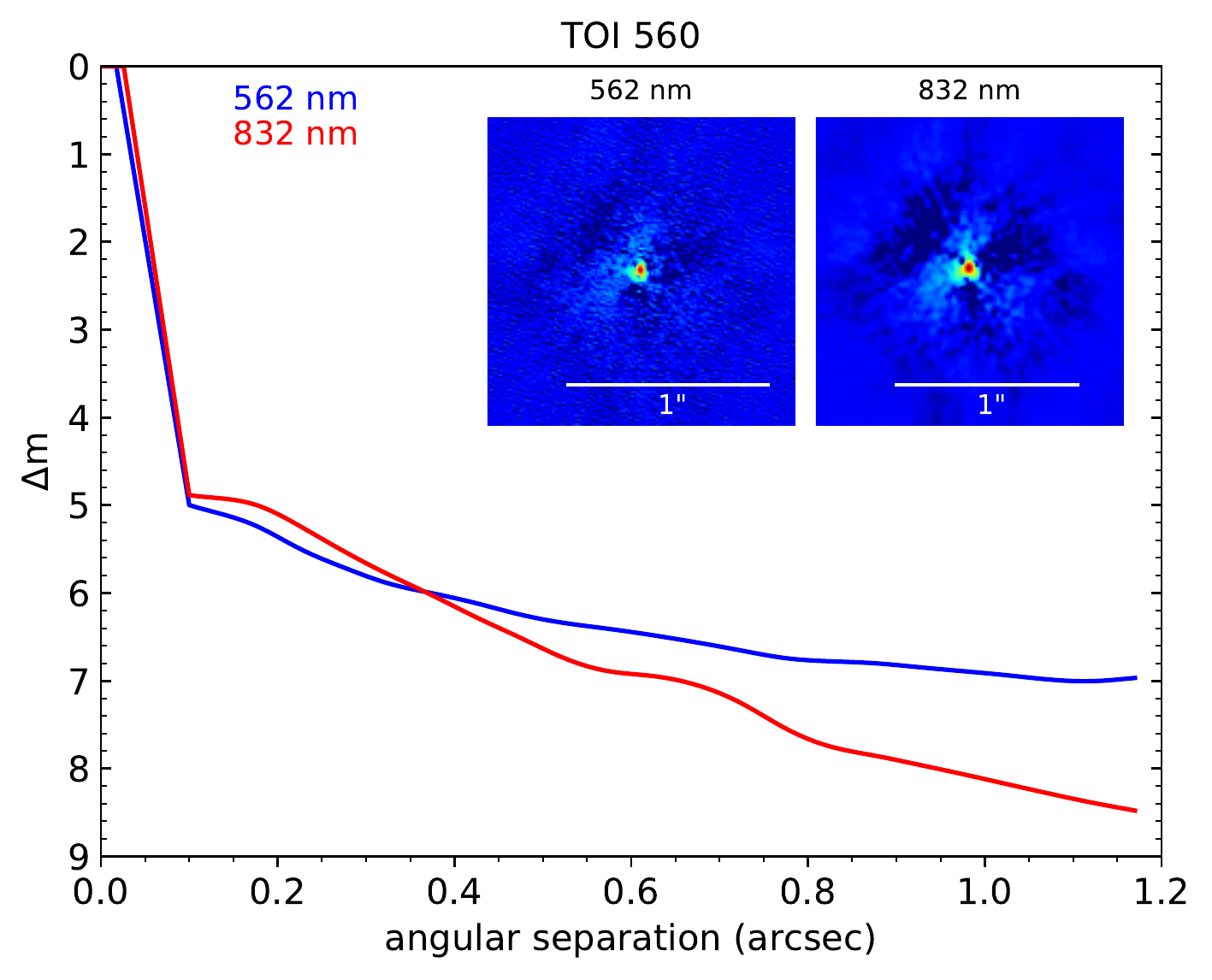}
    \includegraphics[width=.45\textwidth]{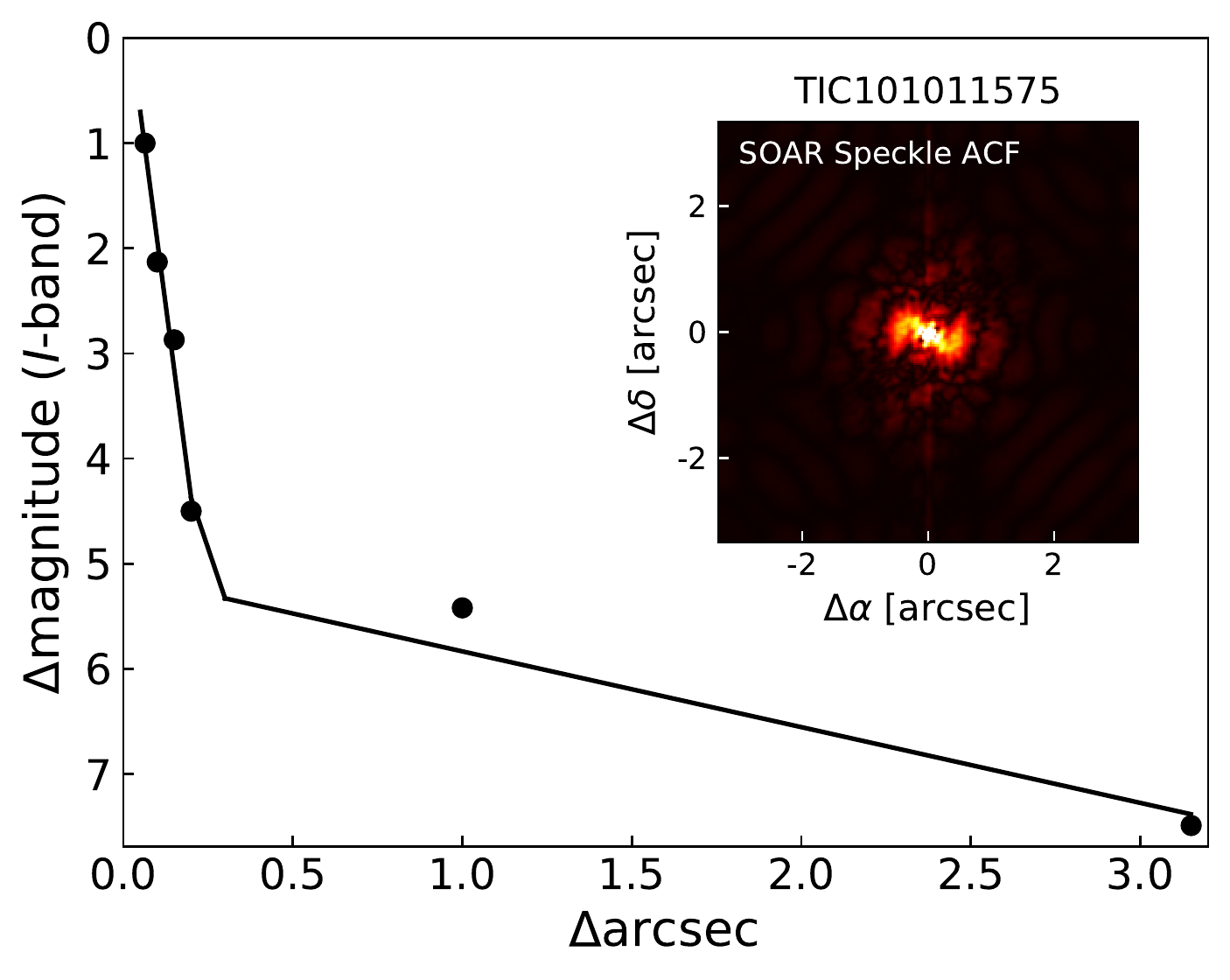}\\
    \includegraphics[width=.45\textwidth]{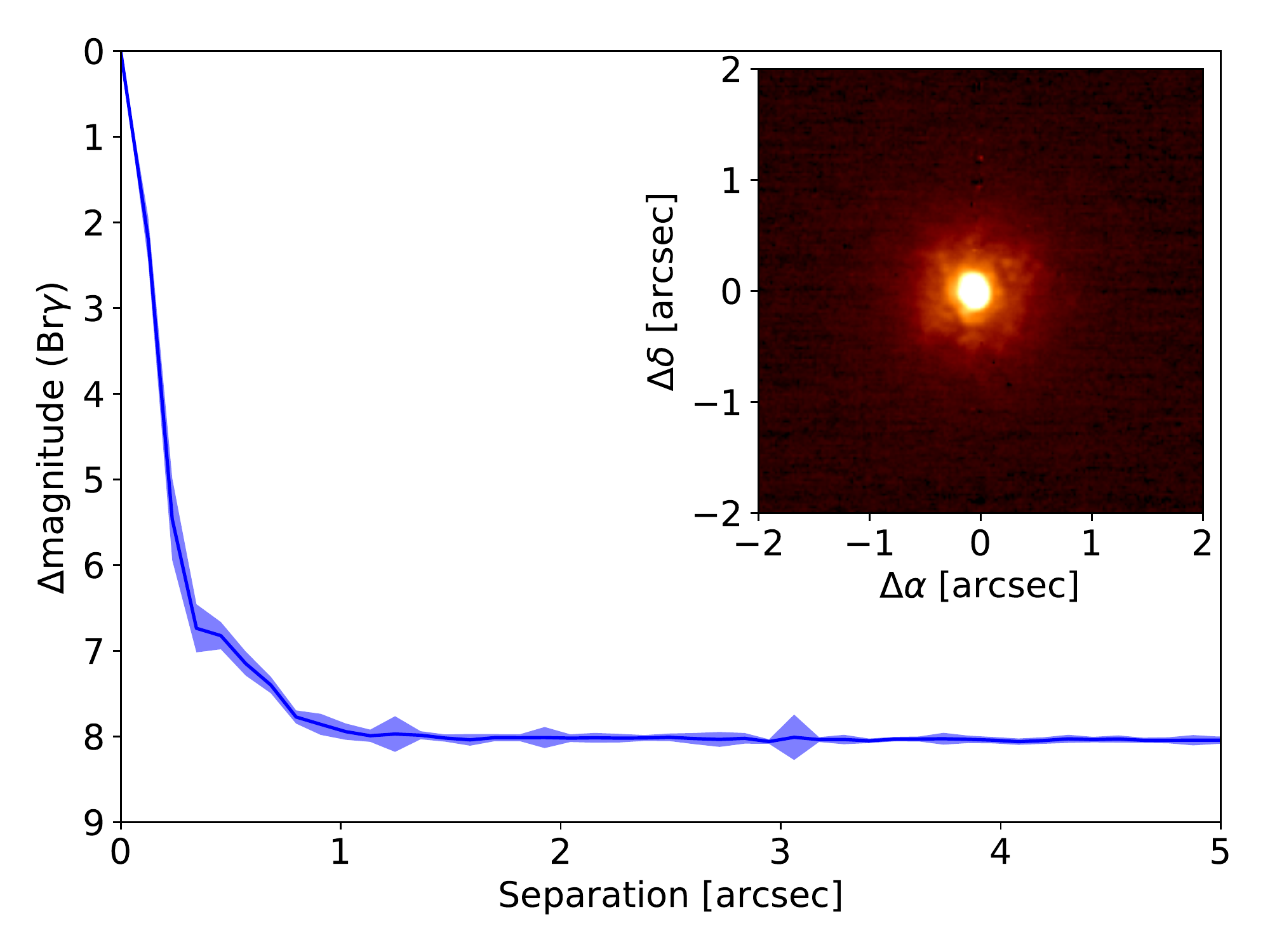}
    \caption{Top-left: Our 5-$\sigma$ contrast curves and the reconstructed speckle images observed using the Zorro speckle instrument on the Gemini South 8-m telescope. Sensitivity is quoted on the vertixal axis in magnitudes relative to the host star, as a function of angular separation on the horizontal axis. \textit{Inset:} a cutout of our image, centered on the target star TOI 560; Top-right: Same as top-left, showing the 5-$\sigma$ detection sensitivity and speckle auto-correlation functions from the SOAR observations; Bottom: Same as the top two panels, showing our sensitivity of the Gemini/NIRI observations to companion stars, as a function of separation from the host star. No visual companions are seen anywhere in all of our high-contrast imaging data, and the star appears single.}
    \label{fig:imaging}
\end{figure*}

For the Gemini/NIRI observations, we carried out the data reduction using a custom set of IDL codes, with which we removed bad pixels, sky-subtracted and flat-corrected the frames, and then aligned the stellar position between frames and co-added the images \citep{2003PASP..115.1388H}. For Gemini South, we utilize only the observations from May 2019, as the March observations had poorer seeing and sky conditions, albeit giving similar results to those obtained about a year earlier. Zorro provides simultaneous speckle imaging in two bands (562 nm and 832 nm) with output data products including a reconstructed image in Figure \ref{fig:imaging} with robust contrast limits on companion detection \citep[e.g.][]{Howell2016}.

For our Gemini/NIRI observation, we did not identify visual companions anywhere in the field of view, which extends to at least $7''$ in all directions from the target star, and TOI 560 appears single to the limit of the Gemini/NIRI resolution. The sensitivity as a function of separation is shown in Figure \ref{fig:imaging}, along with an image of the target. The images are sensitive to companions 8mag fainter than the host star in the background limited regime (beyond $\sim 1''$). For our Gemini South observations, Figure \ref{fig:imaging} shows our final 5-$\sigma$ contrast curves and the reconstructed speckle images. We find that TOI 560 is single to within the contrast limits achieved by the observations. No companion is identified brighter than 5-8 magnitudes below that of the target star from the diffraction limit (20 mas) out to $1''$. At the distance of TOI 560 (d=31 pc) these angular limits correspond to spatial limits of 0.6 to 37 AU. Finally, for our SOAR observation, Figure \ref{fig:imaging} shows the 5$\sigma$ detection sensitivity and speckle auto-correlation functions from the observations. No nearby stars were detected within 3\arcsec of TOI 560 in the SOAR observations.

\subsection{Light curve analysis} 
\label{lcs}
In this section, we vet against false positive scenarios using different tests in ${\S}$\ref{vetting}, and then we
analyze the \textit{TESS} sector 34 light curve to discover TOI 560 c in ${\S}$\ref{discover}. After that, we apply our GP analysis to the light curves in ${\S}$\ref{sect:challenge}, and using \texttt{EXOASTv2} in ${\S}$\ref{exofast}, we jointly analyze the sector 8 and 34 \tess\ light curves as a 2-planet system.

\subsubsection{Vetting against False Positive} 
\label{vetting}
Before investing detailed resources in the RV analysis of a \tess\ candidate planetary system, it is useful to rule out false-positive scenarios caused by eclipsing binaries and other systematics, and there are numerous diagnostics that we employ using the \tess\ light curves alone. We ran two separate vetting analyses on the TOI 560 data gathered by \textit{TESS}.

The first of our vetting tests was performed with the EDI-Vetter Unplugged tool \citep{Zink_2020}\footnote{\href{https://github.com/jonzink/EDI_Vetter_unplugged}{https://github.com/jonzink/EDI\_Vetter\_unplugged}}. \texttt{EDI-Vetter Unplugged} checks for several diagnostics that could be indicative that the target is a false positive. First, EDI-Vetter checks for neighboring flux contamination from nearby stars in the sky, to make sure that the signal is indeed correlated with the expected target; this is necessary due to the large angular size of \textit{TESS} pixels. Second, it examines the abundance of outliers from the transit model, as well as the validity of each individual transit -- e.g. odd/even test, if transits fall into masked data gaps.  Third, it also searches for the statistically significant presence of any secondary eclipses, which could indicate an eclipsing binary.  Other measures that are considered are the similarities between transit signals, the phase coverage of transit signals, and the relative length of transit duration in comparison to period. The results of our EDI-Vetter analysis are in table \ref{tab:edi} in section \ref{sect:resu}.  

Second, we also scrutinized the \tess\ light curves using the Discovery and Vetting of Exoplanets (DAVE) software \citep{Kostov_2019} developed at the NASA Goddard Space Flight Center.  This tool evaluates similar criteria as the EDI-Vetter tool, such as odd/even transit differences and secondary eclipses, but also searches for any photocenter shift, or the difference between the TIC position and the measured photocenter of the transits during transit, which is often also a diagnostic of a blended eclipsing binary.  

In our EDI-Vetter analysis, two of the criteria were flagged as a potential false-positive: the uniqueness of the transit signal and the planet masks (Table \ref{tab:edi}). 
However, the latter is due to the system being a multiplanet system, with two different transiting planets, and the former a transit of TOI 560 c falling into the data download gap of Sector 8. In our DAVE analysis, we show full transit data and phased odd/even, secondary, tertiary, and positive transits for each sector in figures \ref{fig:modshift8} and \ref{fig:modshift34}.  We see little to no variation between the primary odd and even transits, and we see no statistically significant drop in flux during the expected ephemerides for a secondary eclipse or tertiary transit. However, we note the appearance of an odd-even transit depth difference for TOI 560 b. Due to the youth of the star, there is significant out of transit variability from the star that is not filtered out by the DAVE vetting tool. This in turn impacts its transit depth determination, resulting in the even-odd discrepancy. Thus, we effectively ignore this metric present in the DAVE tool as a
potential sign of a false-positive, which is supported by the multi-planet nature of the system.

We also see no statistically significant positive flux variation that could be indicative of lensing between eclipsing binary stars. The evaluation of the photo-center shift during transit helps validate that the transit belongs to the target star, and not a faint companion. Photo-center difference plots for each sector are shown in Figure \ref{fig:photocenter}, and overall show no major variations between the expected and observed locations. Thus, this vetting analysis from DAVE and EDI-Vetter, combined with our high-contrast imaging, rules out a blended or background eclipsing binary companion, and confirms that the transit signals originate from TOI 560, and not another nearby point source.  Further, when combined the increasingly low probability \citep[e.g. $<1\%$, ][]{2012ApJ...750..112L} that multiple blended or background eclipsing binaries would be spatially coincident with TOI 560 and produce two different eclipsing signals near a 3:1 orbital period coincidence, we can conclude that TOI 560 is the host of the two eclipsing companions.  We next turn to presenting our results on constraining the masses of TOI 560 b and c.

\begin{table}
    \centering
    \begin{tabular}{ccc}
    \hline
    Vetting Report & Sector 8 & Sector 34  \\
    \hline
    \hline
        Transit Count & 6 & 6 \\ 
        Flux Contamination & False & False \\
        Too Many Outliers & False & False\\
        Too Many Transits Masked & False & True \\
        Odd/Even Transit Variation & False & False \\
        Signal is not Unique & True & True \\
        Secondary Eclipse Found & False & False\\
        Low Transit Phase Coverage & False & False \\
        Transit Duration Too Long & False & False\\
    \hline
    \hline
        Signal is a False Positive & True & True\\
    \hline
    \end{tabular}
    \caption{Sector 8 and 34 results for false-positive signals analyzed by EDI-Vetter unplugged. The software found no evidence for false positives in this data. The software version used was 0.1.3.} 
    \label{tab:edi}
\end{table}

\subsubsection{The discovery of TOI 560 c}
\label{discover}

\begin{figure*}[!ht]
    \centering
    \includegraphics[width=\linewidth]{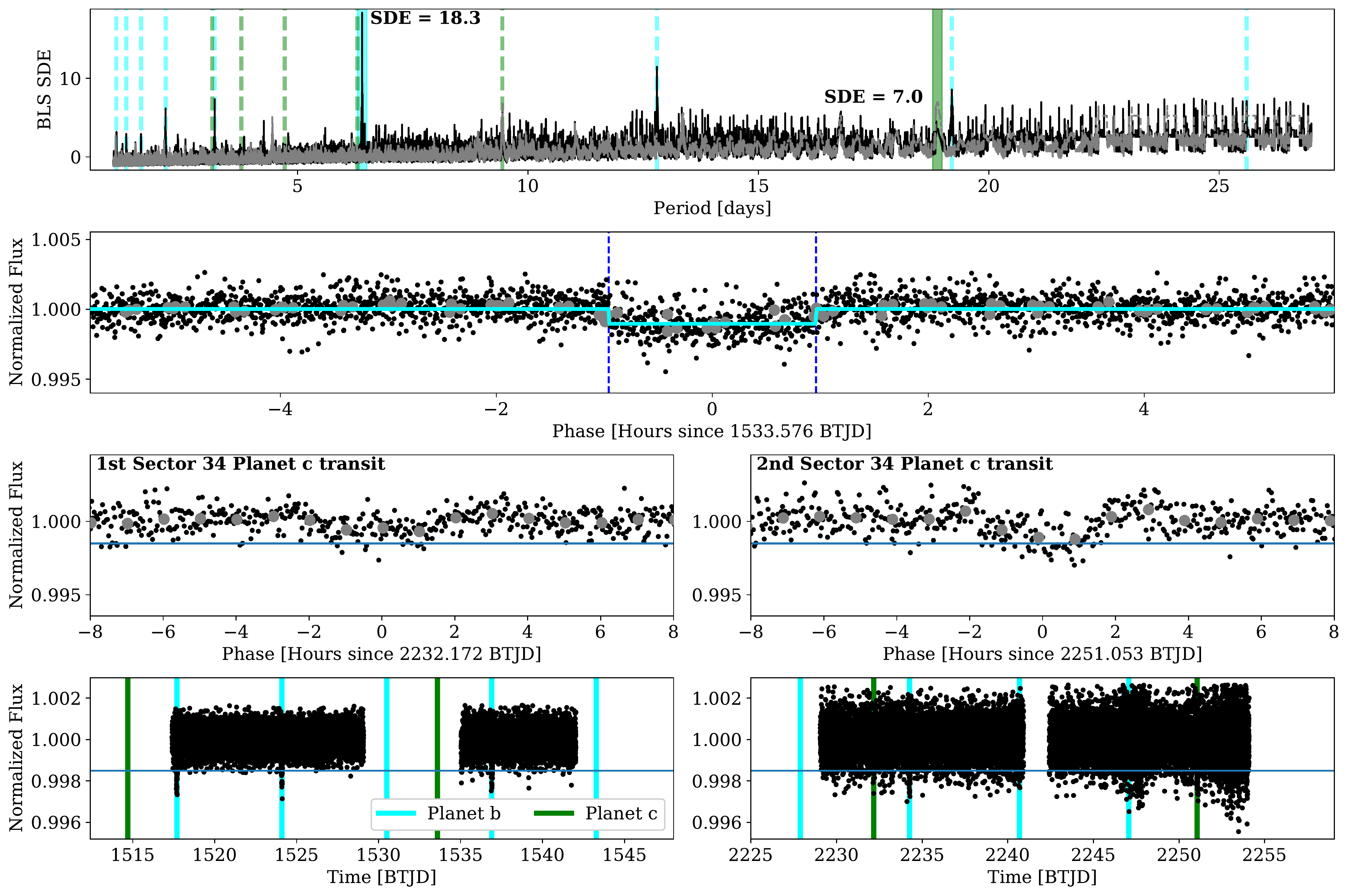}
\caption{Top: Joint BLS periodograms of the Sector 8 and 34 \tess\ light curves, before (black) after (grey) removing the transits of TOI 506 b, plotted as a function of Period in days with the BLS standard deviation statistical significance power on the vertical axis (SDE). The periods of b and c are indicated as teal and green shaded regions respectively. The vertical dashed lines correspond to integer (2x,3x,4x) and integer fraction (1/2, 1/3x, 1/4x, etc.) multiples of these periods in teal and green for b and c respectively as well.} Note: the time units of BTJD are BJD-2457000.0 days. Second Row: Phased light curve for the two transits of TOI 560 c in the Sector 34 \tess\ light curve as unbinned black and binned (one hour) grey dots, plotted as a function of time in hours on the horizontal axis with respect to the transit conjunction time, and the normalized flux on the vertical axis. A teal transit model is overlaid.  The blue dashed vertical lines correspond to the time of conjunction and transit ingress and egress. Third Row: Same as the second row, but the individual transits of TOI 560 c are plotted separately. Fourth Row: The Sector 8 (left) and 34 (right) light curve of TOI 560 with the transit times of TOI 560 b and c marked in teal and green respectively, plotted as a function of time and normalized flux. This shows that \tess\ just missed observing TOI 560 c during this initial sector. The horizontal blue line is a line corresponding to 2 ppt, the approximate depth for TOI 560 c.
    \label{fig:discovery}
 \end{figure*}

With the release of the \tess\ Sector 34 light curve, we readily identified by eye two transits of a second planet candidate with an orbital period of $\sim$18.9 days. Several other teams independently identified this second transiting planet during our analysis, and was identified by the \tess\ mission as TOI 560 c ($P_{c} = 18.8805$ days). After masking the transits of TOI 560 b ($P_{b} = 6.3980$ days), we computed BLS \citep{Kovacs2002} and TLS \citep{Hipkee2019} periodograms and phased LCs of the Sector 34 light curve in Figure \ref{fig:discovery}. Due to unfortunate timing, TOI 560 c did not transit in the Sector 8 \tess\ light curve (Figure \ref{fig:discovery}). We also excluded a period for TOI 560 c of one-half its value, as a third transit of TOI 560 c could have occurred during a data down-link data gap in the \tess\ Sector 34 light curve.  However, under that scenario, a transit of TOI 560 c would have occurred in the Sector 8 light curve, which is ruled out.

The model parameters and prior distributions used in our RV model that considers the transiting b and c planets, as

\subsubsection{GP Analysis of the Light Curves}
\label{sect:challenge}

Stars are not uniform disks; they exhibit RV variations from rotationally-modulated activity, e.g. star spots and plages on the surface of the star \citep[][and references therein]{2011A&A...525A.140D, 2011A&A...527A..82D,2016PASP..128f6001F,Plavchan2015}. Depending on the temperature contrast between magnetically active regions and the surrounding photosphere of the star, these active regions will induce an apparent RV shift that is quasi-periodic as the active regions evolve on time-scales distinct from the rotation period of the star. Gaussian processes (GPs) have been widely and successfully employed in modeling the apparent RVs due to stellar activity \citep{2015PhDT.......193H}. 

The same active regions also give rise to photometric variations from the hemisphere visible to the observer. Thus, we can model the light curves of TOI 560 out of transit with a GP, and use the light curve hyperparameter posteriors to constrain the GP model hyper-parameter priors for our subsequent analysis of the RVs, following the $FF^\prime$ technique \citep{Aigrain2012}. This method allows us to generate what a simulated RV curve would look like from the stellar activity:
\begin{equation} 
    \Delta R V_{spots}(t) = - F(t)F^{'}(t)R_{*}/f 
\end{equation}
\noindent Here, F is the photometric flux, $F^{'}$ is the derivative of F with respect to time, f represents the relative flux drop for a spot at the center of the stellar disk, and $R^{*}$ is the stellar radius. 

Our Gaussian process kernel describes the functional covariance between any two measurements separated in time. The kernel by definition must be a square, symmetric covariance matrix ``K'' of length equal to the number of observations. We follow \citet{2015PhDT.......193H} who introduced the quasi-periodic (QP) Kernel composed of a decay and periodic term: 
\begin{equation}
     K_{QP}(t_{i},t_{j})= \eta^{2}_{\sigma} \overbrace { \exp \left[ - \frac{\Delta t^2}{2\eta^{2}_{\tau}}\right]}^{Decay} \overbrace  { \exp \left[ { - \frac{1}{2\eta^{2}_{l}}}\sin^2 {\left(\pi \frac{\Delta t}{\eta_p}\right)}\right]}^{Periodic} 
     \label{eq1}
 \end{equation} 
\noindent where: $\Delta t = |t_{i}-t_{j}|$, Here, $\eta_{P}$ typically represents the stellar-rotation period, $\eta_{\tau}$ the mean spot lifetime, and $\eta_{\ell}$ is the relative contribution of the periodic term, which may be interpreted as a smoothing parameter (larger is smoother). $\eta_{\sigma}$ is the amplitude of the auto-correlation of the activity signal. The set of $\{\eta_\sigma,\eta_\tau,\eta_P,\eta_L\}$ constitute the set of GP model hyper-parameters that are constrained by the data. There exist families of stellar activity models that generate a particular covariance matrix given by a specific set of hyper-parameters, and thus GPs are a flexible framework for characterizing the non-deterministic time evolution of stellar activity. The $FF^\prime$ analysis does enable us to estimate the expected amplitude of the RV variations at wavelengths comparable to the light curve observations, $\eta_{\sigma}$, but we end up not using that amplitude in our RV modeling.  Instead, we do use the $FF^\prime$ analysis to get estimates of the remaining hyperparameters, particularly $\eta_{\tau}$, $\eta_{p}$,$\eta_{\ell}$ for use in the RV analysis.

We first apply our GP model to the SuperWASP data for the 4 seasons together (Figure \ref{fig:WASP}). We use wide uniform priors on hyper-parameters $\eta_{\sigma}$ $\sim$ $\mathcal{U}(0,50)$, $\eta_{\tau}$ $\sim$ $\mathcal{U}(1,100)$ and $\eta_{\ell}$ $\sim$ $\mathcal{U}(0.02,1)$.  Instead of accounting for intrinsic uncertainties in the light curve, we also fit a jitter term $\eta_{LC}$. For the hyper parameter $\eta_{P}$, we use a prior of $\mathcal{U}(11.5,13)$ based on periodogram analysis of the WASP light curves in ${\S}$\ref{WASP_period}. We recover from the MCMC of the WASP data spot lifetime $\eta_\tau$ of 57.96$^{+23.59}_{-19.39}$ days, a rotation period of $\eta_P$ of 12.03$^{+0.13}_{-0.12}$ days, and a smoothing hyper-parameter $\eta_l = 0.44^{+0.11}_{-0.09}$ which are consistent with our young star and such long spot lifetimes are seen for other young systems \citep[Figure \ref{fig:fits}, e.g.,][]{Cale2021}.  We adopt these three posteriors as priors in constraining the GP model for our RV stellar activity analysis.

\begin{figure}
    \centering
    \includegraphics[width=.45\textwidth]{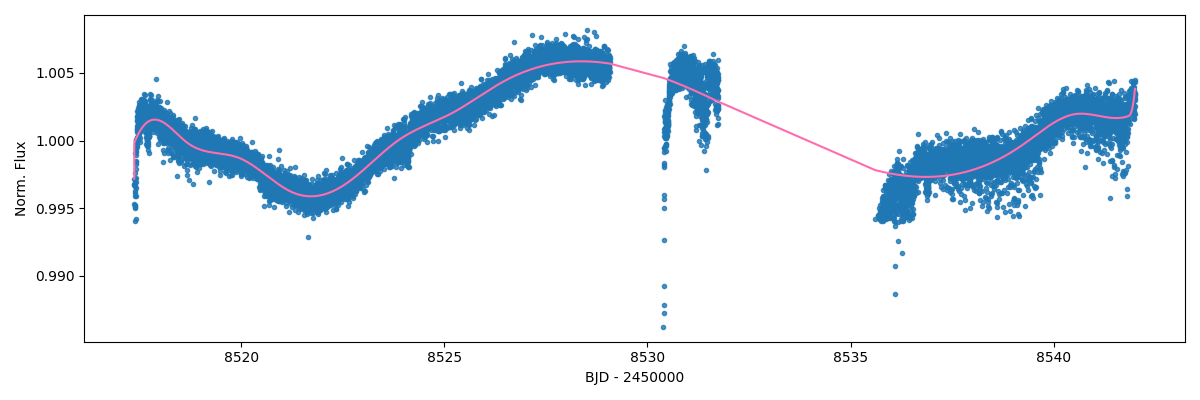}
    \includegraphics[width=.45\textwidth]{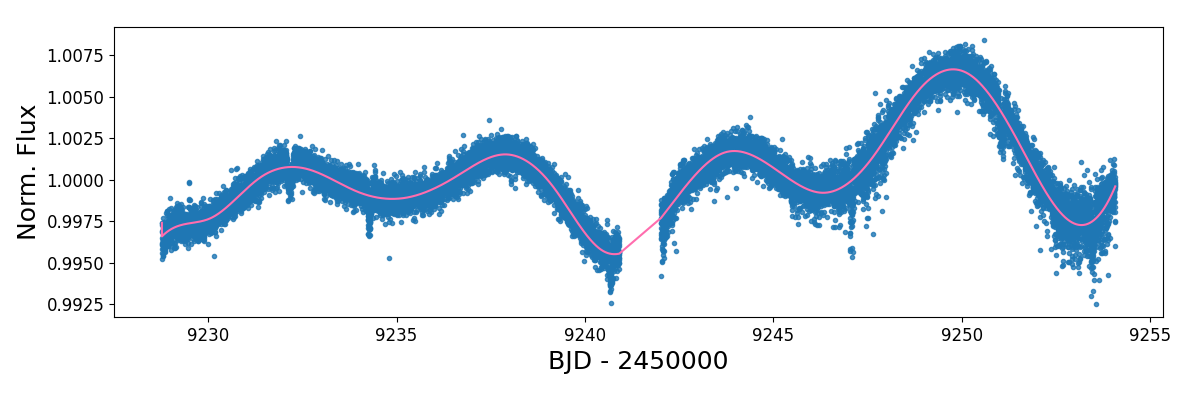}
    \includegraphics[width=.45\textwidth]{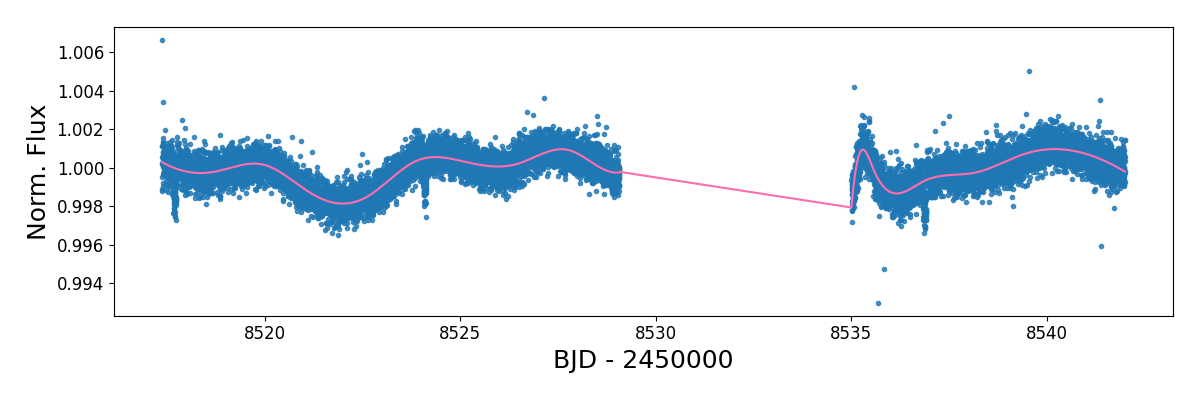}
    \includegraphics[width=.45\textwidth]{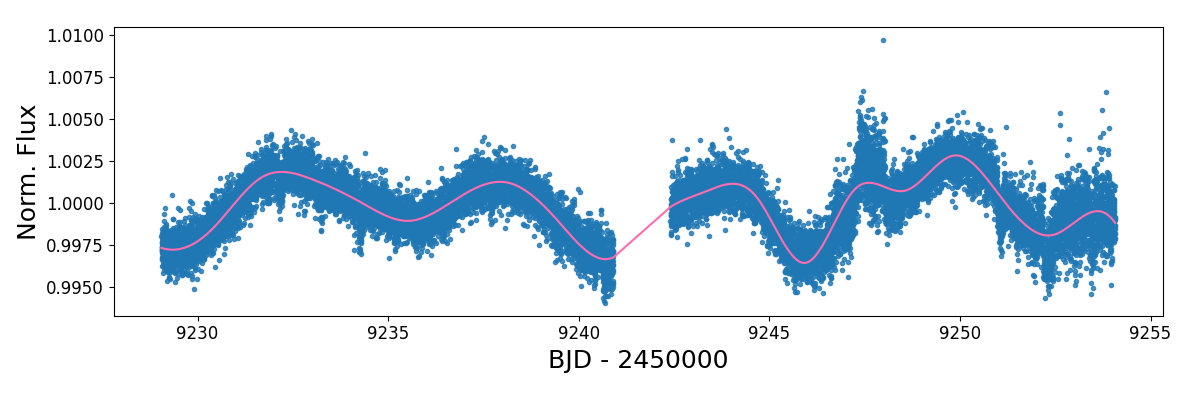}
    \caption{The \tess\ SAP (top two) and PDC-SAP (two bottom) light curves of TOI 560 from Sectors 8 and 34, plotting the normalized flux on the vertical axes as a function of time on the horizontal axes. The light curves are shown as blue data points, and the cubic spline regression is shown as the pink line. The interpolation in the data gap downlink region in the middle of each sector is subsequently discarded in our analysis. Significant photometric modulation due to stellar activity is apparent in both sectors.} 
    \label{fig:lc8+34}
\end{figure}

\begin{figure}
    \centering
    \includegraphics[width=.45\textwidth]{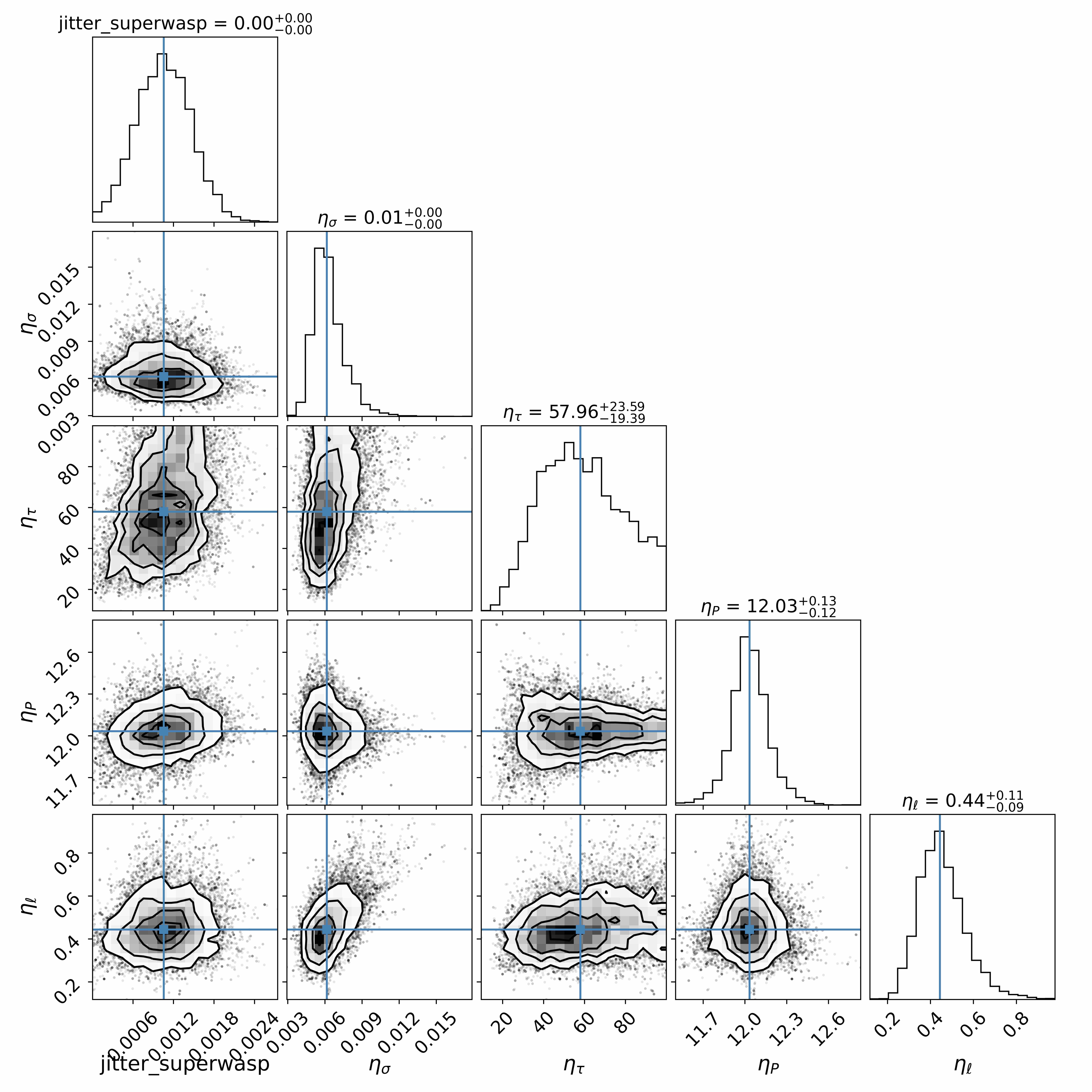}
    \caption{Posterior distributions (along diagonal) and two-parameter covariances (off-diagonal) for the quasi-periodic kernel GP model hyper-parameters 
    of the SuperWASP light curves. The five GP hyper-parameters as described in the text are indicated and median and 68\% confidence interval ranges are displayed at the top of each posterior distribution; median values are also indicated with horizontal and vertical blue lines for the covariance plots, and vertical lines for the posterior distribution. For the covariance plots, 1, 2 and 3$-\sigma$ contours are shown in place of the individual sample values $<$3$-\sigma$ from the medians.}
    \label{fig:fits} 
\end{figure}

 We next analyze both Sectors of TESS PDC-SAP and SAP separately in order to see if we can recover similar GP kernel hyper parameters $\{\eta_\sigma,\eta_\tau,\eta_P,\eta_L\}$ to our SuperWASP analysis. We first median normalize the \tess\ SAP light curve, mask out the transits, and the edges of the light curve data, particularly for Sector 8 where there exists a spacecraft systematic producing a ``ramp-up'' effect at the start of the sector and after the data downlink gap mid-sector. We then fit the remaining light curve via a cubic spline regression (\texttt{scipy.interpolate.LSQUnivariateSpline}; \citet{Virtanen2020}) for each Sector individually with knots sampled in units of 1.3 days (excluding any that happen to fall within the TESS data dump regions). 
The particular value of 1.3 is chosen to be small with respect to the stellar activity time-scales, but long with respect to the cadence, so transits are ``smoothed'' over to produce a ``noiseless'' and smooth representation of the light curves. Then, in Figure \ref{fig:lc8+34} we apply the $FF^\prime$ technique \citep{Aigrain2012} to the spline regression fits.

We used a uniform prior centered on 12.2 days for the stellar rotation period hyper-parameter. As expected due to the relatively short TESS time baselines of two $\sim$13-day observing sequences per sector, the \tess\ light curve does not constrain the stellar rotation period further than the information inferred from the SuperWASP light curve analysis.

For \tess\ SAP sector 8 (34; 8+34) light curve we recover from the MCMC a spot lifetime $\eta_\tau$ of 12.94$^{+23.55}_{-9.03}$ (10.10$^{+23.49}_{-6.62}$ days; 15.97$^{+22.55}_{-11.74}$) days.  In other words, the TESS light curve analysis leads to a significantly shorter spot life-time analysis than inferred from the SuperWASP analysis as well, even for the SAP fluxes, but we discard these results as a consequence of the shorter \tess\ time baseline compared to the SuperWASP light curve, even though the former is at higher photometric precision.  For \tess\ SAP sector 8 (34; 8+34) light curve we recover from the MCMC a smoothing hyper-parameter $\eta_l = 0.25\pm0.03$ ($\eta_l = 0.25\pm0.03$;$\eta_l = 0.25\pm0.03$), and these are all somewhat smaller than the value inferred from the SuperWASP analysis, and this could potentially be due to the higher precision of the \tess\ light curves and improved photometric sampling of the rotation period. Finally, from a MCMC of the \tess\ PDC-SAP light curves, we recover even shorter spot lifetime $\eta_\tau$, indicating that the pre-search data conditioning (PDC) can lead to introducing systematics that lead to inaccurate characterization of the stellar activity (Figure \ref{fig:sapfits2}).

\begin{figure}
    \centering
   \includegraphics[width=.40\textwidth]{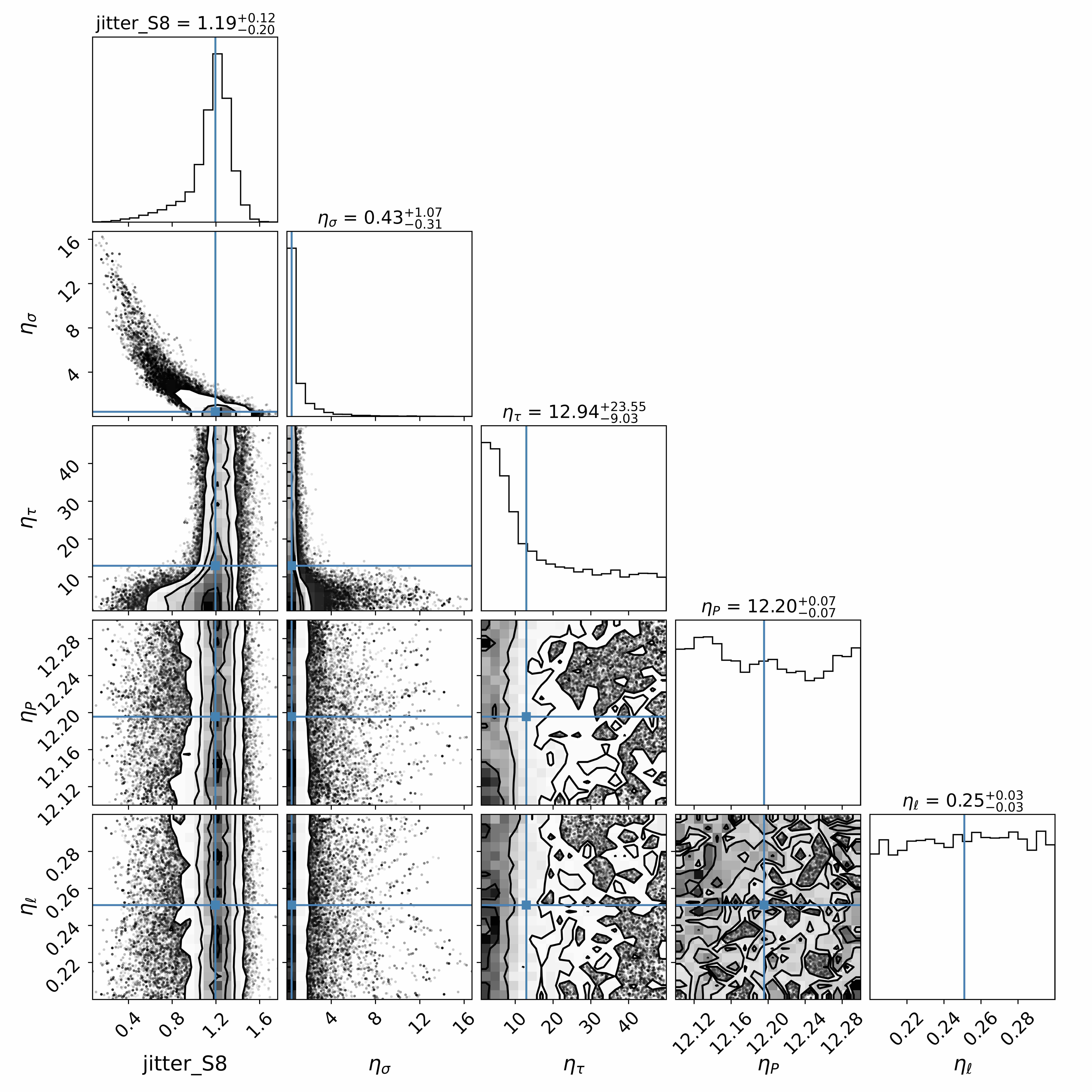}
    \includegraphics[width=.40\textwidth]{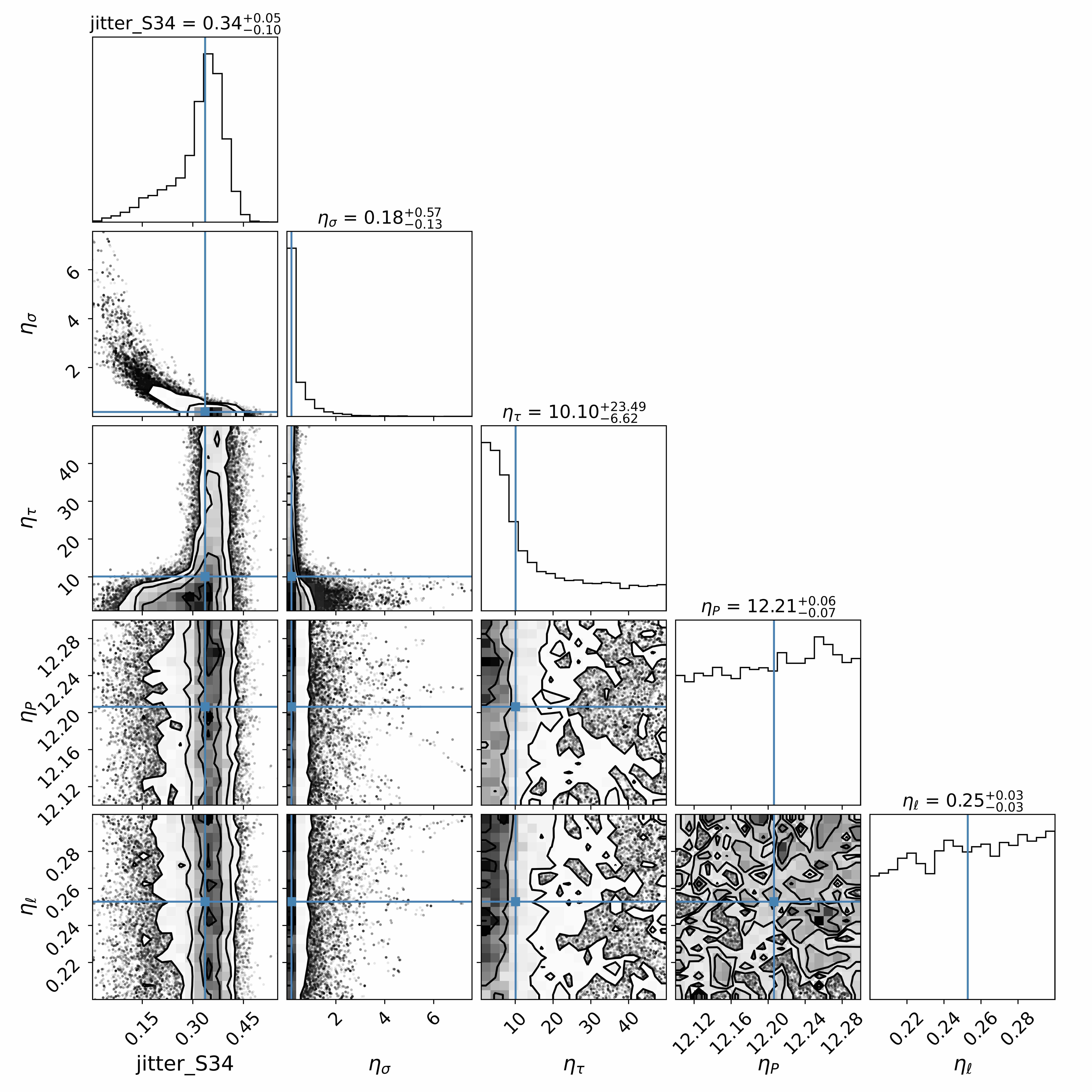}
    \includegraphics[width=.40\textwidth]{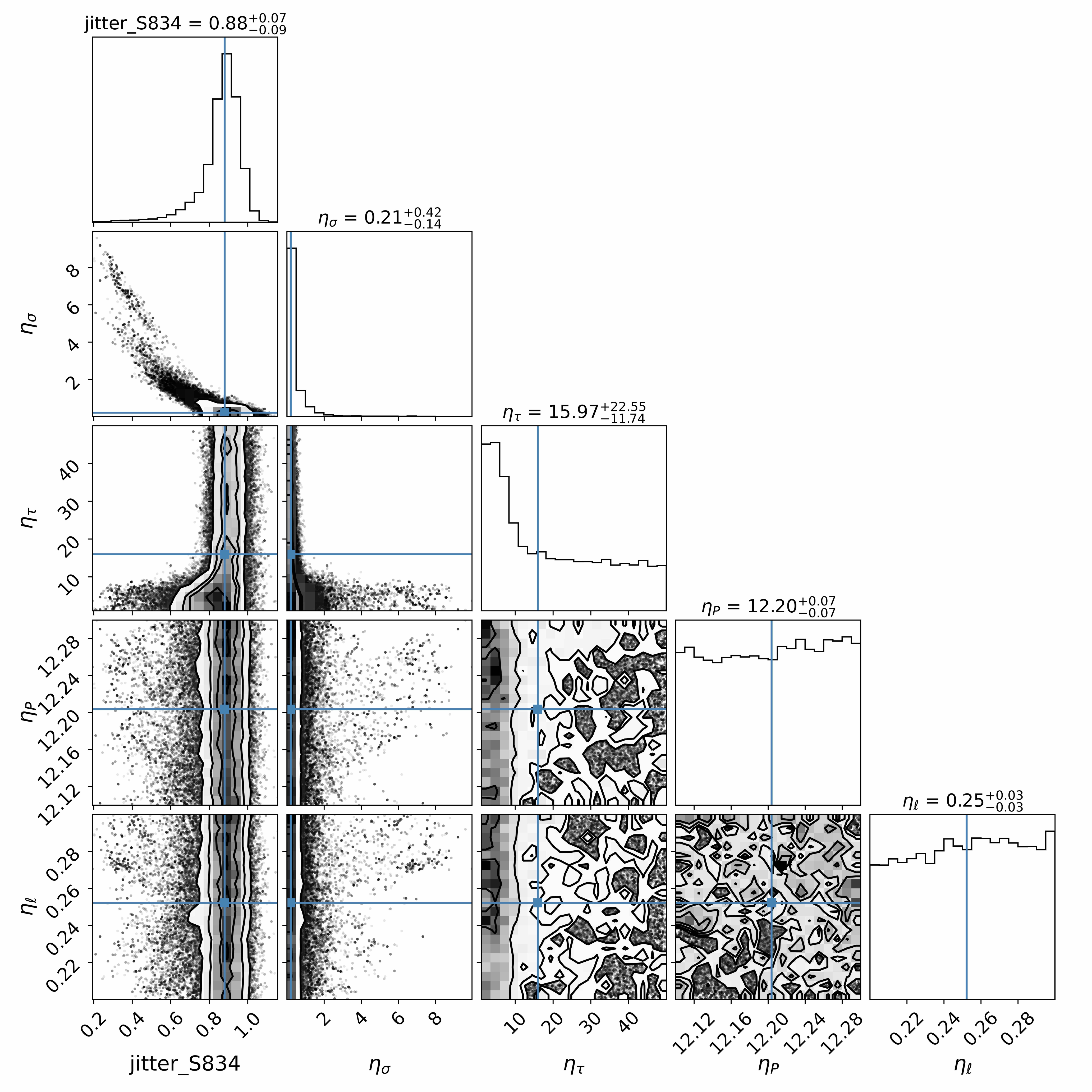}
    \caption{Posterior distributions (along diagonal) and two-parameter covariances (off-diagonal) for the quasi-periodic kernel GP model hyper-parameters of the F$F^{'}$ analysis of the \tess \ SAP light curves. }
    \label{fig:sapfits} 
\end{figure} 

\begin{figure}
    \centering
    \includegraphics[width=.45\textwidth]{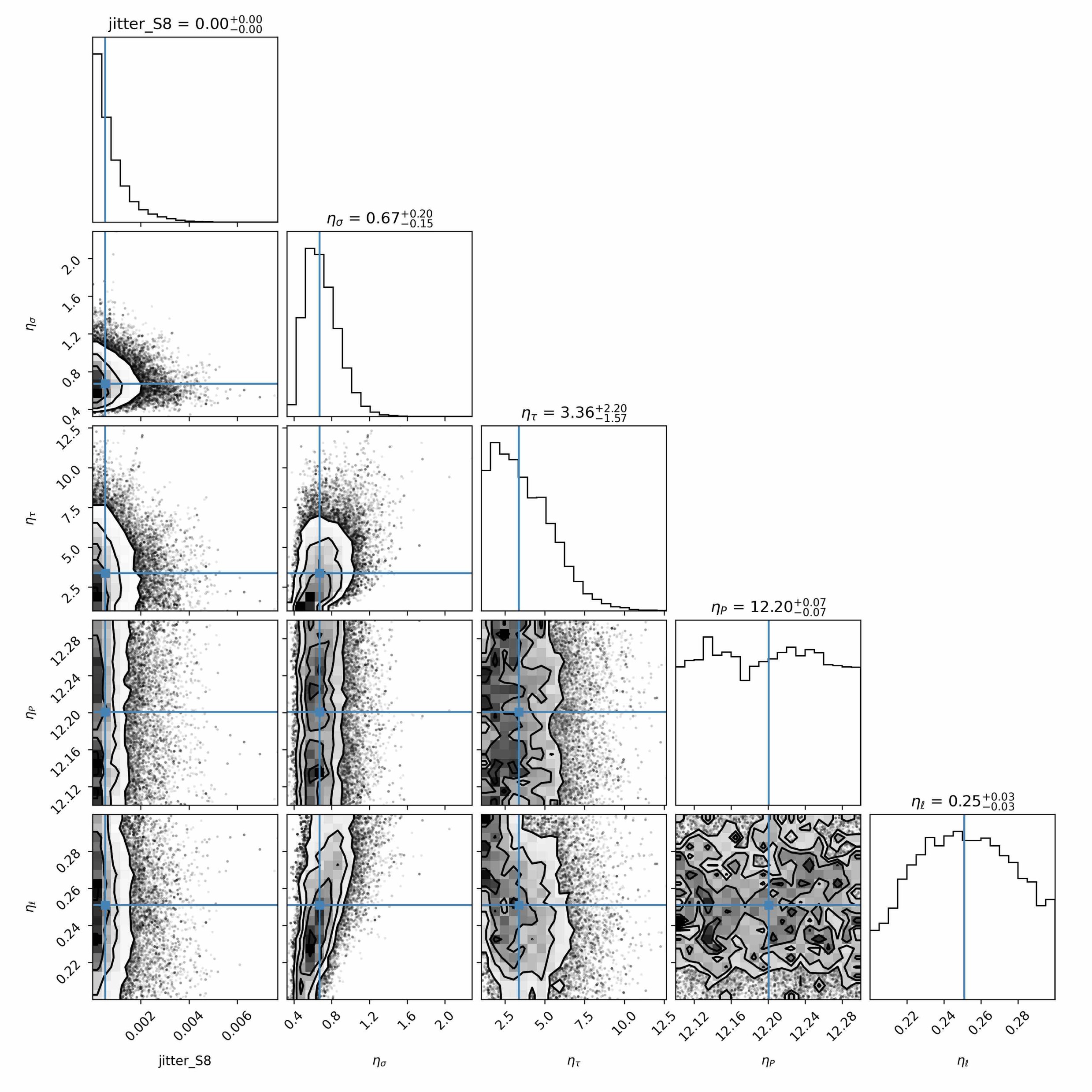}
    \caption{Posterior distributions (along diagonal) and two-parameter covariances (off-diagonal) for the quasi-periodic kernel GP model hyper-parameters of the 
    F$F^{'}$ analysis of the 
    \tess \ PDC-SAP sector 8 light curves.} 

    \label{fig:sapfits2} 
\end{figure}

\subsubsection{ExoFAST} 
\label{exofast}
In this section, we perform an \texttt{EXOFASTv2} analysis of the candidate planets transit light curves.
After normalizing the \textit{TESS} PDC-SAP data as described in section \ref{sect:observations}, we cut out a region of at least 8 hours around each individual transit and create separate data files for each transit and use these as input data for \texttt{EXOFASTv2}.

We jointly model the \textit{TESS}, \textit{Spitzer} and ground-based light curves, and all RVs with \texttt{EXOFASTv2}. The \texttt{EXOFASTv2} RV results are presented in the Appendix in Figure \ref{fig:eastmanrvs}. However, while ExoFAST can jointly model the light curves and RVs, the \texttt{EXOFASTv2} RV model does not account for the stellar activity manifested in the RV measurements. Thus, we perform an independent modeling of the RVs with a stellar activity model in ${\S}$\ref{rvanalysis}.

Our minimal set of priors are detailed in Table \ref{tab:exofasts}, including the period P, time of conjunction, and planetary radius $R_P /R_{*}$ for TOI 560 b and c (the \texttt{EXOFASTv2} full results are in Table \ref{tab:estmantab}). We allow other model parameters, i.e. eccentricity $e$, inclination $i$, and argument of periastron $\omega$ to vary with no imposed priors, starting with circular $e$ and edge-on $i$ values. The posterior values for this initial MCMC run are then used as the initial values for a second iteration run, though we keep the same uniform and Gaussian priors as the initial run. We allow this second run to run longer, and we confirm the second MCMC converges on the same results within 1$\sigma$ to check for the robustness of the MCMC posteriors. Note that the model is parameterized in the EXOFASTv2 standard basis for transit only fits. This model parameterization is $\{ T_{C} , \log P, R_{P}/R_{*}, \cos i, \sqrt e \sin \omega,\sqrt e  \cos \omega \}$ \citep{Eastman2019}.

\begin{table}
    \centering
    \begin{tabular}{c|c|c} 
        \hline
        Parameter & Initial  & Priors \\ 
            $[units]$ & value &  \\ 
                   & [$P_{0}$] & \\ 
        \hline
        \hline
        P$_b$ [days] & 6.3981 &$\mathcal{U}(P_{0} \pm 10\%)$ \\
        TC$_b$ [days] & 2458517.69007&$\mathcal{U}(P_{0}\pm P/3 )$\\ 
        P$_{c}$[days]& 18.881& $\mathcal{U}(P_{0} \pm 10\%)$\\
        TC$_{c}$[days]& 2458533.59092& $\mathcal{U}(P_{0}\pm P/3 )$\\
        $R_{p,b}/R_{*}$& 0.0388 & -- \\
        $R_{p,c}/R_{*}$& 0.0382 & -- \\
        $M_{p,b}/M_{\odot}$& 0.000030 & -- \\
        $M_{p,c}/M_{\odot}$& 0.000030 & -- \\
        $R_{*}$& 0.674 & --\\
        $R_{*,SED}$& 0.679 & --\\
        $T_{eff}$ [K]& 4599.94 & --\\
        $T_{eff,SED}$ [K]& 4591.85 & --\\
        feh & -0.031 & $\mathcal{N}(-0.210,0.080)$\\ 
        initfeh & -0.043 & --\\
        eep& 319.8 & --\\
        log$M_{*}$& -0.145 & --\\
        $A_{V}$& 0.097 & $\mathcal{U}(0,0.143)$\\
        errscale& 1.22& --\\
        distance [pc]& 31.567& --\\
        \hline
    \end{tabular}
    \caption{Planetary and stellar prior probability distributions for our EXOFASTv2 MCMC simulations. 
    $\mathcal{U}(\ell, r)$ signifies a uniform prior with left bound $\ell$ and right bound r. $A_{V}$ is the galactic extinction in the V band, feh is metallicity, initfeh is the initial metallicity at the time of the zero age main sequence, eep stands for equivalent evolutionary point, log$M_{*}$ is the log of the stellar mass, and errscale is the error scale. The eccentricity $e$ and the argument of periastron} $ \omega$ are assumed to be circular, with no priors. 
    \label{tab:exofasts}
\end{table}

\subsection{RV Analysis } 
\label{rvanalysis}
In this section, we jointly analyze our RVs (Table \ref{tab:RVs}) from three spectrographs (iSHELL, PFS, and HIRES). We analyze the RVs as a planetary system with a chromatic GP stellar activity model, and make model comparisons. We constrain our GP model hyper-parameters from the SuperWASP light curve analysis, in particular the stellar rotation period and spot lifetime timescale.

\subsubsection{Planet System Multi-Spectrograph RV Analysis}

To jointly model the RVs from iSHELL, PFS and HIRES, we again use \texttt{pychell}, this time in combination with the co-dependent package \texttt{optimize}, which is a general-purpose Bayesian analysis tool that \texttt{pychell} expands upon with RV-specific MCMC tools.  In Section (\ref{additional}) we show that no statistically significant periodic signals were identified in the raw RVs, and so we proceed herein with the assumption of a two planet model in our Bayesian analysis with the ephemerides from \tess. Each planet in our model consists of five parameters, which comprise a complete orbit basis: the period $P$, time of conjunction $TC$, eccentricity $e$, argument of periastron $\omega$, and RV semiamplitude $K$, with subscripts denoting the planet said parameter is associated with (in this case, b and c).  We next include a GP model for the stellar activity, which is described further in the next sub-section. We also include an absolute RV offset term ($\gamma$) to account for the average overall recessional velocity of the star, and a jitter term ($\sigma$) which quantifies the variational RV amplitude not accounted for by any modeled planets, stellar activity, or instrumental noise that is non-frequency dependent on our time-scales of interest (e.g. white noise).  Analyses of the PDC-SAP \textit{TESS} transit data (Section \ref{T-L-C} - Figure \ref{fig:TESSLC} and Table \ref{tab:exofasts}) shows a period and time of conjunction consistent with those listed on the ExoFOP-TESS database in the \textit{TESS} Object of Interest (\textit{TESS} Project section)\footnote{\href{https://exofop.ipac.caltech.edu/tess/target.php?id=101011575}{https://exofop.ipac.caltech.edu/tess/target.php?id=101011575}} \citep{Akeson_2013} which are $P_b = 6.398069 \pm 0.000015$ days, $P_c = 18.879744 \pm 0.000162$, $TC_b = 2458517.690108 \pm 0.000747$ days and $TC_c = 2458533.620329 \pm 0.005422$, all of which have an uncertainty which are orders of magnitude finer than we can hope to resolve with our RV cadence.  We thus decide to lock both parameters at their nominal values for all analyses in this paper. For all MCMC results quoted, the median value is defined as the 50$^{\mathrm{th}}$ percentile, while the lower and upper uncertainty bounds are the 15.9$^{\mathrm{th}}$ and 84.1$^{\mathrm{th}}$ percentile posterior confidence intervals, respectively. We sample the posterior distributions using the emcee package \citep{Foreman-Mackey} for a subset of models to determine parameter uncertainties, always starting from the MAP-derived parameters (i.e. \texttt{affine invariant} is our actual \texttt{sampler}. We perform a burn-in phase of 1000 steps followed by a MCMC analysis for approximately fifty times the median auto-correlation time (steps) of all chains. The prior, posterior distributions, and results of this model is in section \ref{sect:resu}. Finally, we also perform a model comparison test (Table \ref{tab:compJ2}) by calculating the $\ln \mathcal{L}$, small-sample Akaike Information Criterion \citep[AIC;][]{Akaike_1974, Burnham_2002}, and Bayesian Information Criterion (BIC) for each combination of planets in our model.  These results are shown in Section ${\S}$\ref{sect:rv_mcmcs}.

\begin{table*}
\centering
\begin{tabular}{lrccrr}
    \hline
    Parameter [units] & Initial Value ($P_{0}$) & Priors & MAP Value $J_2$ & MCMC Posterior $J_2$ \\
    \hline
    \hline
    $P_b$ [days] & 6.3980661 & \faLock & -- & -- \\
    $TC_b$ [days] & 2458517.68971 & \faLock &  -- & -- \\
    $e_b$ & 0.294 & $\mathcal{U}(0,1)$; $\mathcal{N}(P_{0},0.13)$ & 0.0.33 & $0.29^{+0.09}_{-0.09}$ \\
    $\omega_b$ & $130\pi/180$ & $\mathcal{U}(P_{0}-\pi,P_{0}+\pi)$; $\mathcal{N}(P_{0},45\pi/180)$& 4.11 & $3.94^{+0.26}_{-0.49}$ \\
    $K_b$ [m\,s$^{-1}$] & 10 & $\mathcal{U}(0, \infty)$ & 6.57 & $6.98^{+1.76}_{-1.82}$ \\
    \hline
    $P_c$ [days] & 18.8805 & \faLock & -- & -- \\
    $TC_c$ [days] & 2458533.593 & \faLock & -- & -- \\
    $e_c$ & 0.093 & $\mathcal{U}(0,1)$; $\mathcal{N}(P_{0},0.13)$ & 0.19 & $0.19^{+0.10}_{-0.10}$ \\
    $\omega_c$ & $-190\pi/180$ & $\mathcal{U}(P_{0}-\pi,P_{0}+\pi)$; $\mathcal{N}(P_{0},45\pi/180)$ & -3.87 & $-3.53^{+0.62}_{-0.063}$ \\
    $K_c$ [m\,s$^{-1}$] & 10 & $\mathcal{U}(0, \infty)$ & 6.80 & $6.64^{+1.36}_{-1.42}$ \\ 
    \hline
    $\gamma_{iSHELL}$ [m\,s$^{-1}$] & $\mathrm{MEDIAN}(RV_{iSHELL})+\pi/100^{a}$ & $\mathcal{N}(P_{0},100)$ & 4.22 & $3.52^{+4.63}_{-4.62}$ \\
    $\gamma_{PFS}$ [m\,s$^{-1}$] & $\mathrm{MEDIAN}(RV_{PFS})+\pi/100^{a}$ & $\mathcal{N}(P_{0},100)$ & -13.28 & -$13.30^{+17.87}_{-17.63}$ \\
    $\gamma_{HIRES}$ [m\,s$^{-1}$] & $\mathrm{MEDIAN}(RV_{HIRES})+\pi/100^{a}$ & $\mathcal{N}(P_{0},100)$ & 4.59 & $7.07^{+13.68}_{-13.48}$ \\ 
    $\eta_{P}$ & 12.03 & $\mathcal{N}(P_{0},0.13)$& 12.02 & $12.00^{+0.09}_{-0.0}$\\
    $\eta_{\ell}$ & 0.44 & \faLock &  -- & --\\
    $\eta_{\tau}$ & 57.96 & \faLock&   -- & --\\
    $\eta_{\sigma,0}$ & 1 & $\mathcal{J}(0.67,50)$ & 23.42 & $27.87^{+5.58}_{-4.30}$\\
    $\eta_{\lambda}$ & 1.17 & $\mathcal{U}(-1,2)$ & 0.61 & $0.66^{+0.30}_{-0.30}$\\
    \hline
\end{tabular}
\caption{The model parameters and prior distributions used in our RV model that considers the transiting b and c planets, as well as the recovered MAP fit and MCMC posteriors for the J2 Kernel. Initial values for orbital period, timing conjunction, eccentricity, and planet radius for both planets as well as stellar mass come from our ExoFAST analysis in ${\S}$\ref{exofast}, and the stellar activity model hyperparameters and prior distributions come from our SuperWASP light curve analysis in ${\S}$\ref{sect:challenge}}. \faLock\ indicates the parameter is fixed.  Gaussian priors are denoted by $\mathcal{N}(\mu, \sigma)$, uniform priors by $\mathcal{U}($lower bound, upper bound$)$, and Jeffrey's priors by $\mathcal{J}($lower bound, upper bound$)$. The $+1$ on the initial gamma values are in case the RVs are already median-subtracted, and to pseudo-randomize this initial parameter. \\ 
$^{a}$ We want the initial value to be the median of the RVs for that spectrograph; the $+1$ is used incase the median is already zero, as Nelder-Mead solvers cannot start at zero.
\label{tab:bc_priors}
\end{table*}

\subsubsection{Stellar Activity RV Model - RVs} 
\label{sect:challenge2}

Rather than applying an independent GP model to each individual RV data set for PFS, HIRES and iSHELL, we extend the kernel in Equation 1 to utilize a single global (joint) GP model and covariance kernel across multiple spectrographs.  We follow \cite{Cale2021} where they already implemented our desired framework in two Python packages. As in \cite{Cale2021}, we first re-parametrize the GP amplitude through a linear kernel, hereafter referred to as the $J_1$ kernel as in \citet{Cale2021} as follows: 

\begin{equation}
    K_{J_1}(t_{i},t_{j}) = \eta_{\sigma,s(i)} \eta_{\sigma,s(j)} \times \exp[...] 
    \label{eq2}
\end{equation}
\noindent Here,  $\eta_{\sigma,s(i)}$,  $\eta_{\sigma,s(j)}$are the effective stellar activity amplitudes for each spectrograph $s$ at times $t_{i}$ and $t_{j}$, respectively, where $s(i)$ represents an indexing set between the observations at time $t_{i}$\footnote{Truly simultaneous measurements with identical mid-point exposure times, i.e., $t_{i}$ = $t_{j}$, would necessitate a more sophisticated indexing set.}. In the $J_{1}$ kernel each instrument is given its own independent amplitude hyper-parameter, $\eta_{\sigma,s(i)}$, but the other three hyper-parameters are shared between all instruments. Also, the covariance kernel is still square, but now has dimensions corresponding to the total number of observations across all spectrographs, and thus represents a ``joint'' covariance matrix.

Second, to first order, we expect the amplitude from stellar activity to be linearly proportional to frequency (or inversely proportional to wavelength) \citep{Cale2021}. This approximation is a direct result of the spot-contrast scaling with the photon frequency (or inversely with wavelength) from the ratio of two black-body functions with different effective temperatures \citep{2010ApJ...710..432R}. Thus, we also consider a variation of this kernel which further enforces the expected inverse relationship between the amplitude with wavelength. As in \cite{Cale2021} we parametrize the kernel to become:
\begin{equation}
    K_{J_2}(t_{i},t_{j},\lambda_{i},\lambda_{j})= \eta_{\sigma,0}^{2} \left( \frac{\lambda_{0}}{\sqrt{\lambda_{i}\lambda_{j}}} \right)^{2 \eta_{\lambda}} \times \exp[...]
    \label{eq3}
\end{equation}
\noindent which hereafter we refer to as the $J_2$ kernel as in \citet{Cale2021}. Here, $\eta_{\sigma,0}$ is the effective amplitude at $\lambda=\lambda_{0}$, and $\eta_{\lambda}$ is an additional power-law scaling parameter with wavelength to allow for a more flexible non-linear (with frequency) relation. $\lambda_{i}$ and $\lambda_{j}$ are the “effective” wavelengths for observations at times $t_{i}$ and $t_{j}$, respectively.  Note for this kernel, we are ignoring the chromatic effects of limb-darkening and convective blueshifts on the RVs which will impact the covariance matrix; however given the flexibility of the GPs, \citet{Cale2021} finds this chromatic kernel effectively recovers the wavelength dependence of stellar activity induced RV variations.  For both eqs. \eqref{eq2} and \eqref{eq3}, the expression within square brackets is identical to that in eq. \eqref{eq1}. 

We apply Kernels $J_1$ and $J_2$ to our RV data, using the priors in Table \ref{tab:bc_priors}, the priors from our FF$^{'}$ analysis in ${\S}$\ref{sect:challenge} for the model hyper-parameters, as well as a set of disjoint quasi-periodic kernels as in Equation 1, one GP for each spectrograph akin to \texttt{RadVel}  \citep{2018PASP..130d4504F}. In the analysis presented herein we fix $\eta_l$ and $\eta_\tau$ to the median values from the SuperWASP light curve analysis posteriors; letting these model hyper-parameters vary in our RV analysis -- with prior minimums of 0.2 and 20 days respectively -- yields quantitatively identical results for the median posterior values for $K_b$ and $K_c$, albeit with slightly larger confidence intervals. We also analyze the RVs using no GP, effectively assuming the stellar activity is not present.

\section{Results} 
\label{sect:resu}

In ${\S}$\ref{results:transits}, we present the transit analyses of the \textit{TESS}, \textit{Spitzer} and ground-based light curves using \texttt{EXOFASTv2}. Then in ${\S}$\ref{sect:rv_mcmcs} we present the RV analysis with \texttt{pychell} and the GP stellar activity model.   

\subsection{Transit Light Curves} 
\label{results:transits}

In this subsection we represent the results of our \texttt{ExoFASTv2} analysis of the light curves and vetting results.

\subsubsection{\tess\ Light Curves} 
\label{T-L-C}

\begin{figure}
    \centering
    \includegraphics[width=0.49\textwidth]{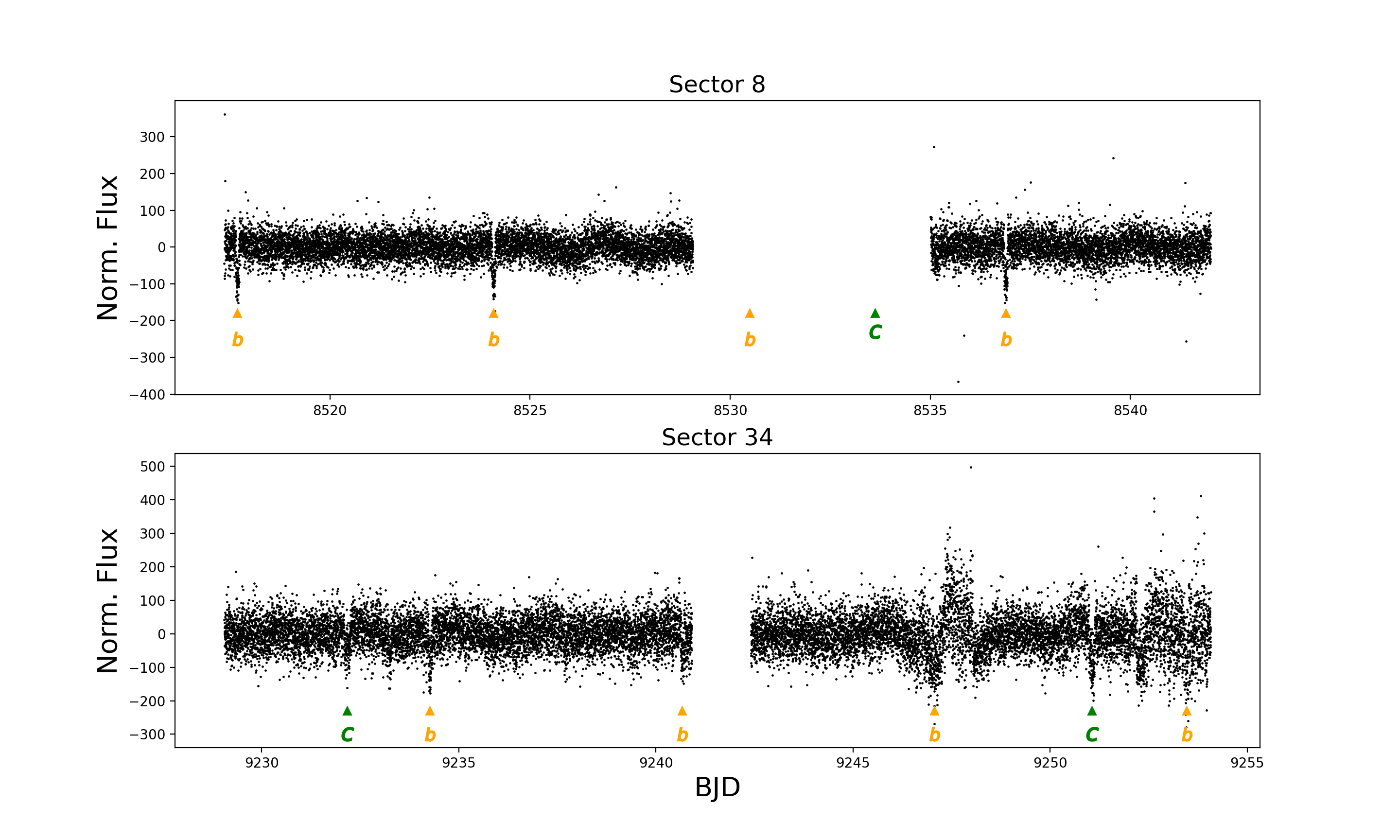}
    \caption{\tess\ Sector 8 (top) and 34 (bottom ) \tess\ light curves, after subtracting from the PDC-SAP time-series the cubic spline fit model for the stellar activity in Figure \ref{fig:lc8+34}, shown as normalized flux on the vertical axis as a function of time in BJD on the horizontal axis. The transit midpoint times of TOI 560 b and c are indicated.} 
    \label{fig:TESSLC} 
\end{figure}

\begin{figure}
    \centering
    \includegraphics[width=0.49\textwidth]{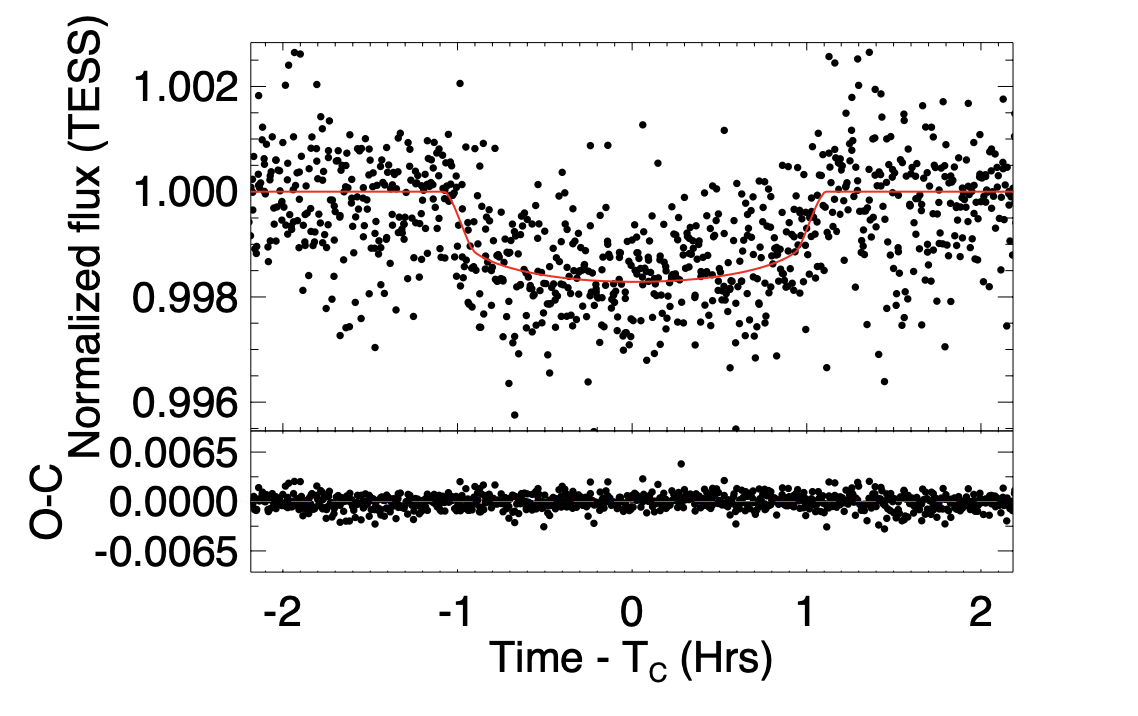}
    \includegraphics[width=0.49\textwidth]{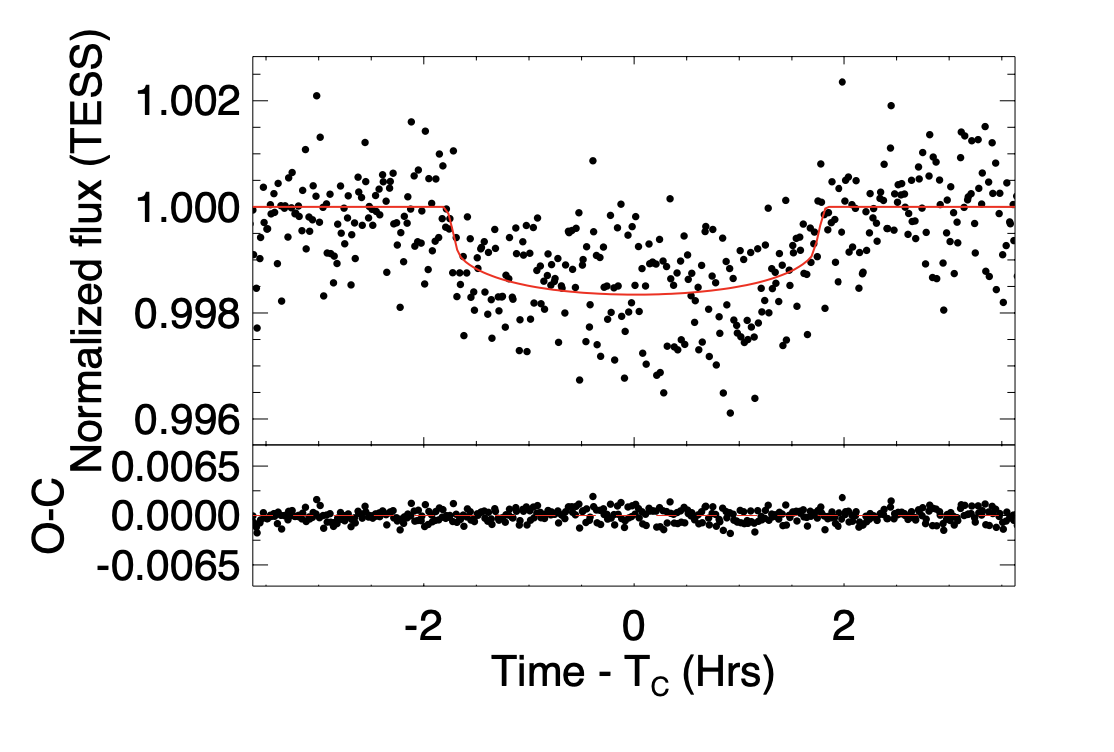}
    \caption{TOI 560 b (top) and TOI 560 c (bottom) \tess\ phase-folded transit light curves, plotted as normalized flux as a function of time since transit conjunction in hours on the horizontal axis. The median MCMC transit model are overlaid as red lines. The bottom panel show the median-model subtracted residuals.} 
    \label{fig:eastmantransits} 
\end{figure}

\begin{figure}
    \includegraphics[width=1.0\linewidth]{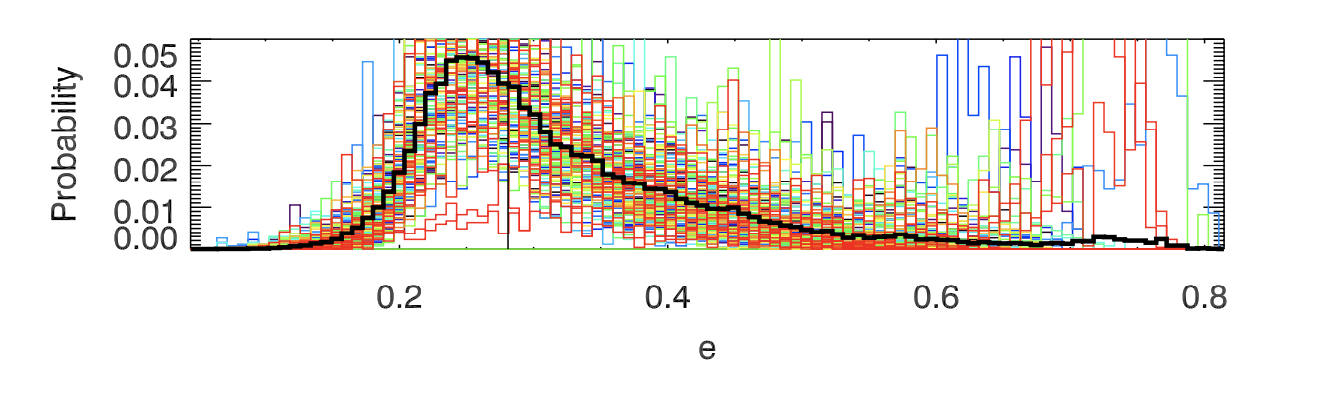}
    \includegraphics[width=1.0\linewidth]{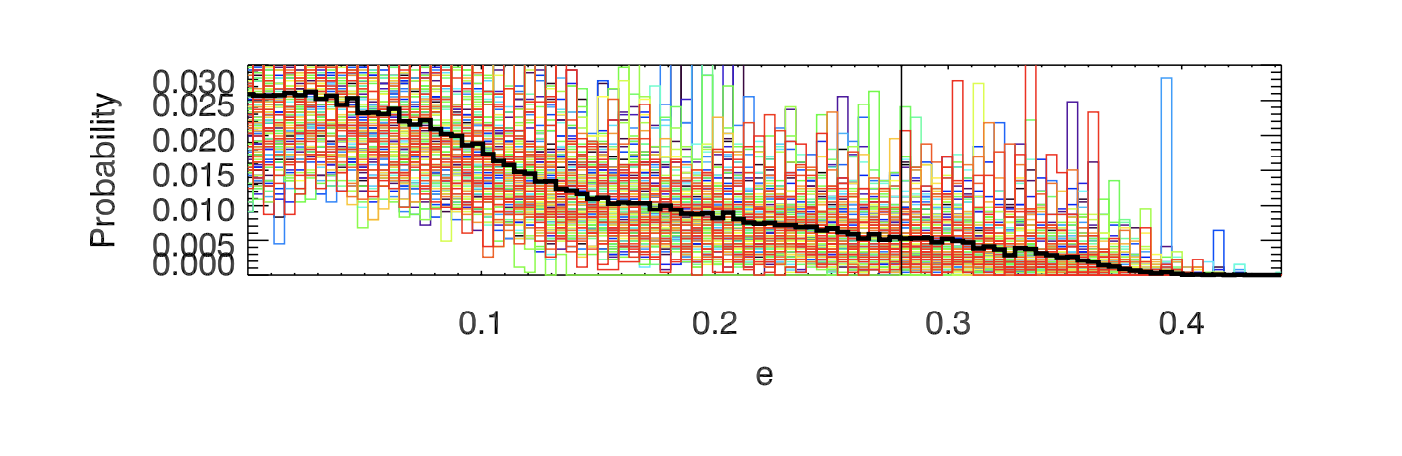}
    \caption{Eccentricity posteriors for TOI 560 b and c from our \texttt{ExoFASTv2} results.  The eccentricity values are on the horizontal axis, and normalized probabilities on the vertical axis.  The colors histograms show individual MCMC walker chains, with the black histogram showing the medians of all walkers. The posterior for TOI 560 b slightly favors a mild eccentricity, whereas for TOI 560 c a circular orbit is preferred.}
    \label{fig:ecc}
\end{figure}

The TOI 560 \tess\ light curves for Sectors 8 and 34 are shown in Figure \ref{fig:TESSLC}, after subtracting off a cubic-spline regression fit for the out-of-transit stellar activity from the PDC-SAP light curves shown in Figure \ref{fig:lc8+34}. The \texttt{ExoFASTv2} analysis reveals clear transits of TOI 560 b and c with expected depths and transit times consistent with the \tess\ mission TOI 560 b  and 560.02 candidates. Table \ref{tab:estmantab} shows the median and confidence intervals for the model and derived planetary parameters.  Figure \ref{fig:eastmantransits} shows the phased \tess\ transits. Of particular note, we confirm a non-zero eccentricity for TOI 560 b at 4.7-$\sigma$; the eccentricity posterior for TOI 560 c is consistent with zero (Figure \ref{fig:ecc}). The statistical significance of the non-zero eccentricity detection for TOI 560 b is primarily driven by the photo-eccentric effect using the \textit{Spitzer} data \citep{2012ApJ...756..122D}, as excluding this particular data set decreases the statistical significance on $e_b>0$ to $1.6-\sigma$, but still with a similar non-zero median eccentricity. Finally, the \textit{Spitzer} light curve is presented in ${\S}$\ref{Spitzerresults}, ground-based PEST, NGTS, and LCO light curves are presented in the Appendix. 

\begin{table*}[h!]
     \begin{center}
    \begin{tabular}{|c|l|r|r|} 
        \hline
        Planetary Parameter & Units  & Planet b & Planet c\\ 
        \hline
        \hline
        P& Period (days)  & $6.3980661^{+0.0000095}_{-0.0000097}$&$18.8805^{+0.0024}_{-0.0011}$ \\
        R$_{P}$&  Radius (R$_{J}$)& $0.253^{+0.014}_{-0.011}$& $0.2433^{+0.011}_{-0.0096}$\\
        R$_{P}$&  Radius (R$_{Nep}$)& $0.74^{+0.04}_{-0.03}$& $0.71^{+0.03}_{-0.03}$\\
        R$_{P}$&  Radius (R$_{\oplus}$)& $2.84^{+0.16}_{-0.12}$& $2.73^{+0.12}_{-0.11}$\\
        T$_{c}$& Time of conjunction$^a$ (BJD$_{TDB}$) & $2458517.68971^{+0.00053}_{-0.00054}$&$2458533.593^{+0.041}_{-0.091}$\\
        T$_{T}$& Time of minimum projected separation$^b$ (BJD$_{TDB}$ ) & $2458517.68973^{+0.00053}_{-0.00054}$&$2458533.593^{+0.041}_{-0.091}$\\ 
        T$_{0}$& Optimal conjunction Time$^c$ (BJD$_{TDB}$)& $2458703.23362^{+0.00044}_{-0.00045}$&$2459251.0510^{+0.0016}_{-0.0014}$\\
        $a$&Semi-major axis (AU)& $0.05995^{+0.00074}_{-0.00072}$&$0.1233 \pm 0.0015$\\
        $i$& Inclination (Degrees) & $89.45^{+0.38}_{-0.50}$&$89.61 ^{+0.26}_{-0.24}$\\
        $e$& Eccentricity & $0.294^{+0.13}_{-0.062}$&$0.093^{+0.13}_{-0.066}$\\
        $\omega_{*}$& Argument of Periastron (Degrees) & $130^{+30}_{-46}$&$-190 \pm 130$\\
        T$_{eq}$& Equilibrium temperature$^d$ (K) & $742.7^{+7.2}_{-7.0}$&$517.8^{+5}_{-4.9}$\\
        $\tau_{cir}$& Tidal circularization timescale (Gyr)& $44^{+69}_{-36}$&$9500^{+15000}_{-7100}$ \\
        $K$& RVsemi-amplitude(m/s)$^e$  & $2.9 ^{+2.7}_{-1.9}$&$1.5^{+1.9}_{-1.1}$\\
        $R_{P}/R_{*}$& Radius of planet in stellar radii & $0.03803^{+0.00063}_{-0.00064}$&$0.0379 \pm 0.0011$\\
        $a/R_{*}$& Semi-major axis in stellar radii  & $19.03^{+0.51}_{-0.48}$&$39.15^{+1.0}_{-0.99}$\\
        $\delta$& $(R_{P}/R_{*})^{2}$& $0.001446 \pm 0.000048$&$.001433^{+0.000086}_{-0.000082}$\\
        $\delta_{\textit{B}}$& Transit depth in B (fraction)& $0.00192^{+0.00059}_{-0.00032}$&$0.00186^{+0.00056}_{-0.00029}$\\
        $\delta_{\textit{R}}$& Transit depth in R (fraction)& $0.00216^{+0.00080}_{-0.00047}$&$0.00208^{+0.00076}_{-0.00043}$\\
        $\delta_{\textit{z'}}$& Transit depth in z' (fraction)& $0.00202^{+0.00085}_{-0.00041}$&$0.00195^{+0.00081}_{-0.00036}$\\
        $\delta_{\textit{4.5}\mu m}$& Transit depth in 4.5 $\mu$m (fraction)& $0.0046^{+0.0023}_{-0.0012}$&$0.0042^{+0.0023}_{-0.0012}$\\
        $\delta_{TESS}$& Transit depth in \tess\ (fraction)& $0.00163^{+0.00021}_{-0.00013}$&$0.0042^{+0.0023}_{-0.0012}$\\
        $\tau$& Ingress/egress transit duration (days) & $0.003232^{+0.00020}_{-0.000078}$&$0.00588^{+0.00098}_{-0.00040}$
        \\ 
        $T_{14}$& Total transit duration (days)& $0.0866 \pm 0.0014$&$0.1510^{+0.0039}_{-0.0032}
        $\\
        $T_{FWHM}$ & FWHM transit duration (days)& $0.0833^{+0.0014}_{-0.0013}$&$0.1448^{+0.0038}_{-0.0031}$\\
        $b$&Transit Impact parameter& $0.134^{+0.13}_{-0.093}$&$0.26^{+0.18}_{-0.17}$\\
        $b_{s}$& Eclipse impact parameter&  $0.19^{+0.18}_{-0.13}$&$0.26^{+0.13}_{-0.17}$\\
        $T_{s}$&Ingress/egress eclipse duration (days)& $0.00500^{+0.00049}_{-0.00066}$&$0.00597^{+0.00047}_{-0.00046}$\\
        $T_{s,14}$&Total eclipse duration (days)& $0.127^{+0.012}_{-0.017}$&$0.152^{+0.013}_{-0.015}$\\
        $T_{s,FWHM}$&FWHM eclipse duration (days)& $0.122^{+0.012}_{-0.016}$&$0.147^{+0.012}_{-0.015}$\\
        $T_{s,2.5 \mu m}$&Blackbody eclipse depth at $2.5 \mu m$ (ppm)&$1.57^{+0.12}_{-0.11}$&$0.0537^{+0.0065}_{-0.0059}$\\
        $T_{s,5.0 \mu m}$&Blackbody eclipse depth at $5.0\mu m$ (ppm)& $26.8^{+1.3}_{-1.2}$&$4.85^{+0.39}_{-0.36}$\\
        $T_{s,7.5 \mu m}$&Blackbody eclipse depth at $7.5\mu m$& $61.4^{+2.7}_{-2.5}$&$18.8^{+1.3}_{-1.2}$\\
        $\langle F \rangle $&Incident Flux ($10^{9}$ erg s$^{-1}$ m$^{-2}$)& $0.0629^{+0.0034}_{-0.0054}$&$0.01600^{+0.00071}_{-0.00078}$\\
        $T_{P}$&Time of Periastron  (BJD$_{TDB}$)& $2458511.67^{+0.15}_{-0.43}$&$2458517.0^{+4.0}_{-5.8}$\\
        $T_{S}$&Time of eclipse(BJD$_{TDB}$)& $2458520.15^{+0.83}_{-0.85}$&$2458524.1^{+1.3}_{-1.7}$\\
        $T_{A}$&Time of Ascending Node(BJD$_{TDB}$)& $2458516.27^{+0.32}_{-0.49}$&$2458528.85^{+0.70}_{-1.0}$\\
        $T_{D}$&Time of Descending Node (BJD$_{TDB}$)& $2458512.20^{+0.27}_{-0.25}$&$2458538.22 \pm 0.80$\\
        $V_{c}/V_{e}$&Equivalent Circular to Measured Eccentric Velocity ratio& $0.791^{+0.029}_{-0.027}$&$0.991^{+0.061}_{-0.048}$\\
        $e \cos \omega_{*}$& Eccentricity times Cosine of the Periastron Angle & $-0.18^{+0.20}_{-0.22}$&$-0.004^{+0.11}_{-0.14}$\\
        $e \sin \omega_{*}$& Eccentricity times Sine of the Periastron Angle & $0.199^{+0.043}_{-0.069}$ & $-0.003^{+0.047}_{-0.066}$\\
        $d/R_{*}$& Separation at mid transit & $14.2^{+1.1}_{-1.2}$&$38.6 \pm 2.7$\\
        $P_{T}$ & A priori non-grazing transit prob & $0.0676^{+0.0061}_{-0.0048}$&$0.0249^{+0.0018}_{-0.0016}$\\
        $P_{T,G}$&A priori transit prob& $0.0730^{+0.0066}_{-0.0051}$&$0.0269^{+0.0020}_{-0.0017}$\\
        $P_{S}$&A priori non-grazing eclipse prob& $0.0437^{+0.093}_{-0.0024}$&$0.0247^{+0.0025}_{-0.0011}$\\
        $P_{S,G}$&A priori eclipse prob & $0.0472^{+0.010}_{-0.0026}$&$0.0267^{+0.0027}_{-0.0012}$\\
        \hline
    \end{tabular}
    \end{center}
    \caption{\texttt{ExoFASTv2} planetary parameters: Median values and 68\% confidence interval created for TOI 560 b and c, \texttt{ExoFASTv2} commit number 101011575.2p.2. See Table 3 in \cite{Eastman2019} for a detailed description of all parameters.\\ $^a$ Time of conjunction is commonly reported as the ”transit time”. $^b$ Time of minimum projected separation is a more correct ”transit time”. $^c$ Optimal time of conjunction minimizes the covariance between TC and Period. $^d$ Assumes no albedo and perfect redistribution. $^e$ The recovered semi-amplitudes for the planets in the \texttt{ExoFASTv2} analysis are significantly smaller than the those recovered in Table \ref{tab:bc_priors} with our RV analysis that includes a stellar activity model, and thus would appear at first to be contradictory. However, they are not statistically significant detections in this \texttt{ExoFASTv2} analysis, as the posteriors are peaked at 0 m/s, and the reported confidence intervals herein are consequently misleading. The corresponding 5-$\sigma$ upper-limits are $\sim$15 and 10 m/s for $K_b$ and $K_c$ respectively.}
    \label{tab:estmantab}
\end{table*} 

\subsubsection{\textit{Spitzer} Light Curve} 
\label{Spitzerresults}

\begin{figure}
    \includegraphics[width=1.0\linewidth]{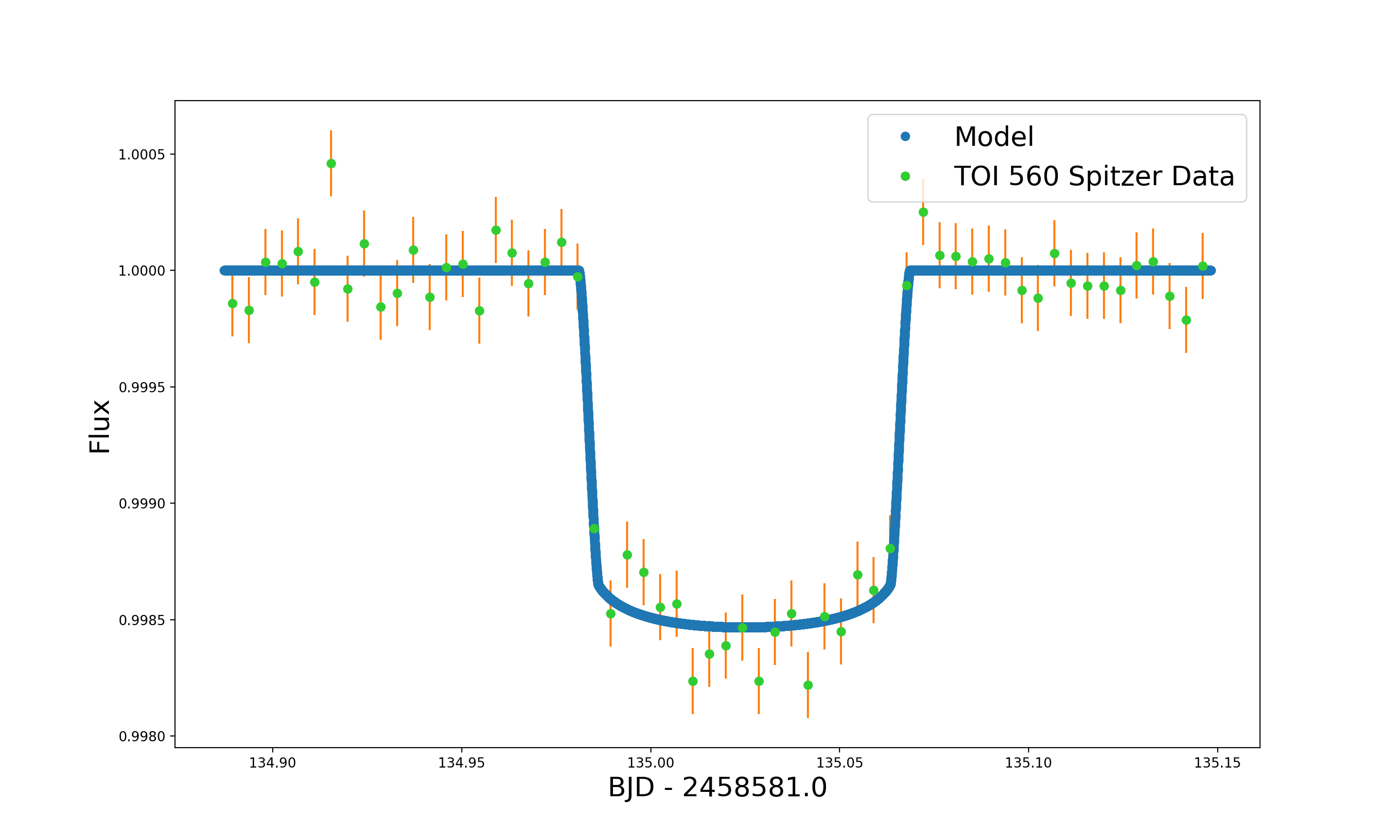}
    \caption{
    The \textit{Spitzer} light curve for TOI 560, with normalized flux plotted as a function of time since mid-transit.The Spitzer data points are binned for clarity, and overlaid with a transit model independently fit} in blue.
    
    \label{label:spitzerlc}
\end{figure}

The \textit{Spitzer} light curve (Figure \ref{label:spitzerlc}) shows a transit that is consistent with the expected timing, duration and depth from \tess for TOI 560 b. This helps rule out false positive eclipsing binary scenarios due to the lack of any observed chromaticity in the depth and shape of the observed transit. Additionally, the \textit{Spitzer} light curve is at higher photometric precision and temporal sampling with respect to \tess  given the larger aperture of \textit{Spitzer}; there is less limb-darkening compared to visible wavelengths, and less photometric variations due to stellar activity. Therefore these observations help further constrain the exoplanet parameters of TOI 560 b  and refine the orbital ephemerides. In particular, the \textit{Spitzer} light curve enables us to detect the eccentricity of TOI 506 b via the photo-eccentric effect as deviating from zero at 4.7$\sigma$.  This eccentricity is marginally detected in the \textit{TESS} transits, but is not statistically significant without the \textit{Spitzer} observations.

\subsection{RV \texttt{pychell} Results}
\label{sect:rv_mcmcs} 

\begin{figure*}
    \centering
    \includegraphics[width=\textwidth]{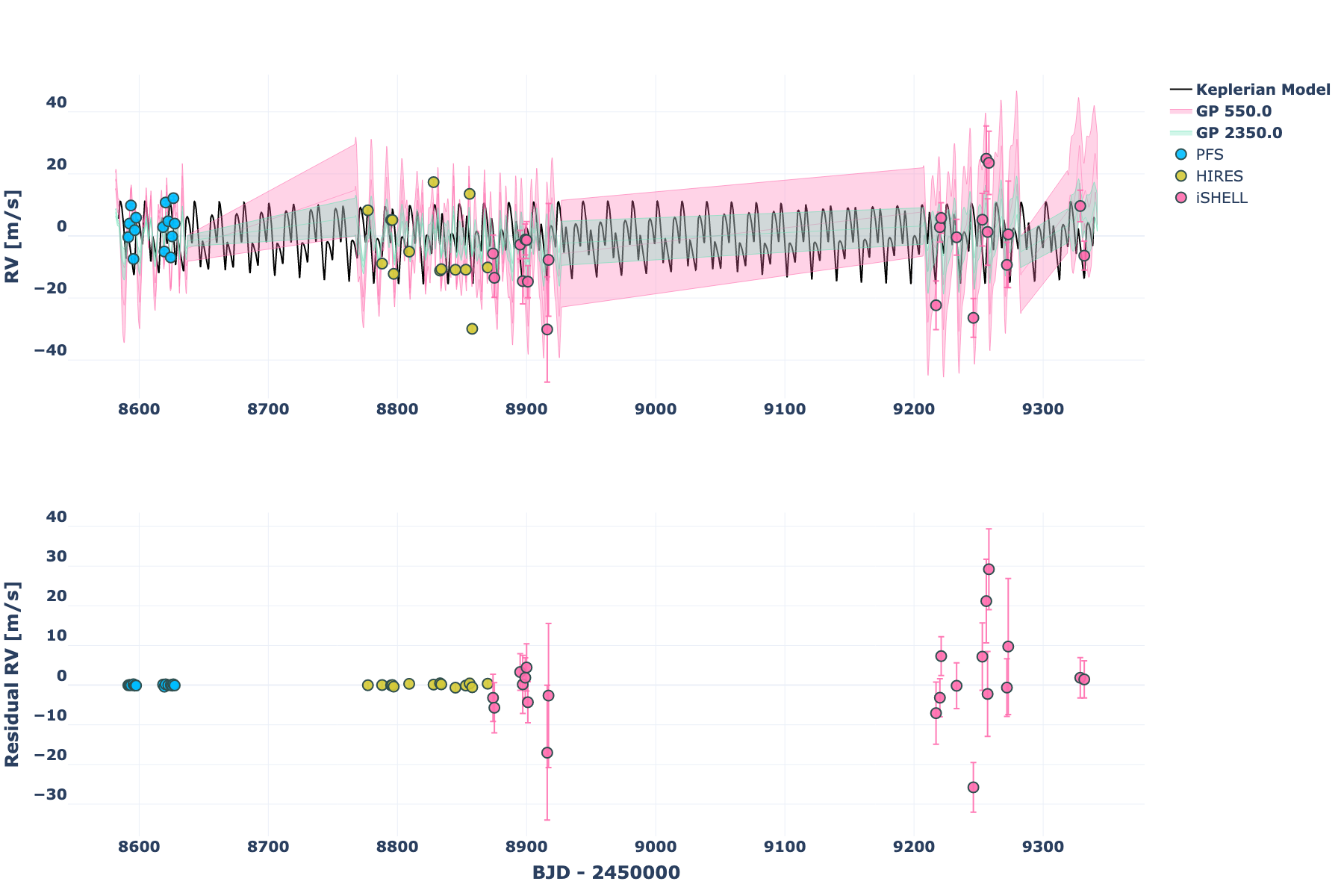}
    \caption{Full RV time-series plot for the joint chromatic GP kernel $J_{2}$ model, with time in BJD on the horizontal axis and RV on the vertical axis in m/s. RV measurements are shown as colored circles for each RV spectrograph. The black solid line is the Keplerian best fit MAP model for TOIs 560 b and c.  The shaded regions are the chromatic GP 1-$\sigma$ uncertainty regions from realizations of the $J_2$ covariance kernel, with the PFS and HIRES sharing the same GP at 550 nm, and a marginally smaller GP amplitude in the NIR for iSHELL.  Residuals (data-model) are shown in the lower plot.}
    \label{fig:j2}
\end{figure*}

\begin{figure*}
    \centering
    \quad
    \includegraphics[width=.45\textwidth]{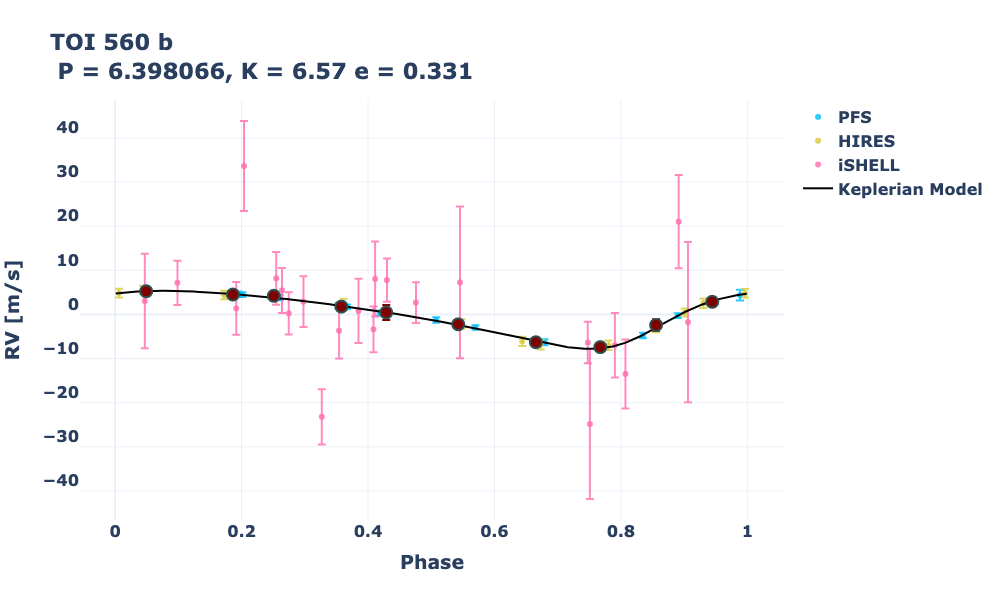}
    \includegraphics[width=.45\textwidth]{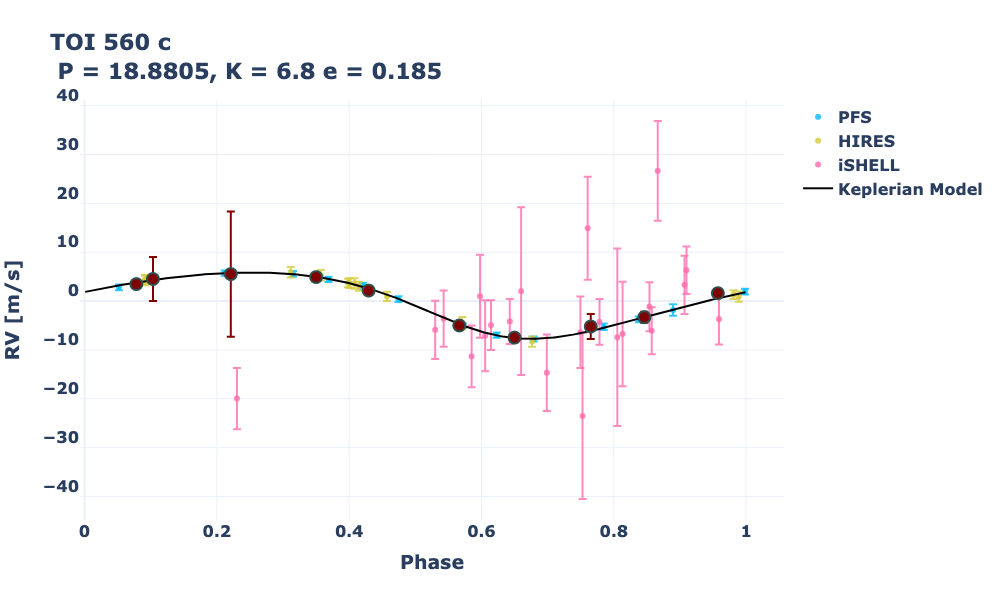}
    \caption{Phased RV time-series plot for the joint chromatic GP kernel $J_{2}$ model, with orbital phase on the horizontal axis and RV on the vertical axis in m/s.  The left panel is phased to TOI 560 b, and the right panel TOI 560 c.  RV measurements, after subtracting the best-fit GP are shown as small colored circles for each RV spectrograph. The maroon points are binned RVs every 0.1 in orbital phase}. The best fit MAP Keplerian model is shown as the black curve, with a fixed circular orbit, velocity semi-amplitude $K$, and orbital period $P$ indicated. 
    \label{fig:j2-1}
\end{figure*}

\begin{figure*}
    \centering
    \includegraphics[width=1.05\textwidth]{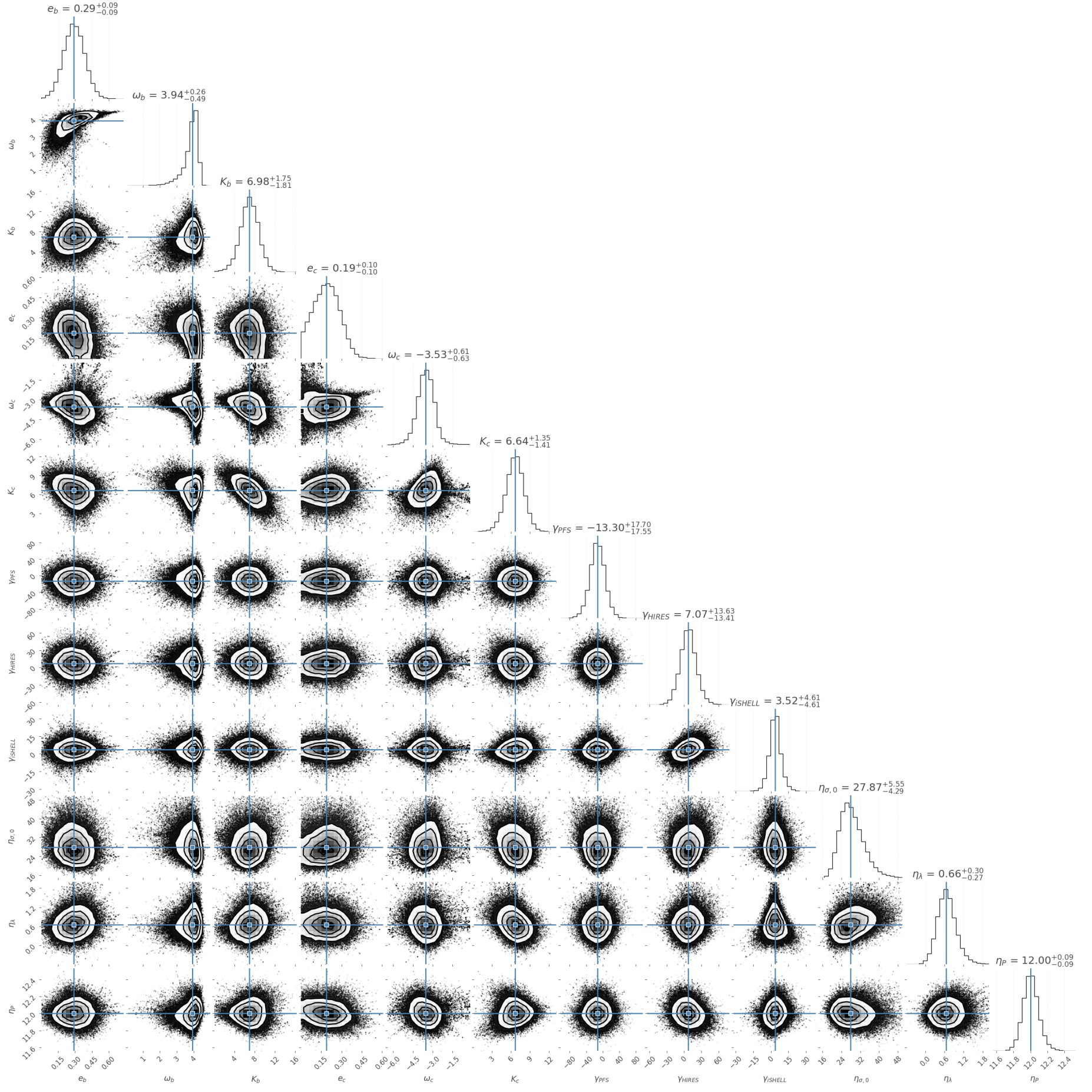}
    \caption{MCMC cornerplot for our joint chromatic GP kernel $J_{2}$ RV model. Posterior distributions are along the diagonal, and two-parameter covariances are shown off the diagonal. The model paramter median and 68\% confidence interval ranges are displayed at the top of each posterior distribution; median values are also indicated with horizontal and vertical blue lines for the covariance plots, and vertical lines for the posterior distribution. For the covariance plots, 1, 2 and 3$-\sigma$ contours are shown in place of the individual sample values $<3-\sigma$ from the medians.} 
    \label{fig:j2-2} 
\end{figure*}

Using our RV model with \texttt{pychell} we produce consistent and stable MCMC results for all GP kernels considered that provide detections for the masses of TOI 560 b and c, establishing their nature as a pair of Neptune-mass exoplanets consistent with their radii (Table \ref{tab:bc_priors}). For our adopted final result using the $J_2$ kernel, the RV models with MAP values and MCMC cornerplot are shown in Figures \ref{fig:j2}, \ref{fig:j2-1}, and \ref{fig:j2-2}. The best MAP fit, and MCMC median and confidence intervals are presented in Table \ref{tab:bc_priors}. The table of exoplanet mass and density measurements are derived from our RV analysis are in Table \ref{tab:masses}. In the Appendix in Figure \ref{fig:mass-radius}, we present the mass-radius diagram for known transiting exoplanets with measured masses \citep{Akeson_2013}, with the detections for TOI 560 b and c shown to be consistent with the older Neptune-mass regime. 

We present in the Appendix the results of the RV analyses for the $J_1$ joint GP kernel, the disjoint ``classsic'' quasi-period GP kernel where there is a separate GP for every spectrograph, and for the RV analysis with no stellar activity assumed in Tables \ref{tab:bc_priors3} and \ref{tab:bc_priors2} respectively. All yield similar detections or upper-limits to the exoplanet velocity semi-amplitudes and consistent model hyper-parameters for the stellar activity.

A model comparison in Table \ref{tab:compJ2} shows that the RVs support the detection of both planets b and c combined and also planets b and c individually. The 1 and 2 planet models are statistically indistinguishable. The no-planet scenario is ruled out.
Additionally, we do recover $3.8-\sigma$ ($4.7-\sigma$) detections of TOI 560 b (c) with our $J_2$ kernel.
For the other kernels presented in the Appendix, we find qualitatively similar marginal and $>$3-$\sigma$ recoveries of $K_b$ and/or $K_c$. Our $J_2$ kernel model estimates the masses of planets b and c to be $0.94^{+0.31}_{-0.23}M_{Nep}$ and $1.32^{+0.29}_{-0.32} M_{Nep}$. 


\begin{table} 
    \centering
    \begin{tabular}{ccc}
        \hline
        Planet & Mass & Density (gcm$^{-3}$)\\ 
        \hline
        \hline
        b & $15.9^{+5.3}_{-3.9}$ M$_{\oplus}$, $0.94^{+0.31}_{-0.23}$ M$_{Nep}$ & $3.8^{+1.4}_{-1.1}$  \\ 
        \hline 
        c & $22.5^{+5.0}_{-5.5}$ M$_{\oplus}$, $1.32^{+0.29}_{-0.32}$ M$_{Nep}$ & $6.1^{+1.6}_{-1.7}$  \\ 
        \hline
    \end{tabular}
    \caption{Mass and density detections from our joint chromatic $J_2$ GP kernel RV model.} 
    \label{tab:masses}
\end{table}

\begin{table} 
    \centering
    \begin{tabular}{cccccc}
        \hline
        Planets & $\ln \mathcal{L}$ & $\Delta$ AICc & $\Delta$ BIC&  N free & $\chi^2_{\mathrm{red}}$  \\
        \hline
        \hline
        b, c &  -197.0& 0.0 & 1.5 & 12 & 1.1 \\
        b &  -202.1& 0.3 & 0.0 & 9 & 0.9 \\
        c &   -204.5& 5.1  & 4.7 & 9 & 0.9 \\
        None & -233.9 & 55.3 & 51.7 & 6 & 1.4 \\ 
        
        \hline
    \end{tabular}
    \caption{
    A model comparison test for planets b and c generated through the joint J$_2$ kernel model, which support the detection of at least one planet, b or c individually, or both planets with indistinguishable AICc values. The no-planet scenario is ruled out.} 
    \label{tab:compJ2} 
\end{table}

\section{Discussion} 
\label{sect:discussion}

In the previous sections we have validated the presence of a two-planet system orbiting a young, $\sim$600 Myr host star TOI 560, detected the individual planet masses at $> 3-\sigma$, establishing the planets as Neptune analogs. 
In this discussion, in ${\S}$\ref{sect:dynamical_stability}, we present a dynamical stability analysis of the system, which appear stable over the 10 Myr timescale explored. Next we consider the near 1:3 orbital resonance of TOI 560 b and c, and what implications that may have for additional planets and formation in \ref{middled}. Finally, in \ref{kempton} we discuss the suitability of the TOI 560 system for future atmospheric characterization.

\subsection{Dynamical Stability: Two-planet \texttt{REBOUND} Simulation}
\label{sect:dynamical_stability}

\begin{figure}
    \centering
    \includegraphics[width=.45\textwidth]{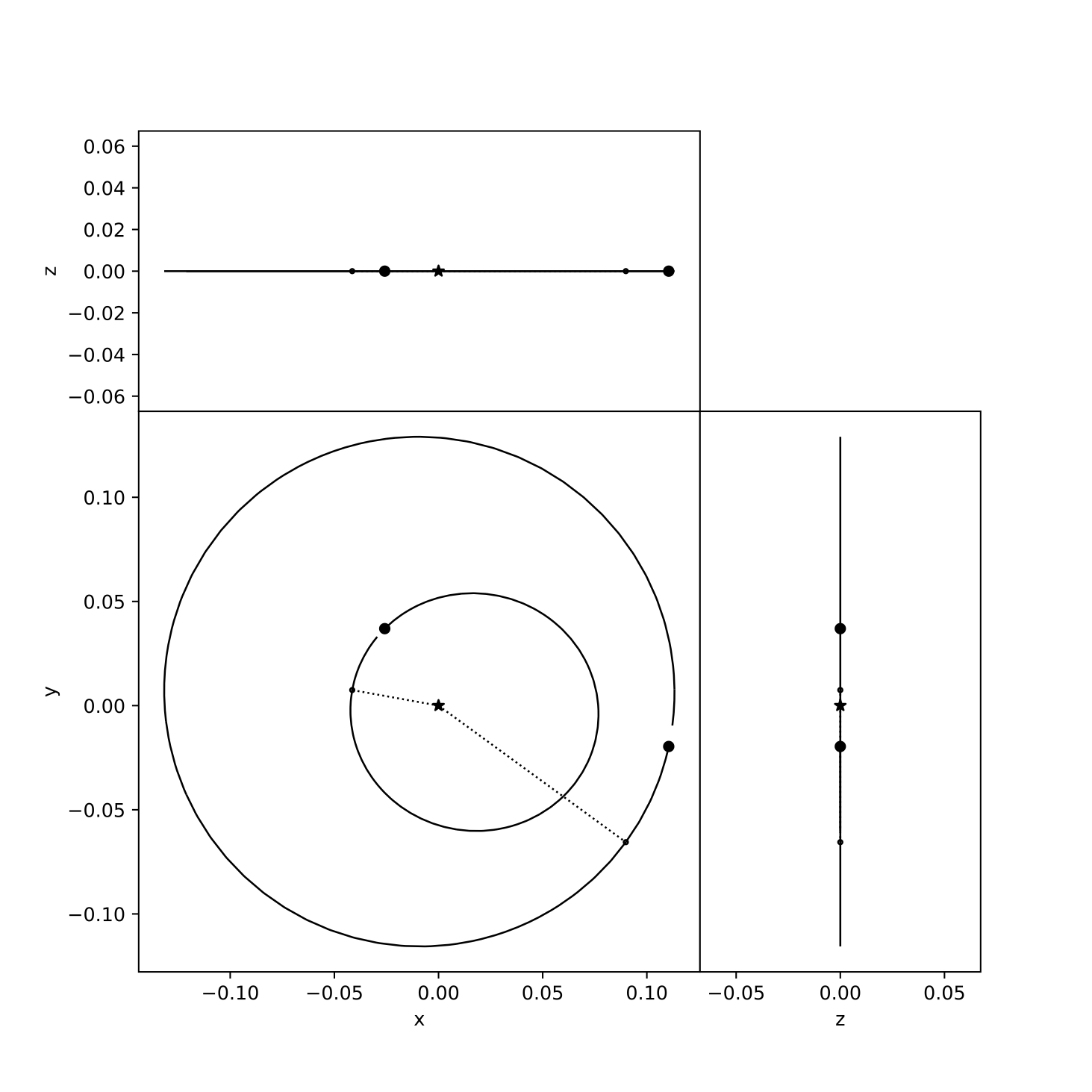}
    \includegraphics[width=.45\textwidth]{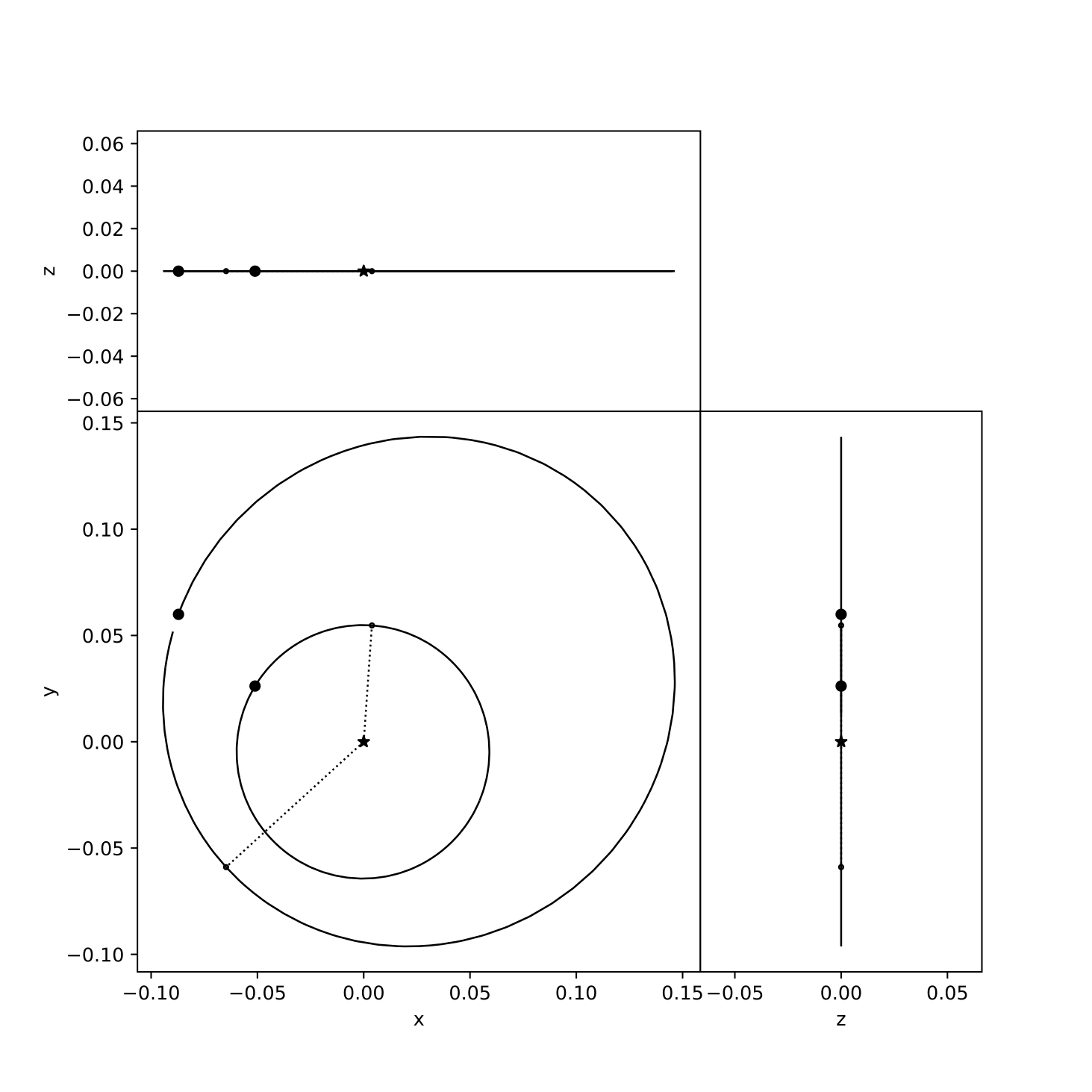}
    \caption{Overhead and edge-on diagrams showing the initial (top) and final (bottom)} orbital configurations of our two-planet model in \texttt{REBOUND} with inclination estimated from \texttt{EXOFASTv2}. The $x$,$y$ and $z$ axes are in AU.  The planet positions are noted as black dots, and the orbits as circles. The periastron vectors are marked with dashed lines starting from the star (star symbol) at the origin of the orbits, and show significant evolution from the start to the beginning of the simulation.
    \label{fig:tpf2}
\end{figure}

\begin{figure*}
    \centering
\includegraphics[width=0.45\textwidth]{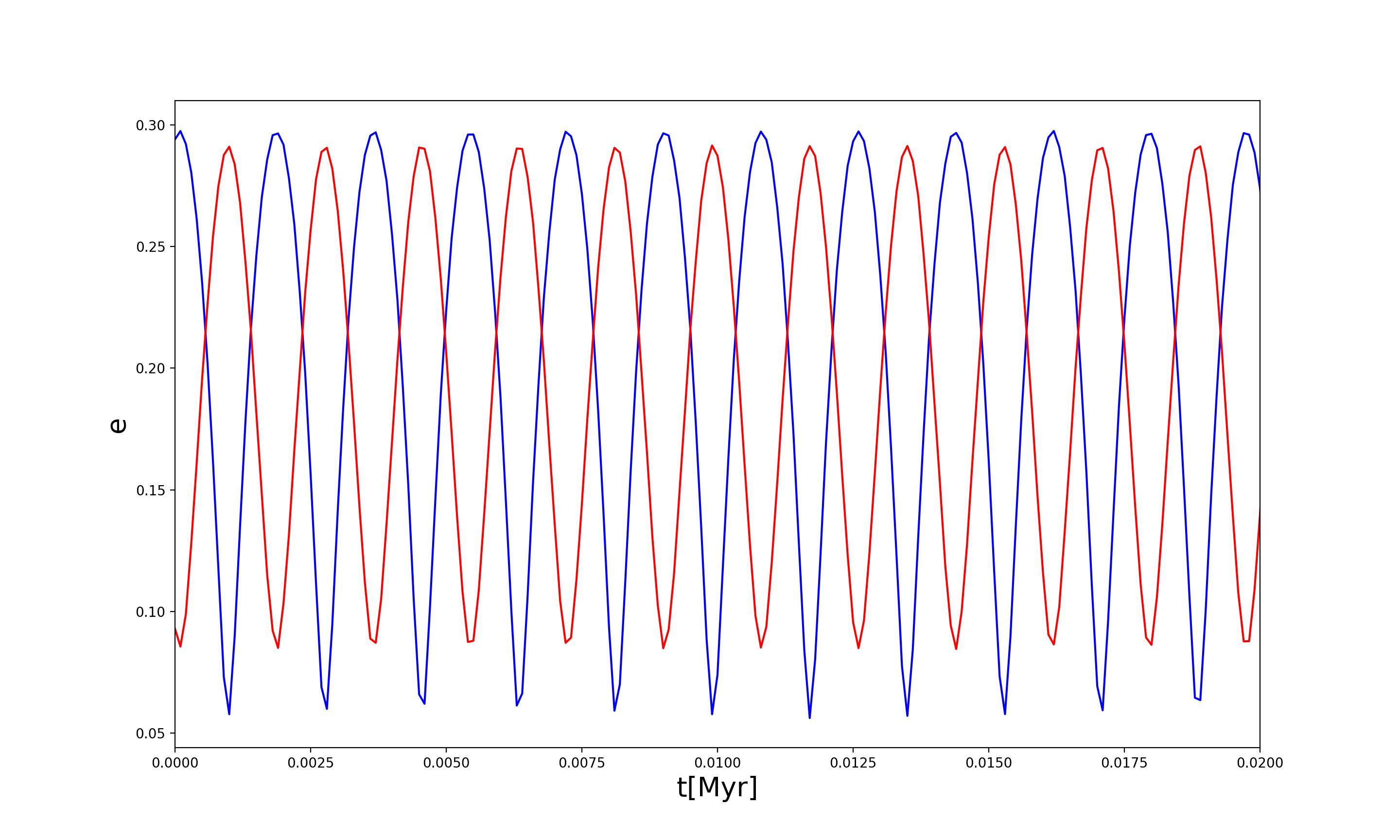}
\includegraphics[width=0.45\textwidth]{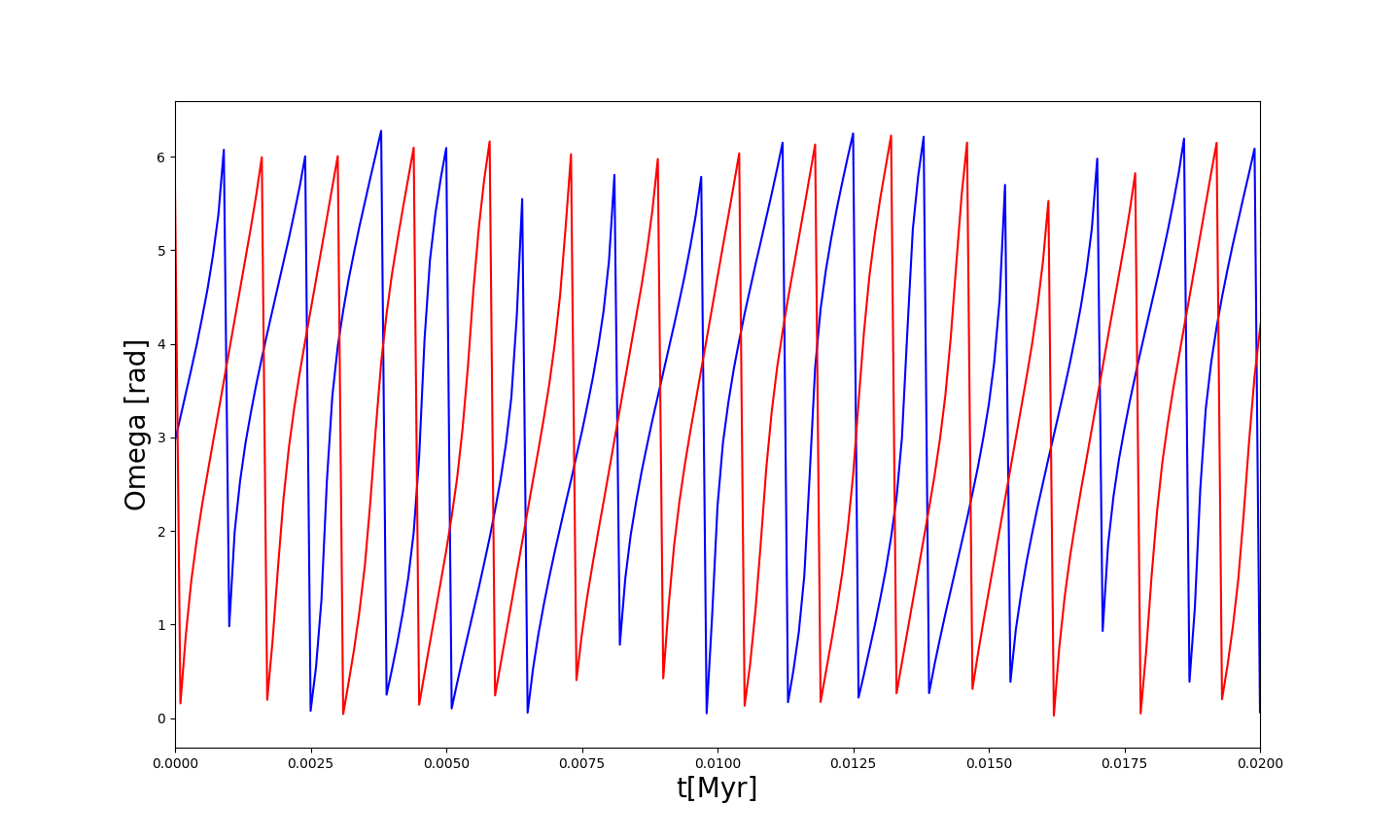}        \includegraphics[width=0.45\textwidth]{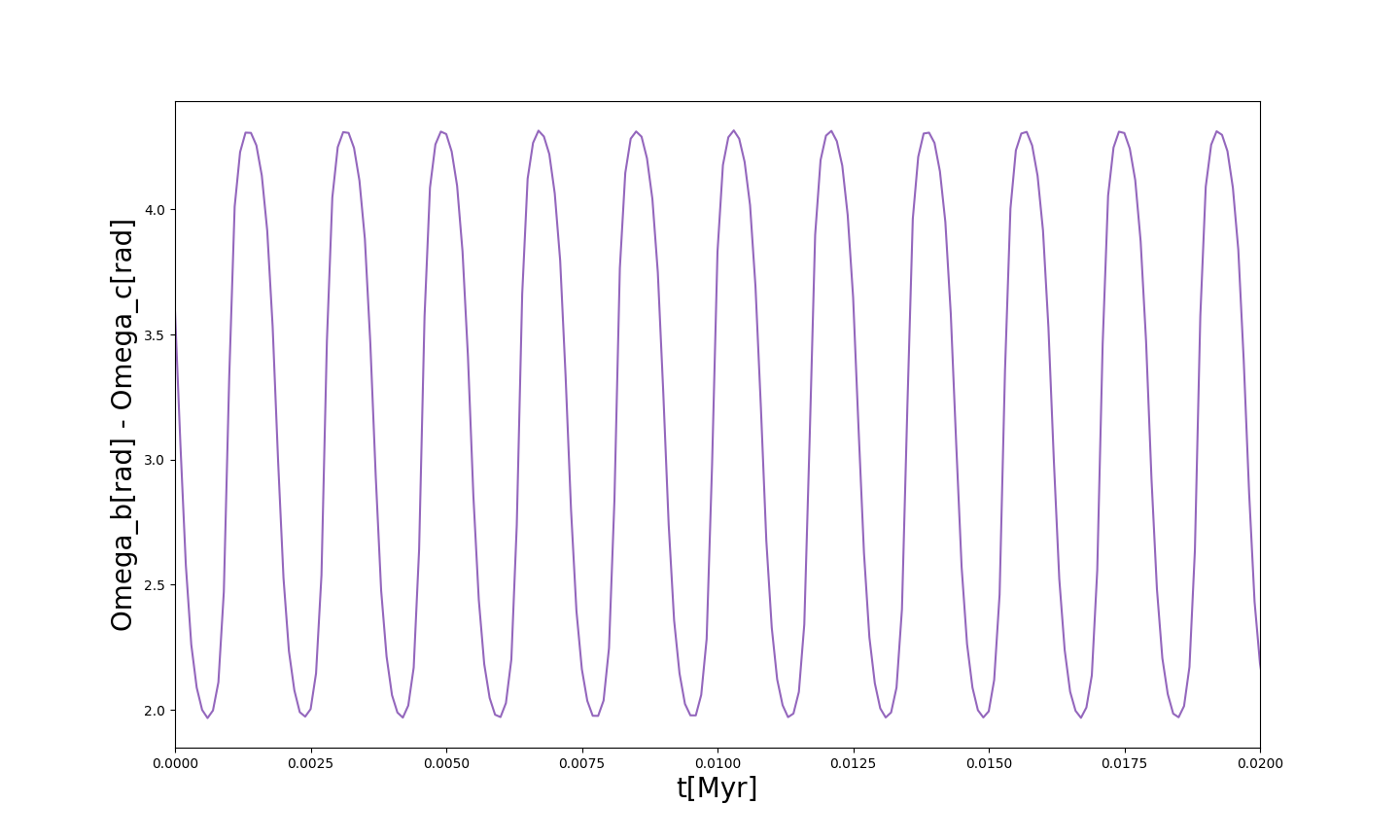}
            
    \caption{\texttt{REBOUND} simulations showing the evolution of the orbital elements of TOI 560 b (blue) and c (red).  The semi-major axis evolution is stable (not shown), and the planets are assumed coplanar so there is no inclination variation (not shown). The first panel shows the significant temporal evolution of the eccentricities for TOI 560 b and c, consistent with the moderate eccentricity we detect for TOI 560 b.  The second panel shows the longitudes of periastron (in degrees) on the vertical axes as a function of time on the horizontal axes, showing significant dynamical precession. The bottom panel shows the difference in the longitudes of periastron between the two planets, showing that the system exhibits a libration in the longitudes of periastron. Only the first 20,000 years of orbital evolution are shown, but the system behaves similarly for the duration of the 10 Myr simulation.}
    \label{fig:3rebounds2}
\end{figure*}

We perform dynamical stability tests of the TOI 560 system over a duration of 10 Myr using \texttt{REBOUND}, an N-body gravitational integrator built in \texttt{C} and usable in \texttt{python} \citep{rebound, reboundias15}.  To define each orbit, we use the $P$, $T_P$, eccentricity, and inclination median values from our \texttt{EXOFASTv2} MCMC analysis in Table \ref{tab:estmantab}, and we use the median planet masses $M_p$ from our J2 kernel model values from Table \ref{tab:masses}. We choose an integration time-step shorter than $1/20^{\mathrm{th}}$ of the shortest orbital period ($dt = 0.32$ days). The results of this dynamical simulation are shown in Figures \ref{fig:tpf2} and \ref{fig:3rebounds2}. The simulations show the dynamical interactions between TOI 560 b  and c, with a mean motion resonant libration precession of the longitudes of periastron. This demonstrates that the detected eccentricity for TOI 560 b could be a consequence of the dynamical interaction with TOI 560 c. The orbits of both planets are stable and possessing moderate eccentricities over the entire 10 Myr simulation duration. We did not run the simulation to $\sim$1 Gyr, the estimated lifetime of the system, because of the near 1:3 orbital resonance; a mean-motion orbital resonance (MMR) can either stabilize or destabilize the system within a timescale of a few hundred orbits, so if the system was unstable, the instability would have been apparent in the simulation well within a few Myr.


\subsection{The Search for Additional Candidates}

\label{additional}

Fourier-like periodograms can be useful in determining prominent frequencies in unevenly sampled RV time-series. First, we calculate Generalized Lomb-Scargle (GLS) (\citealt{Zechmeister2009,Scargle1982}) periodograms. The GLS periodogram power at a given trial period directly correlates with the Keplerian velocity semi-amplitude of a circular-orbit planet signal at that trial period. Multiple periodic signals are identified by subtracting a best-fit periodic signal from the data, and then repeating the periodogram calculation on the residuals; this process is repeated until no more statistically significant signals are identified.

For TOI 560 in particular, we compute a GLS periodogram utilizing a stellar activity + two planet model for b and c. We first remove the nominal GP and zero-points; this periodogram is used to see if we can identify the RV signal of b or c, or any additional planet(s) present in the system as a periodogram peak. Then we further remove best-fit RV signal for TOI 560 b, and re-compute the periodogram; this periodogram is used to see if we can identify the RV signal of TOI 560 c or any additional planet(s) present in the system. Third, subtract off the best fit model for planet c, and re-compute the periodogram of the residuals; this periodogram is used to search for any additional planets in the system. In all cases, the orbital period and time of conjunction and thus orbital phase of either TOI 560 b and/or c are fixed in fitting the planet signals to the RV time-series data, and only the velocity-semi-amplitude, eccentricity, and angle of periastron are free (fitted) parameters. We calculate false-alarm probabilities using both the standard analytic formula derived from the periodogram power (\citealt{Zechmeister2009,Scargle1982}), and empirically through a 1000-trial bootstrap process of: scrambling (re-ordering) the times of observations while holding the ordering of the RV values fixed (or conversely, scrambling the ordering of the RV values while holding the ordering of the times of observations fixed), then recomputing the periodogram, and then investigating the empirical distribution of the top power values. While GLS periodograms have historically been successful in identifying genuine multi-planet signals in RV time-series data, this approach is also fraught with the identification of false positive signals because of the overly simplistic assumptions \citep[e.g.,][]{Robertson2014,Robertson2015}; it can be difficult to discern whether GLS periodogram peaks are real planet signals, or some form of systematic red noise \cite[e.g. a cadence alias, imperfect signal subtraction in the iterated residuals, stellar rotational modulation of activity, etc. ][]{2016MNRAS.459.3565V,2010ApJ...722..937D}. We also compute the window function of the RV time-series by setting all RV values equal to zero and computing a Lomb-Scargle periodogram to identify any periodogram peaks that are associated with temporal sampling cadence aliases, such as integer fraction multiples of one day or one year.

Second, Bayesian model-based  $\ln \mathcal{L}$ periodograms enable statistically-robust comparisons that jointly capture the full model complexity of potentially multi-planet systems with models for stellar activity embedded in the noisy, unevenly sampled RVs \citep[e.g.,][]{Anglada-Escude2016,Tuomi2014}. First, we compute a maximum-likelihood fit to the RVs for a model with stellar activity and a first ``trial'' planet of unknown phase and velocity semi-amplitude at a trial orbital period.  We follow the same implementation as in \citet{Cale2021}. We compute this one-planet model $\ln \mathcal{L}$ over a set of trial periods to generate the first  $\ln \mathcal{L}$ periodogram, which is used to see if we can identify the RV signal of b or c, or any additional planet present in the system as a periodogram peak. Second, we add to our model a planet at the orbital period and time of conjunction of TOI 560 b, in addition to the trial planet.  We compute maximum-likelihood fits to this two-planet model over the same set of trial periods to compute the second $\ln \mathcal{L}$ periodogram, which is used to see if we can identify the RV signal of TOI 560 c, or any additional planet present in the system as a periodogram peak. Third, we add to our model a third planet at the orbital period and time of conjunction of TOI 560 c to see if we can identify the RV signal for any additional planet present in the system besides TOI 560 b or c.

\begin{figure*}
    \centering
    \includegraphics[width=\textwidth]{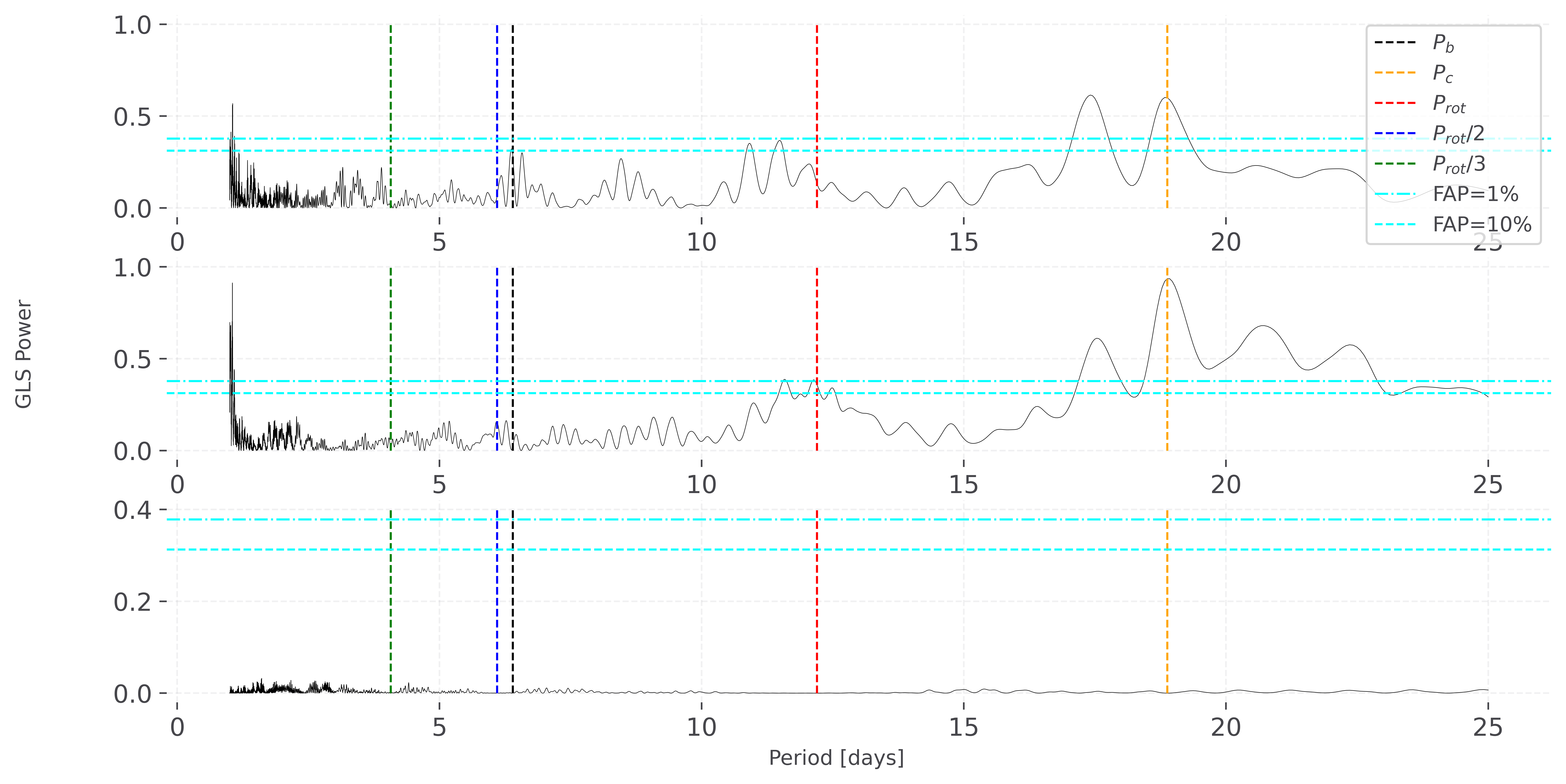}
    \caption{A series of GLS periodograms examining the signals in our TOI 560 RV time-series as described in the text including iSHELL, HIRES and PFS. The top panel is a periodogram of the RVs after subtracting a stellar activity model fit. The middle panel additionally subtracts a model fit for TOI 560 b and stellar activity to search for a second planet, and the bottom panel subtracts a model fit for both TOI 560 b and c and stellar activity to search for a third planet. The periods of b \& c are marked with dashed black and orange vertical lines and also the rotation period, one half of it, and one third of it are marked with dashed red, blue, and green vertical lines, in each panel. The bootstrap false alarm probabilities (FAPs) of 1\% and 10\% were marked with cyan horizontal dashed lines. Note, it remains challenging to find non-transiting RV planets at orbital periods near the stellar rotation period or one of its prominent integer fraction aliases (e.g. 1/2, 1/3), as the GP stellar activity model could potentially ``absorb'' such a Keplerian signature \citep{Vanderburg2016,Kossakowski2022}.}

    \label{fig:gls_iterations}
\end{figure*}

\begin{figure*}
    \centering
    \includegraphics[width=\textwidth]{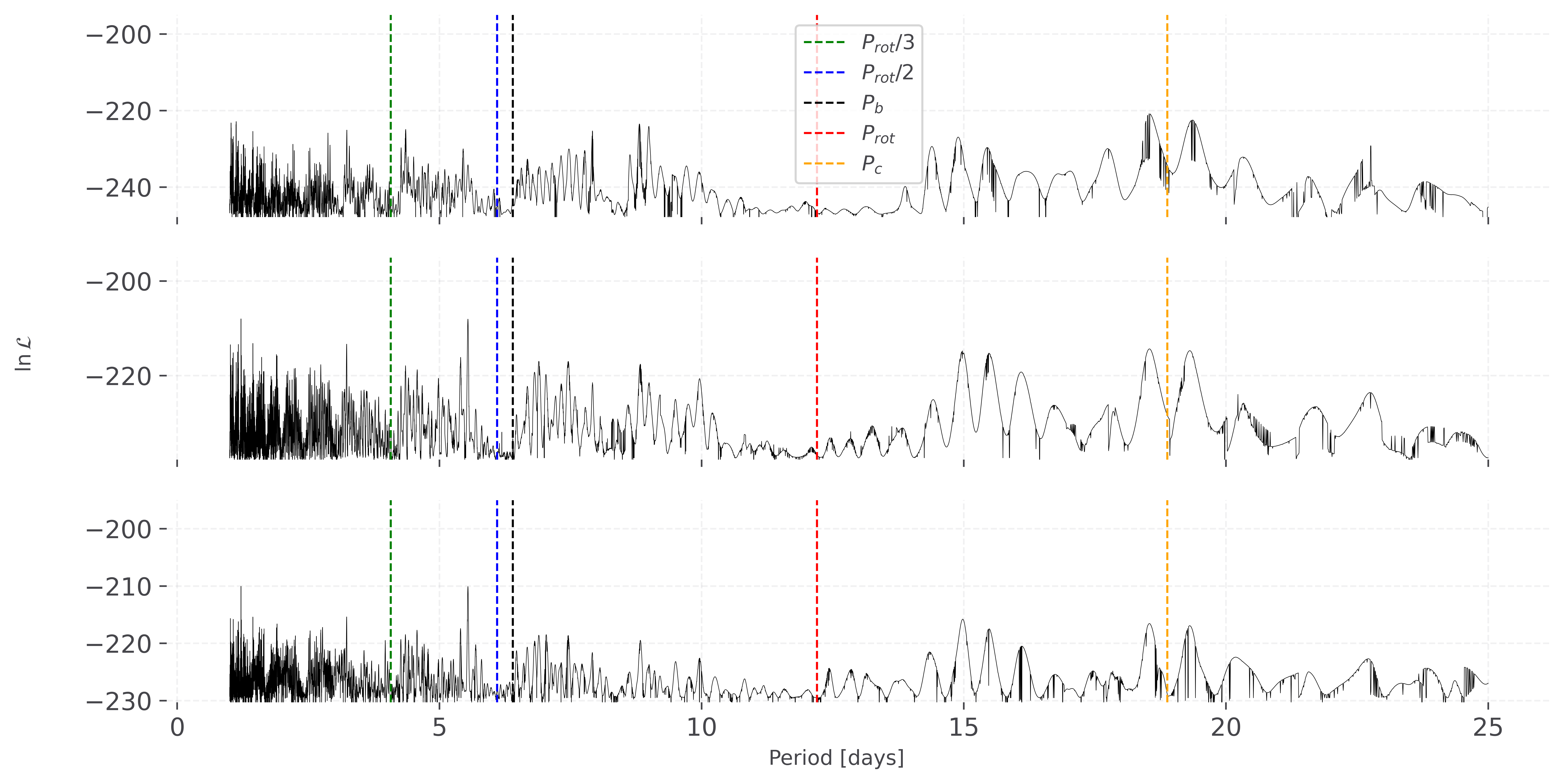}
    \caption{Similar to the previous figure, a series of $\ln \mathcal{L}$ periodograms examining the signals in our TOI 560 RV time-series. All three panels depict a single, circular planet search with a floating TC. The top panel includes no extra planets, the middle panel models out TOI 560 b to search for a second planet, and the bottom panel models out both TOI 560 b \& c to search for a second planet. The periods of b \& c are marked with dashed black and orange vertical lines and also the rotation period, one half of it, and one third of it are marked with dashed red, blue, and green vertical lines, in each panel. }

    \label{fig:lnL_iterations}
\end{figure*}

The GLS periodograms are shown in Figure \ref{fig:gls_iterations}, and the $\ln\mathcal{L}$ periodograms are shown in Figure \ref{fig:lnL_iterations}. We find two $\ln\mathcal{L}$ periodogram peaks that are persistent with a $\Delta \ln \mathcal{L} >$ 15 at 4.57 and 5.54 days, particularly for the periodograms where TOI 560 b is included in the model. The 5.54 day period would likely not be dynamically stable with the TOI 560 b at 6.40 days.  However, the 4.57 periodogram peak suggests a possible non-transiting candidate that would be near a 2:3 orbital resonance with TOI 560 b. This period is also not near an alias of the stellar rotation period. Furthermore, the 5.54 day peak is close to the frequency difference between the 4.57 day peak and the cadence alias of 5 days and is thus potentially a cadence alias of the 4.57 day peak. Consequently, we perform a model comparison with a third non-transiting interior planet at this orbital period of 4.57 days.  However, the model comparison disfavors the detection of this planet. Thus our existing data is insufficient to confirm the possibility of an additional non-transiting candidate planet in this system.


\subsection{Implication of the near 1:3 orbital period resonance of TOI 560 b and c}
\label{middled}

Given the youth of TOI 560, neither planet would have been tidally circularized by the star if they started with an initial high eccentricity formation mechanism \citep{1996ApJ...470.1187R,2003ApJ...592..555B}. Our RVs are not sufficient to constrain the eccentricities of either planet, given that we have only a marginal mass detection of TOI 560 b and a non-detection of TOI 560 c, and also due to the presence of the stellar activity. However, our \texttt{ExoFASTv2} analysis of the \tess\, \textit{Spitzer} and ground transits of TOI 560 b confirms a moderate eccentricity of 0.29 (Figure \ref{fig:ecc}). Both planets show that a high eccentricity is disfavored. This implies that TOI 560 b and c likely did not form in a high-eccentricity migration scenario, and more likely formed in-situ, unless the migration took place when a significant gas disk was still present to dampen the orbital eccentricities \citep{2013ApJ...769...86P,1996Natur.380..606L,1997Icar..126..261W}. The orbital eccentricity of TOI 560 b will be an important parameter to further constrain with future observations, such as may be possible with secondary eclipse observations and additional high precision RVs. 

The orbital period near-commensurability of TOI 560 b and c to a 1:3 ratio also has interesting implications for planet formation. For one, such an orbital resonance is rare among mature exoplanet systems identified by \textit{Kepler} \citep[Figure 4,][]{2014ApJ...790..146F}. Far more common among mature compact multi-exoplanet systems are the 2:3 and 1:2 near-commensurability orbital period resonances. This raises the interesting possibility that there could be a third middle planet in the TOI 560 system that is non-transiting with an orbital period of $\sim$12.6 days that would establish the TOI 560 system in a 1:2:3 orbital resonant chain at an age of $<$1.5 Gyr. Such systems of transiting exoplanet systems orbiting mature stars with middle non-transiting companions are more common than perceived -- e.g. \citet{christiansen2017}, \citet{buchhave2016}, \citet{sun2019}, and \citet{osborn2021} identified similar exoplanet configurations for the HD 3167, Kepler-20, Kepler-411, and TOI-431 planetary systems respectively. 

The 22 Myr AU Mic system is possibly also in a resonant 4:6:9 orbital chain with a middle non-transiting planet candidate \citep[][]{Cale2021,Wittrock2022}, and the 20 Myr v1298 Tau system is in a 2:3:6:10 resonant chain \citep{2019ApJ...885L..12D,2021arXiv211108660F,2021arXiv211109193S}. Similarly, it is an interesting coincidence that TOI 560 b and c are near 1:2 and 2:3 orbital period resonant near-commensurability with the stellar rotation period of TOI 560 of 12.2 days. It has recently been shown that AU Mic b is in a 4:7 orbital period commensurability with the rotation period of the AU Mic host star \citep{2021A&A...654A.159S}. These period commenurabilities are perhaps coincidental. However, if indeed resonant chains are common for young exoplanet systems, and if there are further instances of stellar spin -- planet orbital period commensurabilities found in the future for young exoplanet systems, this could have important implications for planet formation mechanisms.  It will be important to conduct future Rossiter-McLaughlin observations of TOI 560 b and c to determine if their orbits are also aligned with the stellar spin axes, as is the case for AU Mic b and v1298 Tau b \citep{2021AJ....162..137A,2020A&A...643A..25P,2020A&A...641L...1M,2020ApJ...899L..13H,2021arXiv211010707J,2022MNRAS.509.2969G}.  The expected amplitude for the R-M effect for TOI 560 b and c is $\sim$4 m/s, calculated with a $v$sin$i$=2.8 km/s from the known rotation period of 12.2 days, and our $R_p/R_*$ and $R_*$ from our \texttt{ExoFASTv2} analysis (Table \ref{tab:estmantab}), and assuming the stellar rotation axis is in the planet of the sky, which is consistent with the observed constraints on $v$sin$i$ from TRES and NRES ${\S}$\ref{sect:recon}.

Finally, we do investigate the dynamical stability of one test case of a middle, non-transiting planet for TOI 560 -- simulated with masses 9.2, 4.2 and 1 $M_\oplus$ for TOI 560 b,c and the hypothetical middle d at $P=12.6$ days respectively.  However, we find that this particular test case was not dynamically stable. Exploring the full dynamical stability of possible scenarios is beyond the scope of this work, but the mild eccentricity of TOI 560 b may preclude the dynamical stability of any middle planets. It would also be incredibly challenging to detect such a middle planet with RVs alone given that the rotation period of the star is 12.2 days; stellar activity will potentially hide any such RV signal given our current stellar activity modeling tools \citep{2016MNRAS.459.3565V}.

\subsection{The suitability for the TOI 560 planets for atmospheric characterization}
\label{kempton}

\begin{figure}[!ht]
    \centering
    \includegraphics[width=0.75\linewidth]{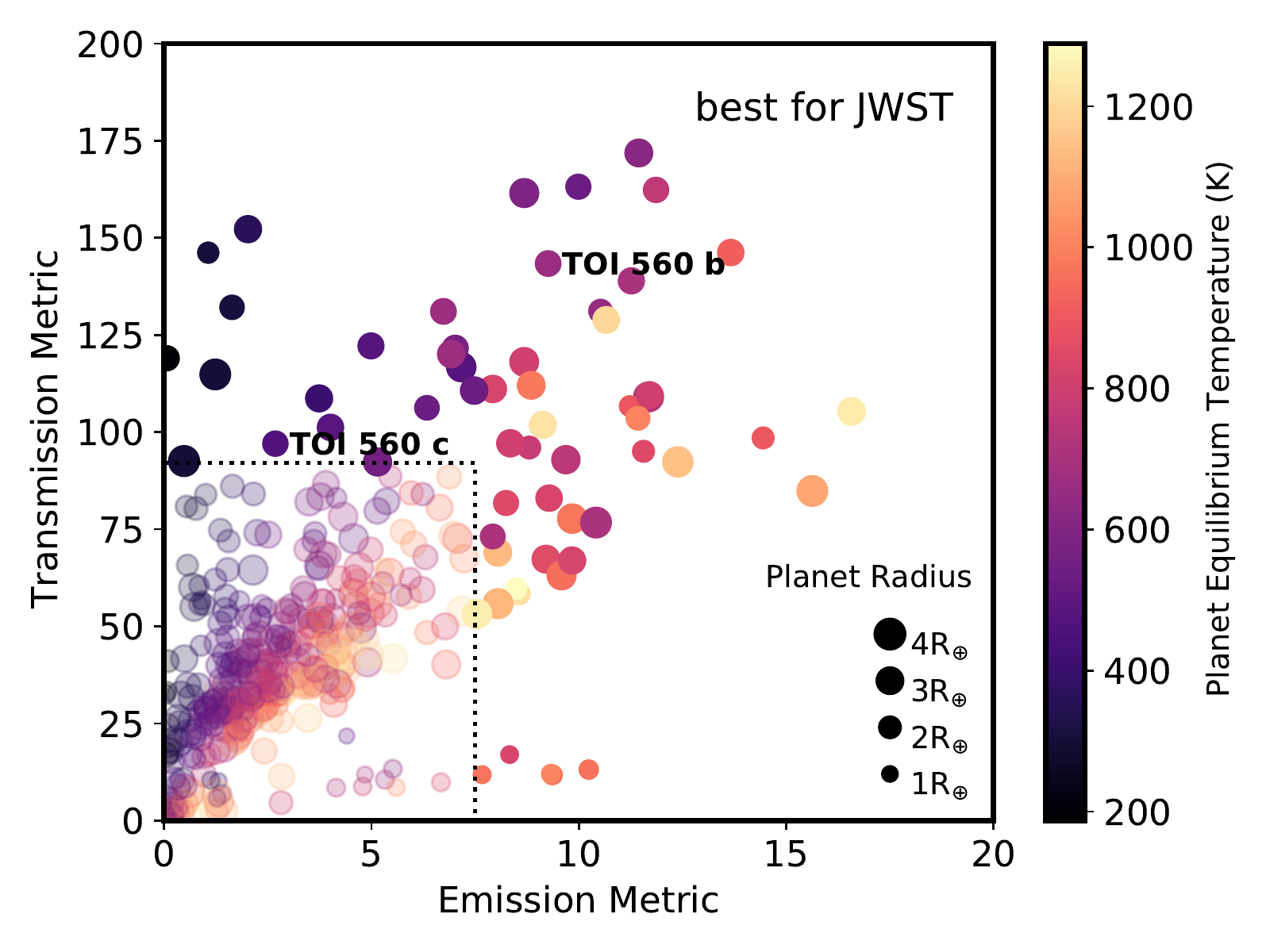}
\caption{\cite{Kempton2018} metrics of signal-to-noise for hypothetical observations of exoplanet atmospheres in transmission (during primary transit) on the vertical axis and emission (during secondary eclipse) on the horizontal axis for a subset of TOIs detected as of the end of September 2021 \citep[NASA Exoplanet Archive,][]{Akeson_2013}. Only planets smaller than Neptune, with Teq $<$ 1300K, and that are predicted to impart a Doppler RV signal K $>$3 ms$^{-1}$ are shown. The points’ sizes are scaled with radius and the color scaled to estimated equilibrium temperature, with the colorbar on the right. Increased metric means higher SNR and the dashed lines indicate the boundary above and to the right of which systems are suitable for JWST observations (\cite{Kempton2018}). TOIs 560 b and c are labeled immediately to the right of their data points.} 
    \label{fig:kempton}
\end{figure}

Finally, we evaluate the suitability of TOI 560 b and c for atmospheric characterization. We compute the transmission spectroscopy metrics (TSM) and emission spectroscopy metrics (ESM) \citep{Kempton2018} for a set of TOIs in Figure \ref{fig:kempton}, including TOI 560 b and c of 151.9,11.2 and 102.8,3.4 for TSM and ESM respectively:

\begin{equation}
    TSM = \mathrm{(Scale\ factor)} \times \frac{R_P^3 T_{\rm eq}}{M_P R_*^2} \times 10^{-m_J/5}
\end{equation}
\begin{multline}
    ESM = 4.29 \times 10^6 \times \frac{B_{7.5}(T_{\rm day})}{B_{7.5}(T_{eff})} \\
    \times \bigg(\frac{R_P}{R_*}\bigg)^2 \times 10^{-m_K/5}
\end{multline}

\noindent where $R_P$ and $M_P$ are in Earth masses, $T_{\rm eq}$ is in Kelvin, $R_*$ is in solar units, $B_{7.5}$ is the Planck function at a wavelength of 7.5 $\mu$m, $T_{\rm day} = 1.1T_{\rm eq}$, the scale factor for planets with $2.75 < R_P < 4$ $R_\oplus$ is 1.28, and $m_J$ and $m_K$ are the magnitudes of the host star in the J and K bands.  Note, to maintain consistency with ExoFOP-TESS, we use the \citet{ChenKipping2016} mass-radius relation for estimating the planet masses in computing the TSM and ESM values.  However, using the \citet{Wolfgang2016} exoplanet mass-radius relation -- which includes a different treatment of sample selection in deriving their mass-radius relation -- or our median planet masses yields similar results. The TSM and ESM uncertainties are large due to the current planet mass constraints.

TOI 560 b  in particular is among the best TOIs identified to date for both transmission and emission spectroscopy characterization, and TOI 560 c is also suitable for transmission spectroscopy with JWST.  The significance of TOI 560 b and c for atmospheric characterization is in part due to its youth and relative brightness at NIR wavelengths, which coincidentally also makes it a priority candidate for NIR RVs. This system is also useful for comparative planetology with other well-studied sub-Neptunes. For example, TOI 560 c is similar size and temperature to GJ 1214b, which is well known for its thick clouds/hazes \citep{Kreidberg2014}. This provides an opportunity to test whether the same is true for a planet with a different host star type, and a different system architecture. In addition, the inner planet is hotter, so can be used to test the prediction that the atmospheres of hotter planets are less affected by clouds/haze \citep{Crossfiled&Kreidberg}. Another nice feature of TOI 560 is that the star is not too bright, so can be observed by every JWST instrument. 

Further, since TOI 560 is a multiplanet system, JWST observations could be optimized to capture both transits of b and c with a single telescope pointing and perform comparative planetology; such an observation would be achievable by timing the observations so that the egress of one planet occurs $\sim$1 hour before the ingress of the second.  Using the NASA Exoplanet Archive \citep{Akeson_2013}, for example, there are 129 transits of TOI 560 b and c visible by \textit{JWST} between 6/1/2022 and 5/31/2027. Of these, there are 6 (9) pairs of transits of b and c that have their transit-midpoints within 6 (10) hours of one-another visible by JWST, or about one (two) per year.  Three of these six closest transit pairs are overlapping, and the other three have predicted time separations of 9 minutes -- 2.5 hours.  We summarize these transits in Table \ref{tab:JWST}.

\begin{table*} 
    \centering
    \begin{tabular}{lllrr}
        \hline
        Planet & $T_c$ (JD- & $T_c$ & $\Delta T_c$ & $T_1^\prime - T_4$  \\
         & 2460000) & (UT) & (hr) & (hr)  \\
        \hline
        \hline
        b & 251.5643$\pm$0.0047 & 11/3/2023 01:33 & \nodata & \nodata \\
        c & 251.67$\pm$0.02 & 11/3/2023 04:15 & 2.71$\pm$0.50 & -0.17$\pm$0.94 \\
        \hline
        c & 270.56$\pm$0.02 & 11/22/2023 01:22 & \nodata & \nodata \\
        b & 270.7584$\pm$0.0047 & 11/22/2023 06:12 & 4.84$\pm$0.50 & 1.96$\pm$0.94 \\
        \hline
        b & 629.0054$\pm$0.0056 & 11/14/2024 13:12 & \nodata & \nodata \\
        c & 629.20$\pm$0.02 & 11/14/2024 18:32 & 5.33$\pm$0.58 & 2.45$\pm$1.02 \\
        \hline
        c & 648.15$\pm$0.02 & 12/3/2024 15:38 & \nodata & \nodata \\
        b & 648.2440$\pm$0.0056 & 12/3/2024 17:51 & 2.22$\pm$0.58 & -0.66$\pm$1.02 \\
        \hline
        b & 1025.6710$\pm$0.0065 & 12/16/2025 05:30 & \nodata & \nodata \\
        c & 1025.67$\pm$0.03 & 12/16/2025 05:55 & 0.41$\pm$0.66 & -2.47$\pm$1.10 \\
        \hline
        b & 1403.2150$\pm$0.0073 & 12/28/2026 17:10 & \nodata & \nodata \\
        c & 1403.34$\pm$0.03 & 12/28/2026 20:12 & 3.03$\pm$0.74 & 0.15$\pm$1.18 \\
        \hline
        
    \end{tabular}
    \caption{Upcoming Transit Pairs of TOI 560 b and c visible by JWST, with mid-point time separations of $<$6 hours. A negative Egress-Ingress separation between the two planets ($T_1^\prime - T_4$) indicates an overlapping double-transit.} 
    \label{tab:JWST} 
\end{table*}

Given that both planets are nearly equal in size, the impact of stellar insolation on atmospheric evolution will be amenable to investigation. The youth of the TOI 560 system also makes these planets touchstones for constraining the temporal evolution of Neptune-size exoplanet atmospheres.  It will be particularly interesting to constrain atmospheric escape rates for the TOI 560 system to understand the role atmospheric escape may play in the radius distribution of exoplanets at short orbital periods orbiting older main-sequence stars \citep{2019ApJ...883L..15P}, and the implications it may have for terrestrial planet occurrence rates \citep{2015ApJ...798..112M,2020ApJ...893..122D,2019ApJ...874...81F,2018AJ....156...24M,2013PNAS..11019273P}.

\subsection{Comparisons to Contemporaneous Works}

During the preparation of this manuscript, additional papers were written presenting an independent HARPS RV analysis of this system in \cite{Barragan2022} as well as transit spectroscopy in \cite{Zhang2022}.  Our analyses arrived at similar conclusions to \citet{Barragan2022} for the youth of the TOI 560 system, primarily supported by our common analysis of the archival SuperWASP light curve to identify the stellar rotation period. Our detections of the RV semi-amplitudes are consistent to within the uncertainties of TOI 560 b and c with the work presented in \cite{Barragan2022}.  Our analysis also included the PFS and HIRES RVs, but not the HARPS and CORALIE RVs, and a joint chromatic RV analysis combining these data sets with our iSHELL RVs may be warranted with continued future RV monitoring of this system to further constrain the masses and confirm the mild eccentricities from the RV data alone.

Differentiating our two works, we had evidence for a moderate eccentricity orbit for planet b from the \textit{Spitzer} light curve photo-eccentric effect, and therefore we proceeded our analysis with an eccentric model whereas in \cite{Barragan2022} they assumed a circular orbit model. \citet{Barragan2022} also presents a more detailed investigation of the suitably of the planets for atmospheric characterization and the search for hydrogen and helium escape, the latter of which a repeated detection of helium escape is reported in \citet{Zhang2022}.
\section{Conclusions and Future Work} 
\label{sect:conclusions}
In this work, we have presented a validation of the TOI 560 system orbiting a young $\sim$0.15--1.4 Gyr active K4 star, based on three seasons of non-contemporaneous RV measurements from iSHELL, PFS, and HIRES, photometric data from \tess\, \textit{Spitzer}, and ground-based follow-up observations from PEST, NGTS and LCO, and high-resolution images from Gemini South, North, and SOAR. The system consists of two nearly equal-sized transiting Neptune-sized planets ($P=6.3981,18.8865$ days, $R_p=0.74,0.71 R_{Nep}$, $M_{b}=0.94^{+0.31}_{-0.23}M_{Nep}$, $M_{c}=1.32^{+0.29}_{-0.32}$) in a near 1:3 orbital resonance, both of which are suitable for comparative atmospheric characterization with JWST in a single observation sequence (e.g. a double transit and/or double eclipse). Additionally with the aid of the \textit{Spitzer} light curve, we confirm a moderate eccentricity for TOI 560 b via the photo-eccentric effect. The youth, orbital dynamics, and suitability of TOI 560 b and c for atmospheric characterization make it a touchstone system for characterizing and constraining the dynamics and atmospheric formation and evolution of compact multi-Neptune planetary systems.

In the future, additional visible and NIR precise and high cadence RVs are necessary to further constrain the stellar activity, dynamical masses of the planets, to assess if the planets are under-dense relative to exoplanets orbiting older main-sequence stars, and to search for additional candidates in the system. Future contemporaneous and high cadence precise RVs at visible and NIR wavelengths will enable more stringent constraints.

\begin{acknowledgements}

MM and PPP acknowledge support from NASA (Exoplanet Research Program Award \#80NSSC20K0251, TESS Cycle 3 Guest Investigator Program Award \#80NSSC21K0349, JPL Research and Technology Development, and Keck Observatory Data Analysis) and the NSF (Astronomy and Astrophysics Grants \#1716202 and 2006517), and the Mt Cuba Astronomical Foundation.

This paper includes data collected by the NASA TESS mission that are publicly available from the Mikulski Archive for Space Telescopes (MAST). Funding for the TESS mission is provided by NASA’s Science Mission Directorate. We acknowledge the use of public TESS data from pipelines at the TESS Science Office and at the TESS Science Processing Operations Center.

The authors wish to recognize and acknowledge the very significant cultural role and reverence that the summit of Maunakea has always had within the indigenous Hawaiian community, where the iSHELL and HIRES observations were recorded. We are most fortunate to have the opportunity to conduct observations from this mountain. The authors also wish to thank the California Planet Search (CPS) collaboration for carrying out the HIRES observations recorded in 2020 presented in this work.

This work includes observations obtained at the international Gemini Observatory, a program of NSF’s NOIRLab, which is managed by the Association of Universities for Research in Astronomy (AURA) under a cooperative agreement with the National Science Foundation. Some of the observations in the paper made use of the High-Resolution Imaging instrument Zorro obtained under Gemini LLP Proposal Number: GN/S-2021A-LP-105. Zorro was funded by the NASA Exoplanet Exploration Program and built at the NASA Ames Research Center by Steve B. Howell, Nic Scott, Elliott P. Horch, and Emmett Quigley.
On behalf of the Gemini Observatory partnership: the National Science Foundation (United States), National Research Council (Canada), Agencia Nacional de Investigaci\'{o}n y Desarrollo (Chile), Ministerio de Ciencia, Tecnolog\'{i}a e Innovaci\'{o}n (Argentina), Minist\'{e}rio da Ci\^{e}ncia, Tecnologia, Inova\c{c}\~{o}es e Comunica\c{c}\~{o}es (Brazil), and Korea Astronomy and Space Science Institute (Republic of Korea). Data were collected as part of program GN-2019A-LP-101.

Based on data collected under the NGTS project at the ESO La Silla Paranal Observatory.  The NGTS facility is operated by the consortium institutes with support from the UK Science and Technology Facilities Council (STFC)  projects ST/M001962/1 and  ST/S002642/1. This work has made use of data from the European Space Agency (ESA) mission {\it Gaia} (\url{https://www.cosmos.esa.int/gaia}), processed by the {\it Gaia} Data Processing and Analysis Consortium (DPAC, \url{https://www.cosmos.esa.int/web/gaia/dpac/consortium}). Funding for the DPAC has been provided by national institutions, in particular the institutions participating in the {\it Gaia} Multilateral Agreement.

This paper is based on observations obtained from the Las Campanas Remote Observatory that is a partnership between Carnegie Observatories, The Astro-Physics Corporation, Howard Hedlund, Michael Long, Dave Jurasevich, and SSC Observatories.

MINERVA-Australis is supported by Australian Research Council LIEF Grant LE160100001, Discovery Grant DP180100972, Mount Cuba Astronomical Foundation, and institutional partners University of Southern Queensland, UNSW Sydney, MIT, Nanjing University, George Mason University, University of Louisville, University of California Riverside, University of Florida, and The University of Texas at Austin. We respectfully acknowledge the traditional custodians of all lands throughout Australia, and recognise their continued cultural and spiritual connection to the land, waterways, cosmos, and community. We pay our deepest respects to all Elders, ancestors and descendants of the Giabal, Jarowair, and Kambuwal nations, upon whose lands the Minerva-Australis facility at Mt Kent is situated.

This work makes use of observations from the LCOGT network. Part of the LCOGT telescope time was granted by NOIRLab through the Mid-Scale Innovations Program (MSIP). MSIP is funded by NSF.


E.A.P. acknowledges the support of the Alfred P. Sloan Foundation. 

L.M.W. is supported by the Beatrice Watson Parrent Fellowship and NASA ADAP Grant 80NSSC19K0597. 

A.C. is supported by the NSF Graduate Research Fellowship, grant No. DGE 1842402.

D.H. acknowledges support from the Alfred P. Sloan Foundation, the National Aeronautics and Space Administration (80NSSC19K0379), and the National Science Foundation (AST-1717000).

I.J.M.C. acknowledges support from the NSF through grant AST-1824644.

P.D. acknowledges support from a National Science Foundation Astronomy and Astrophysics Postdoctoral Fellowship under award AST-1903811. 

A.B. is supported by the NSF Graduate Research Fellowship, grant No. DGE 1745301.

R.A.R. is supported by the NSF Graduate Research Fellowship, grant No. DGE 1745301.

C. D. D. acknowledges the support of the Hellman Family Faculty Fund, the Alfred P. Sloan Foundation, the David \& Lucile Packard Foundation, and the National Aeronautics and Space Administration via the TESS Guest Investigator Program (80NSSC18K1583).  

J.M.A.M. is supported by the NSF Graduate Research Fellowship, grant No. DGE-1842400. J.M.A.M. also acknowledges the LSSTC Data Science Fellowship Program, which is funded by LSSTC, NSF Cybertraining Grant No. 1829740, the Brinson Foundation, and the Moore Foundation; his participation in the program has benefited this work.

T.F. acknowledges support from the University of California President's Postdoctoral Fellowship Program.

C.A.B notes that some of the research described
in this publication was carried out in part at the Jet Propulsion
Laboratory, California Institute of Technology, under a contract
with the National Aeronautics and Space Administration.
\end{acknowledgements}

\facilities{}
NASA IRTF, Keck Observatory, Magellan Telescope, Gemini North, Gemini South, Fred L. Whipple Observatory, TESS, ESO La Silla Paranal Observatory, Las Cumbres Observatory, SOAR telescope
\software{}
Python: \textit{pychell} \citep{Cale2019}, AstroImageJ \citep{AstroimageJ2016}, EDI-Vetter Unplugged \citep{Zink_2020},DAVE \citep{Kostov_2019}, tpfplotter \citep{Aller2020}, REBOUND \citep{Rein2012,Rein2015}, NumPy \citep{Harris2020},  SciPy \citep{Virtanen2020}, Matplotlib \citep{Hunter2007}, AstroPy \citep{Robitaille2013}, Numba \citep{Lam2015}, EXOFASTv2 \citep{Eastman2013}.

\bibliography{ref}{} 
\bibliographystyle{aasjournal}

\clearpage
\appendix
\label{sect:appen}

In this Appendix, we present the results of additional analyses not included in the main text.  
\vspace{1cm}
\section{Reconnaissance Spectroscopy}

Herein we present the detailed reconnaissance spectroscopy from NRES and TRES.

\begin{table*}[htp]
    \centering
    \begin{tabular}{|l|c|c|c|c|}
        \hline
        Parameter & 2019-05-12 & 2019-10-29 & 2019-11-04 &   Average \\
        \hline
        \hline 
       $T_\mathrm{eff}$ [K] & $4626 \pm 100$ &  $4641 \pm 100$& $4650 \pm 100$&  $4639 \pm 12$ \\ 
        $\log g$ & $4.6 \pm 0.1$ &  $4.7 \pm 0.1$& $4.7 \pm 0.1$ & $4.7 \pm 0.1$\\ 
        $Fe/H$ & $-0.08 \pm 0.06$  & $-0.04 \pm 0.06$& $-0.13 \pm 0.06$ & $0.08 \pm 0.05$\\
        v$ \sin i $ [km$s^{-1}$] & $\leq 2$  & $\leq 2$ & $\leq 2$  & $\leq 2$\\
        $M_{*}^a$ & $0.702 \pm 0.031$ M$\odot$  & $0.71 \pm 0.03$ M$\odot$& $0.69 \pm 0.03$ M$\odot$& $0.7 \pm 0.01$\\
        $R_{*}^a$ & $0.677 \pm 0.024$ R$\odot$ & $0.686 \pm 0.024$ R$\odot$& $0.664 \pm 0.023$ R$\odot$ & $0.676 \pm 0.01$\\
        \hline 
     \end{tabular}
     \caption{The average results for the stellar parameters and the internal error rms deviation for the three NRES observations (e.g. the quoted uncertainties are not propagated). \\
     $^a$Estimated from NRES ExoFASTv2 analysis}.
     \label{tab:starr-2}
\end{table*}

\begin{table}[htp]
    \centering
    \begin{tabular}{|l|r|r|r|}
        \hline
        Parameter & 2019-04-16  & 2019-04-19 & SPC\\
        \hline
        \hline 
        ccf$^a$ & 0.97 &  0.97& 0.97 \\
        SNRe$^b$ & 39.7 & 29.9& 34.8 \\
       $T_\mathrm{eff}$ [K] & $4689 \pm 50$ &  $4688 \pm 50$& $4689 \pm$ 50$^c$\\ 
        $\log g$ & $4.67 \pm 0.1$ &  $4.69 \pm 0.1$& $4.68 \pm 0.1$\\ 
        $V_{rot} [Km/s]$ & $2.4 \pm 0.5$ &  $3.7 \pm 0.5$& $3.1 \pm 0.5$\\
        $[m/H]$ & $-0.25 \pm 0.08$  & $0.17 \pm 0.08$& $-0.21 \pm 0.08$\\  
        \hline 
     \end{tabular}
     \caption{The stellar parameters for the two TRES observations.  The first two columns are from the TRES analysis of each individual spectra, and the SPC analysis of both nights is presented in the third column, with the uncertainties quoted derived from the internal error rms deviation (see ${P\S}$\ref{sect:recon}). \\
     $^a$ Peak value of the cross-correlation function.\\$^b$ Effective SNR per spectral pixel.\\ $^c$ The formal error is only 1 K, but we adopt a systematic noise floor of 50 K as commonly used for TRES spectra.} 
     \label{tab:starr-1}
\end{table}

\clearpage 
\section{Mass-Radius Relation}
Figure \ref{fig:mass-radius} below shows the mass-radius relationship of the TOI 560’s 3-$\sigma$ masses plotted in green, compared to all other exoplanets. 
\begin{figure*}[htp]
     \centering 
    \includegraphics[width=\textwidth]{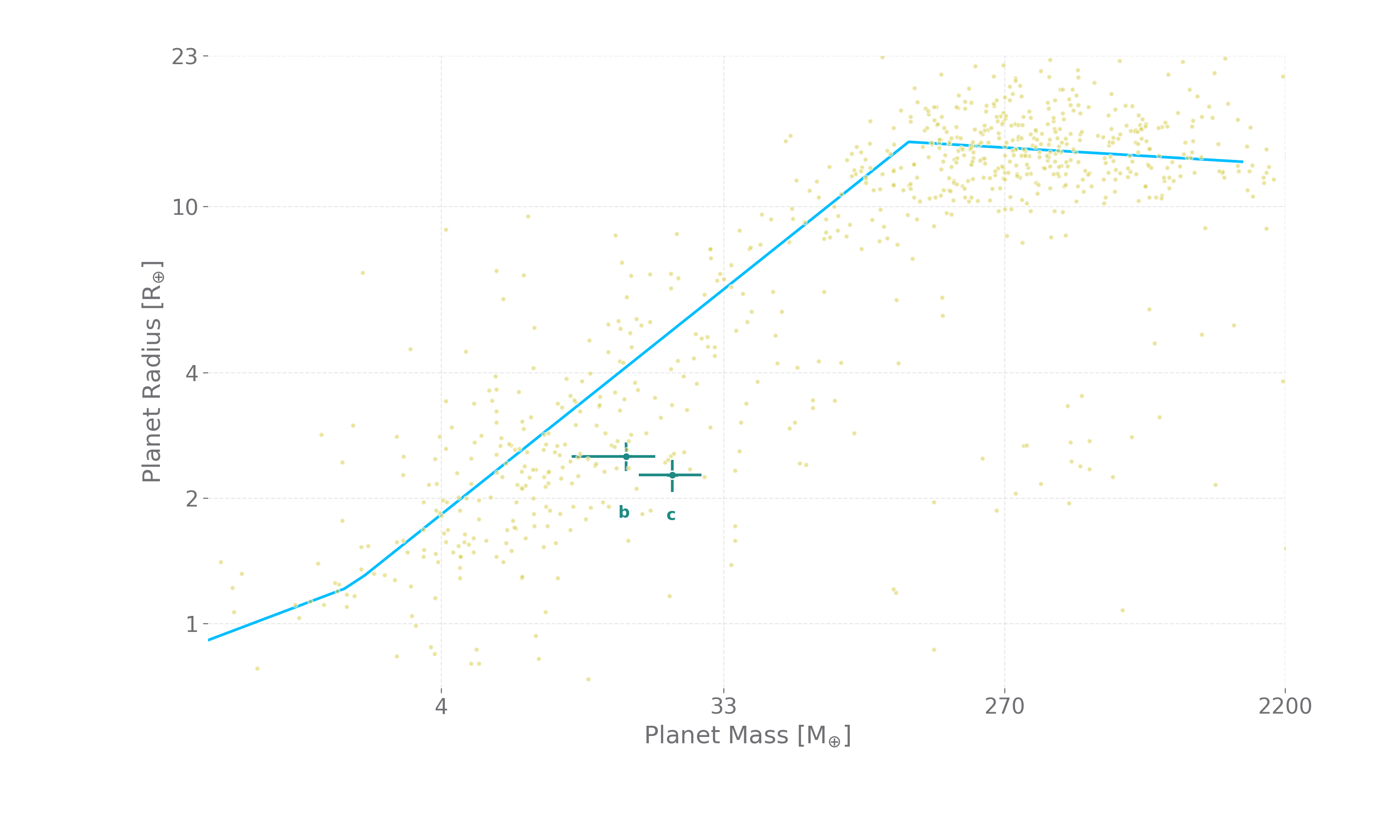} 
     \caption{The mass-radius diagram for all exoplanets with provided radii and masses from the NASA Exoplanet Archive in yellow. TOI 560’s b and c masses are plotted in green with $1-\sigma$ error bars}. 
     \label{fig:mass-radius} 
 \end{figure*}
 
\clearpage
\section{Ground-Based Transit Light Curves} 
\begin{figure*}[htp] 
   \centering
    \includegraphics[width=0.75\linewidth]{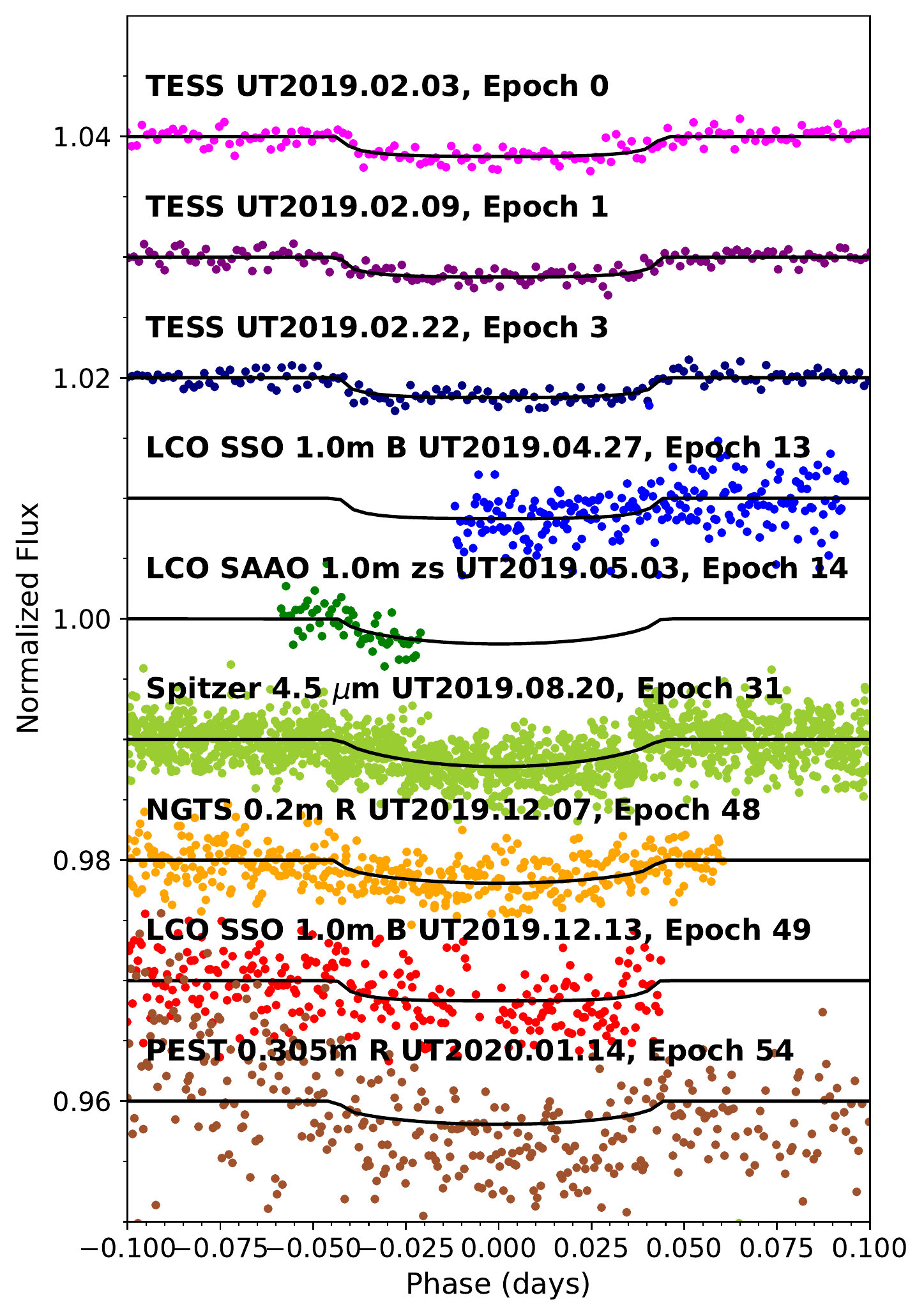}
\caption{TOI 560 b light curves from \tess\, LCO SSO, LCO SAAO, \textit{Spitzer}, NGTS, LCO SSO, and PEST observatories as labeled, on the UT dates and in the filters labeled, plotted as a function of time since mid-transit on the horizontal axis and normalized flux with relative arbitrary offsets on the vertical axis. The ground-based and \textit{Spitzer} data show clear transit detections consistent with the predicted ephemerides from \tess.}
    \label{fig:lcs_b1}
\end{figure*}

\begin{figure*}[htp] 
   \centering
    \includegraphics[width=0.9\linewidth]{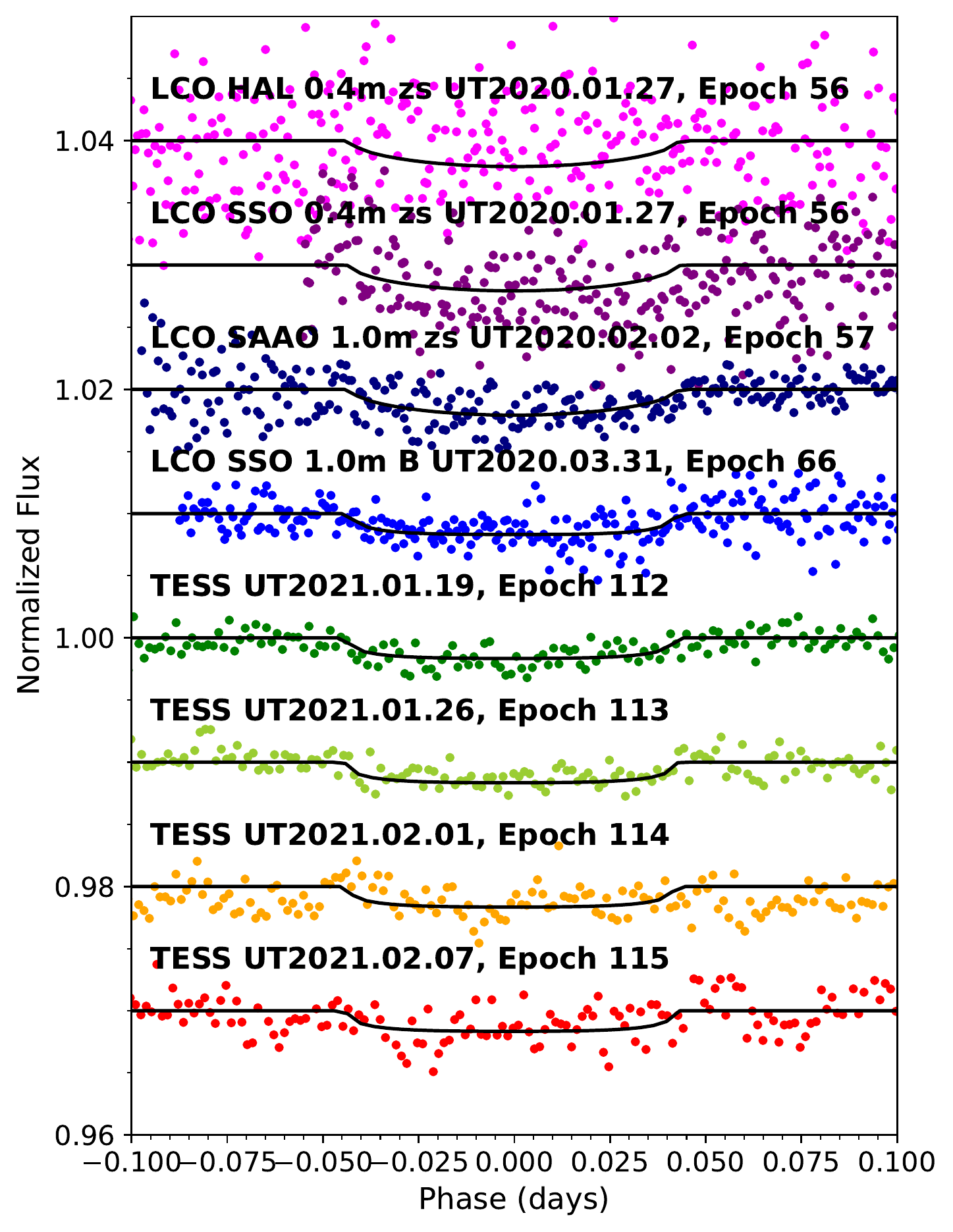}
\caption{TOI 560 b light curves from \tess\, and LCO HAL, LCO SSO, LCO SAAO and LCO SSO observatories as labeled, on the UT dates and in the filters labeled, plotted as a function of time since mid-transit on the horizontal axis and normalized flux with relative arbitrary offsets on the vertical axis. }
    \label{fig:lcs_b2}
\end{figure*}

\begin{figure*}[htp] 
  \centering 
    \includegraphics[width=0.7\linewidth]{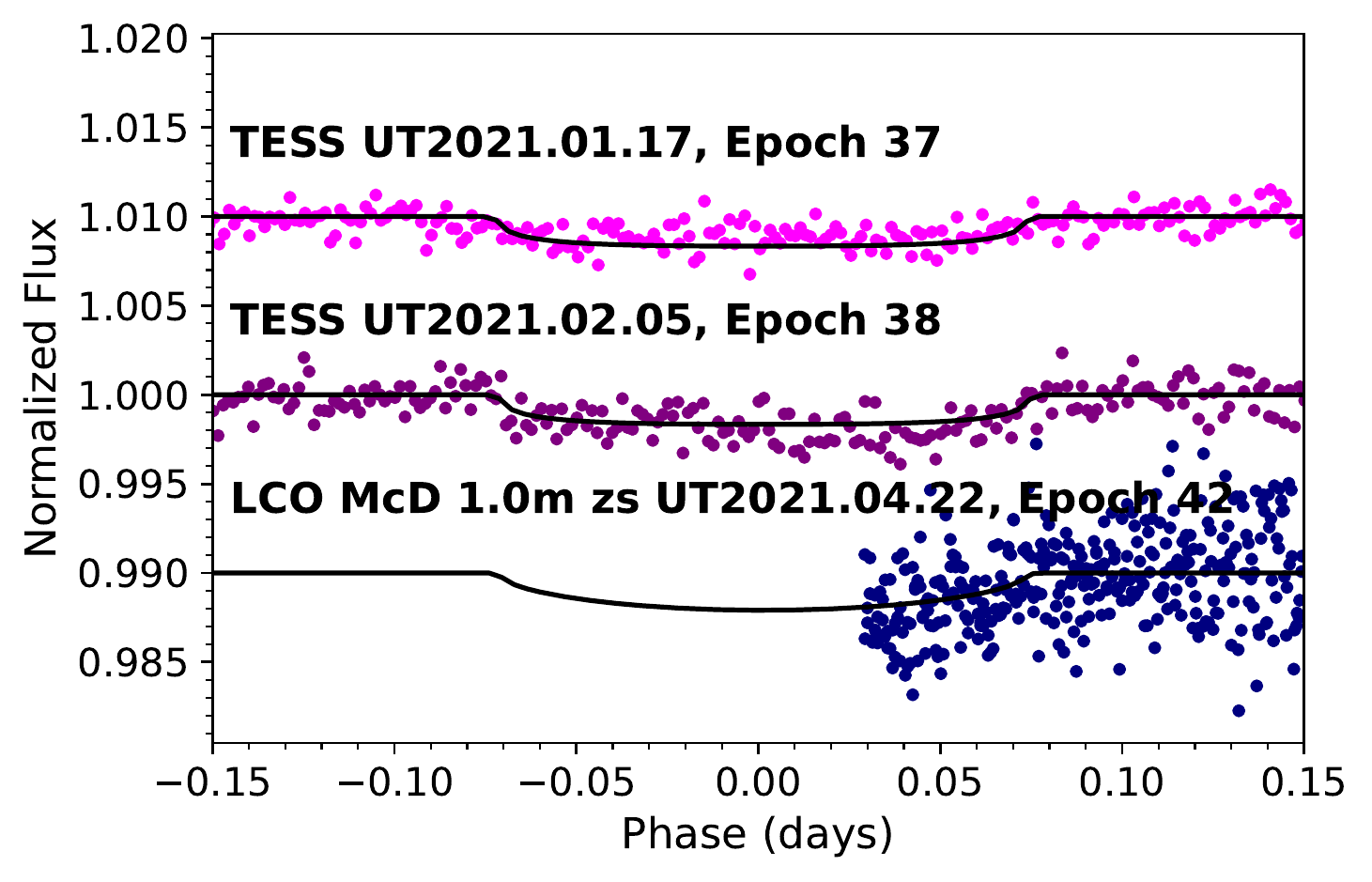}
\caption{TOI 560 c light curves from \tess\, LCO McD observatories as labeled, on the UT dates and in the filters labeled, plotted as a function of time since mid-transit on the horizontal axis and normalized flux with relative arbitrary offsets on the vertical axis.} 
    \label{fig:lcs_c}
\end{figure*}

The transit light curves of TOI 560 b are shown in Figures \ref{fig:lcs_b1}, \ref{fig:lcs_b2}, and \ref{fig:lcs_c}.

\clearpage 

\clearpage 

\clearpage 
\section{RV Models} 
In this section we show the results if other models that we considered in our RV analysis,

\subsection{No Gaussian Process Model for Stellar Activity}
Here we present the results of an RV model with no GP applied to account for the stellar activity.

\begin{figure*}[htp]
    \centering
    \includegraphics[width=0.8\textwidth]{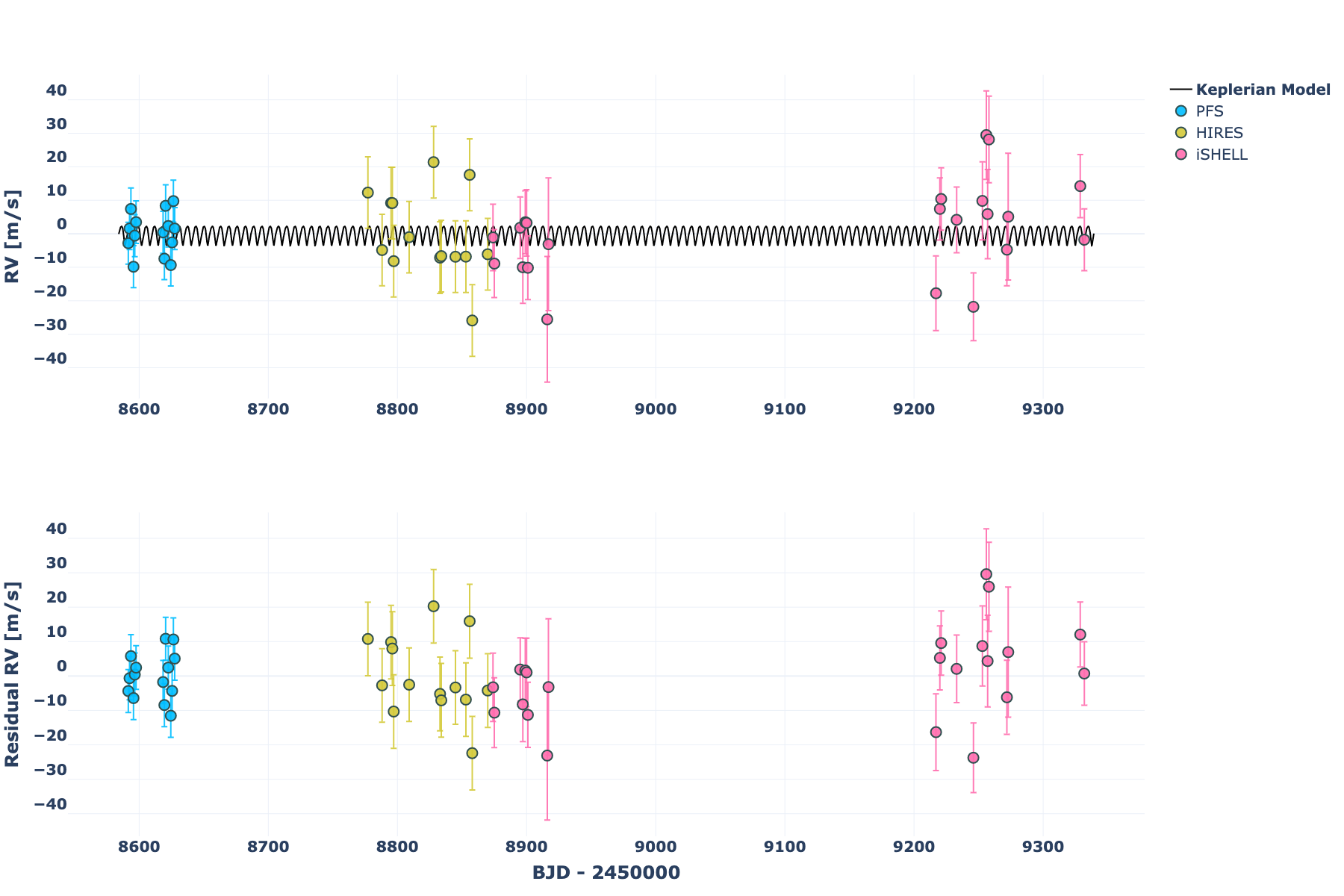}
    \caption{Full RV time-series plot, with the black line representing our Keplerian model of the b and c planets.  Pink, yellow, and blue datapoints are nightly iSHELL, HIRES, and PFS RVs respectively.  The top plot shows the RVs over the full time baseline of observations, while the bottom plot shows the residuals (data $-$ model).}
    \label{fig:bc_fullrvs}
\end{figure*}

\begin{figure*}[htp]
    \centering
    \quad
    \includegraphics[width=.45\textwidth]{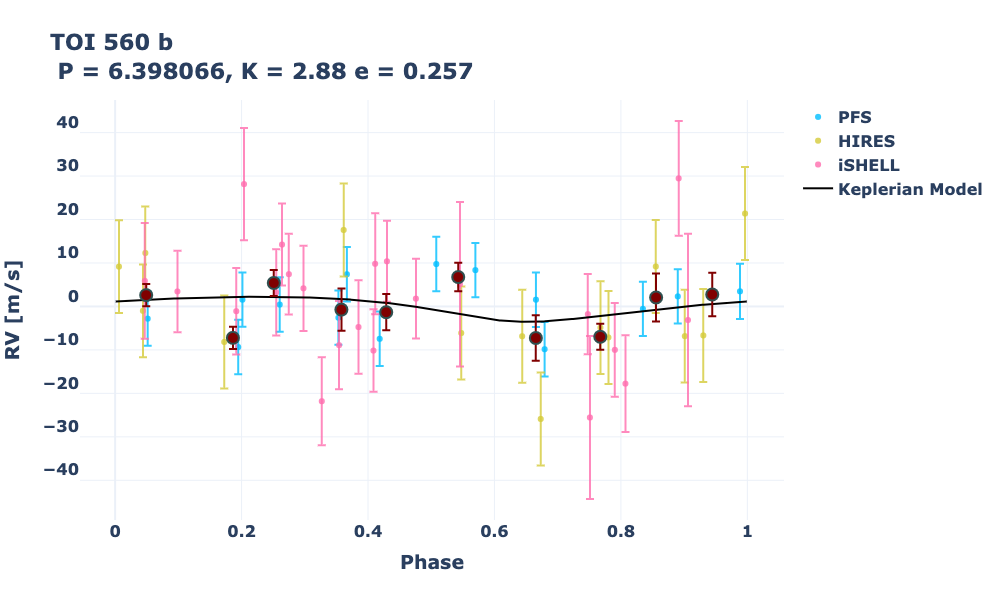}
    \includegraphics[width=.45\textwidth]{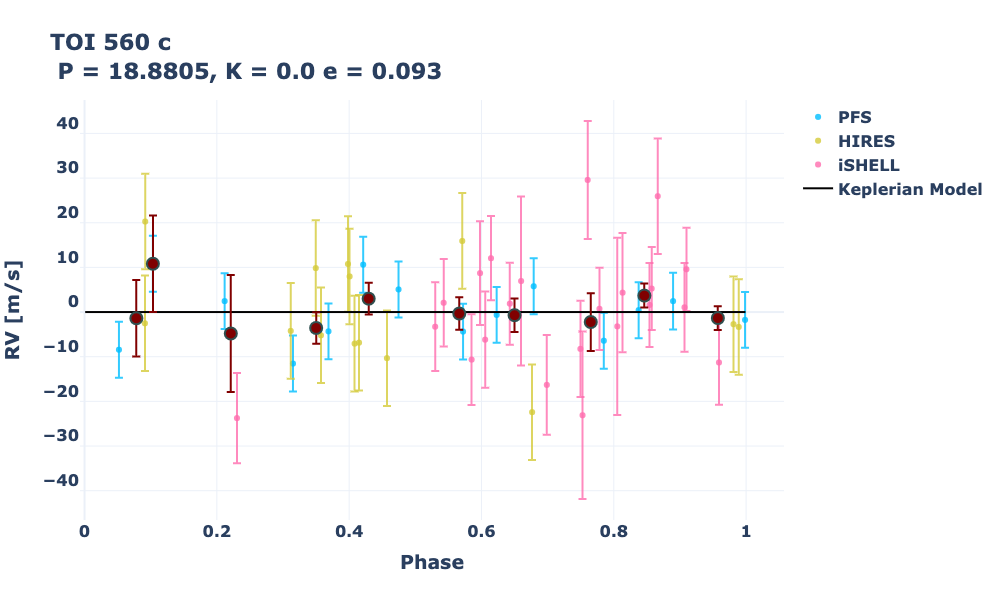}
    \caption{RV time-series plots phased to the period of b (left) and c (right), with the black models representing each individual planet signal, after subtracting the other planet signal. Pink,Yellow, and blue data points are nightly iSHELL, HIRES, and PFS RVs, and red points are binned nightly RVs.} 
    \label{fig:bc_phasedrvs}
\end{figure*}
\begin{figure*}[htp]
    \centering
    \includegraphics[width=1.05\textwidth]{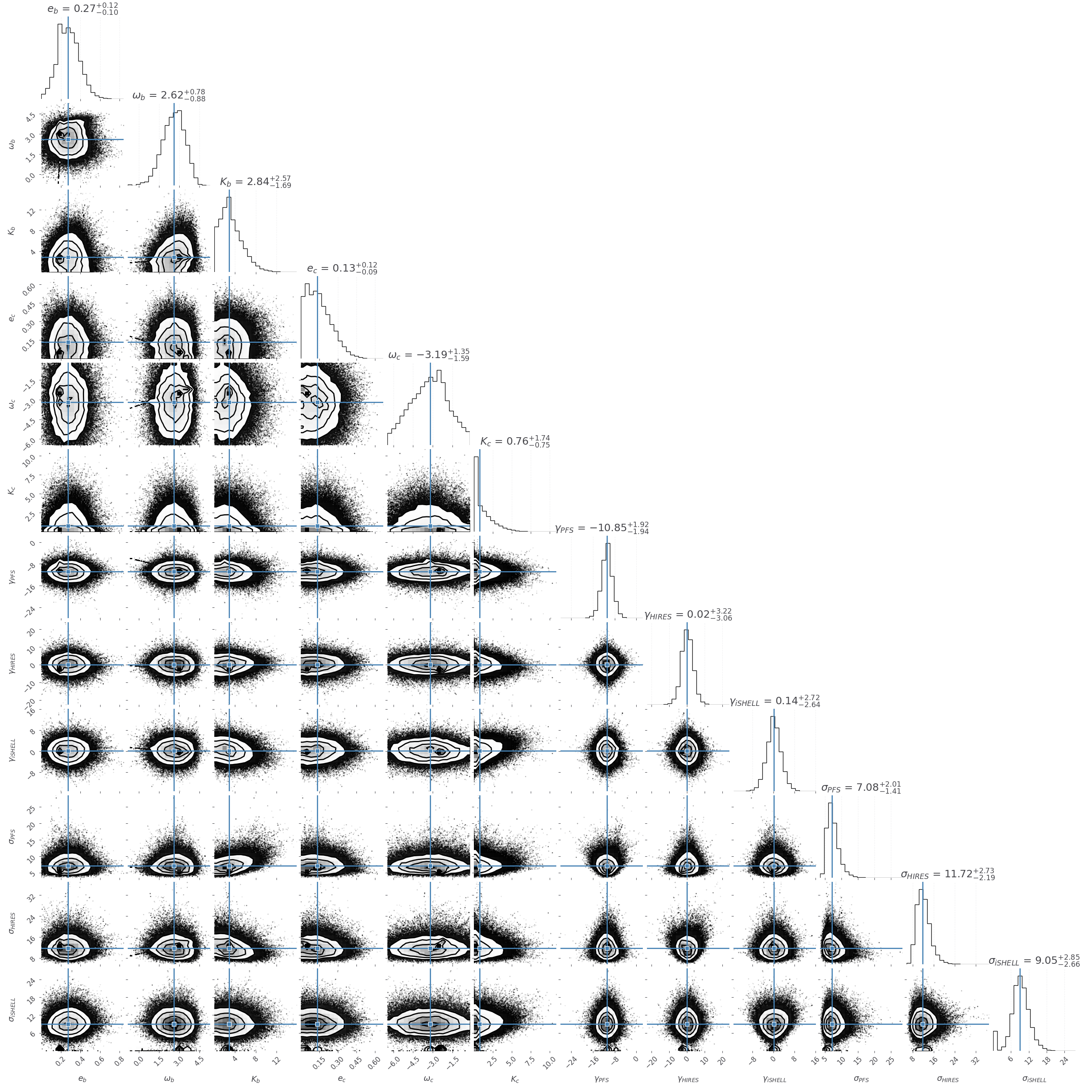}
    \caption{MCMC cornerplot of our two-planet RV model (iSHELL+HIRES+PFS), showing the posterior distributions and covariances of each model parameter that we allowed to vary.}
    \label{fig:bc_cornerplot} 
\end{figure*} 

\clearpage 
\subsection{Joint GP first chromatic kernel J$_{1}$ model}
Here we present the results of an RV model using the joint GP chromatic Kernel J$_{1}$ model parametrize the stellar activity amplitude through a linear kernel where each amplitude is a free parameter, as in \citet{Cale2021}. 
\begin{figure*}[htp]
    \centering
    \includegraphics[width=\textwidth]{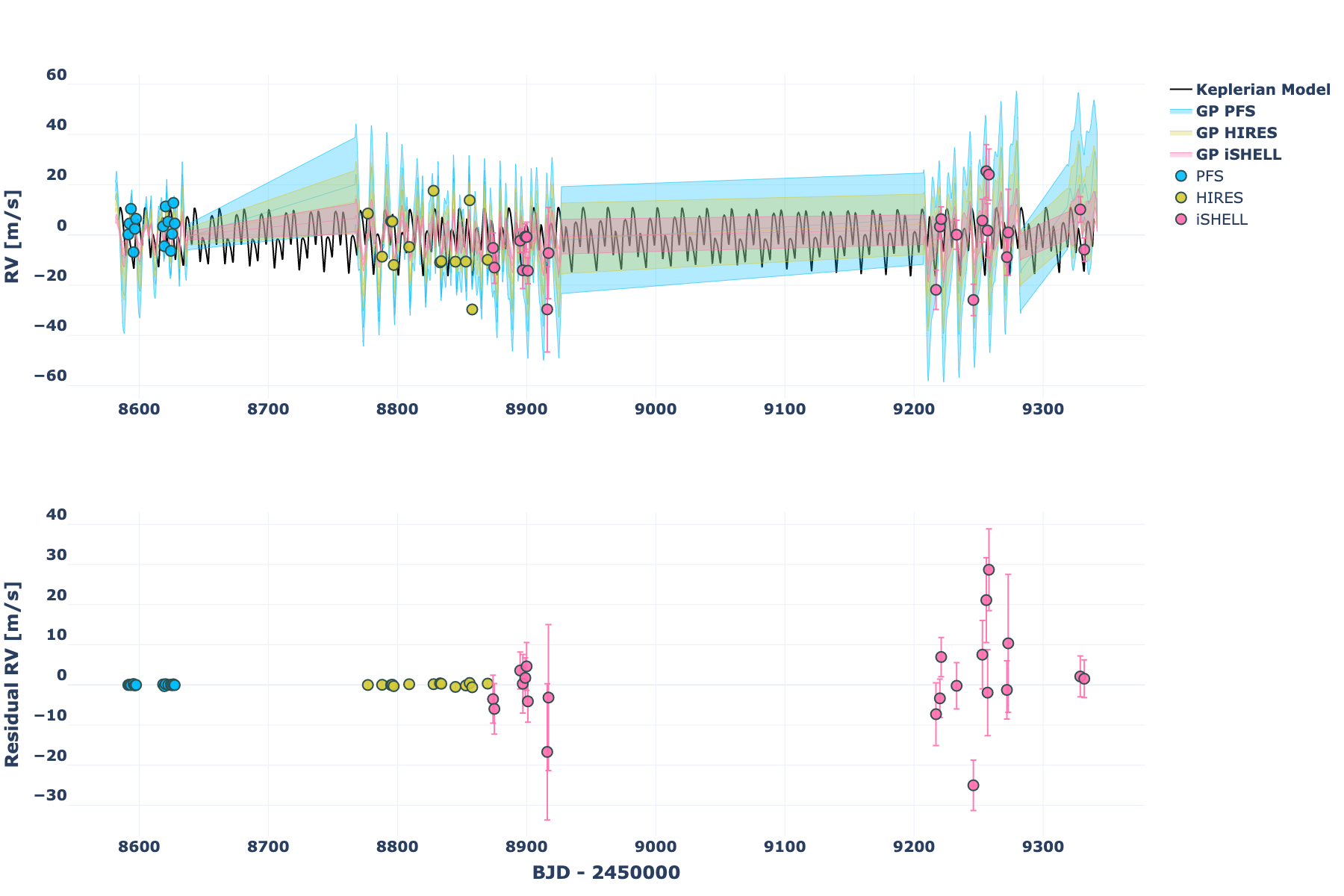}
    \caption{Full, unphased RV time-series plot for the joint GP first chromatic Kernel J$_{1}$ model with the 12.2 day prior on $\eta_{P}$. Residuals (data-model) are shown in the lower plot.  The stellar activity GP model appears to be flexible enough to over-fit the HIRES and PFS RVs}  
    \label{fig:j1}
\end{figure*}
\begin{figure*}[htp]
    \centering
    \quad
    \includegraphics[width=.45\textwidth]{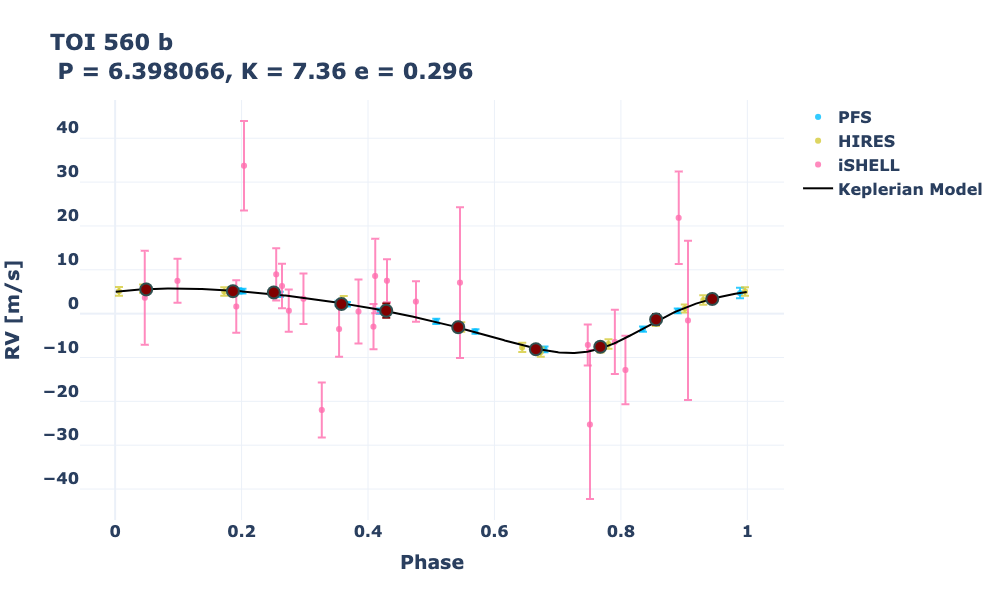}
    \includegraphics[width=.45\textwidth]{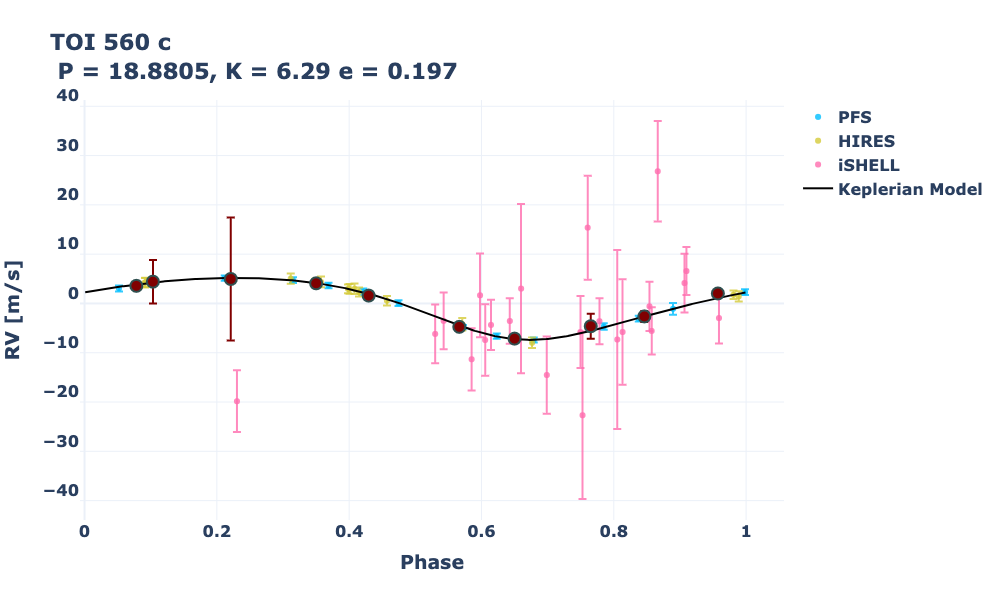}
    \caption{RV time-series plot for the joint GP first chromatic Kernel J$_{1}$ model with the 12.2 day prior on $\eta_{P}$, phased to the period of b and c respectively, after subtracting the best-fit stellar activity model and the other planet.}
    \label{fig:j1-1} 
\end{figure*}
\begin{figure*}[htp] 
    \centering
    \includegraphics[width=1.05\textwidth]{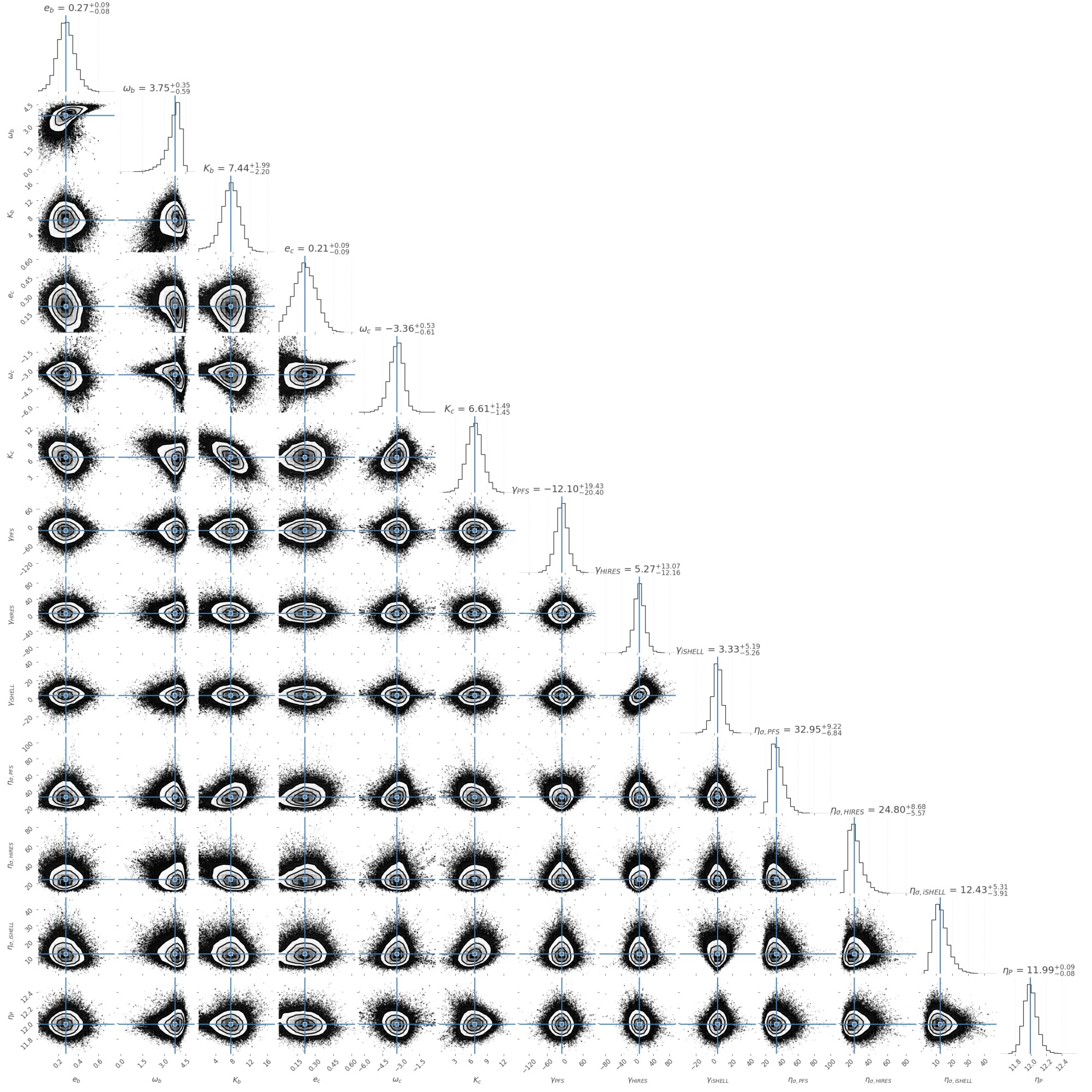}
    \caption{MCMC cornerplot of our joint GP first chromatic Kernel J$_{1}$ model with the 12.2 day prior on $\eta_{P}$, showing the posterior distributions and covariances of each model parameter that we allowed to vary.} 
    \label{fig:j1-2}
\end{figure*}

\clearpage 
\section{Results of our Disjoint Model}
 This model is using an independent (disjoint) GP to model each RV data set, akin to \texttt{RadVel}  \citep{Fulton_2017}.  Given the lack of overlap between RV data sets, and the relatively sparse RV cadence sampling, this RV model yields similar overfit results to our $J_1$ joint kernel analysis in the previous sub-section.
 
 \begin{figure*}[htp]
    \centering
    \includegraphics[width=\textwidth]{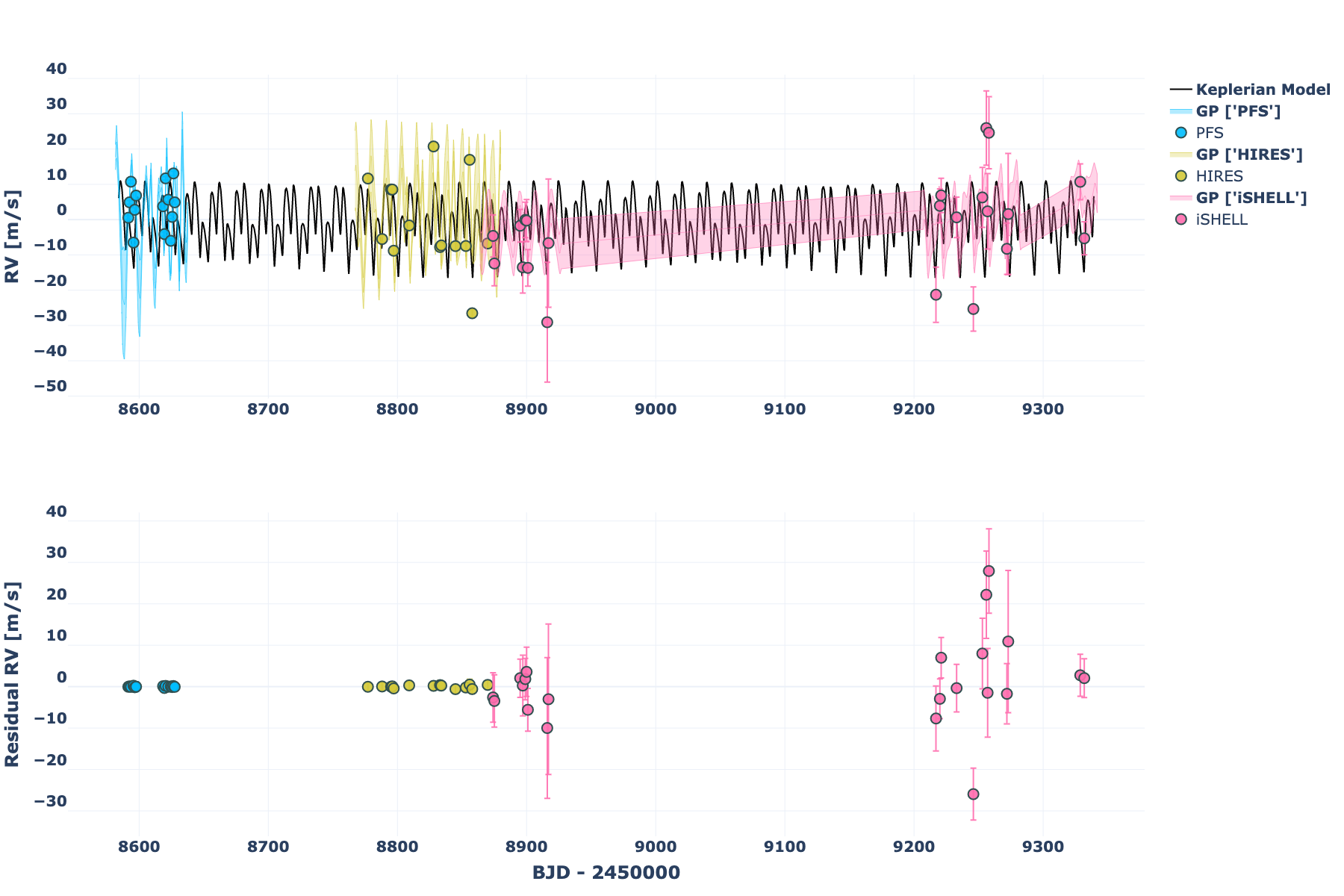}
    \caption{Full, unphased RV time-series plot for the disjoint GP model with the 12.2 day prior on $\eta_{P}$. Residuals (data-model) are shown in the lower plot as in previous figures.} 
    \label{fig:disjoint1}
\end{figure*}
\begin{figure*}[htp]
    \centering
    \quad
    \includegraphics[width=.45\textwidth]{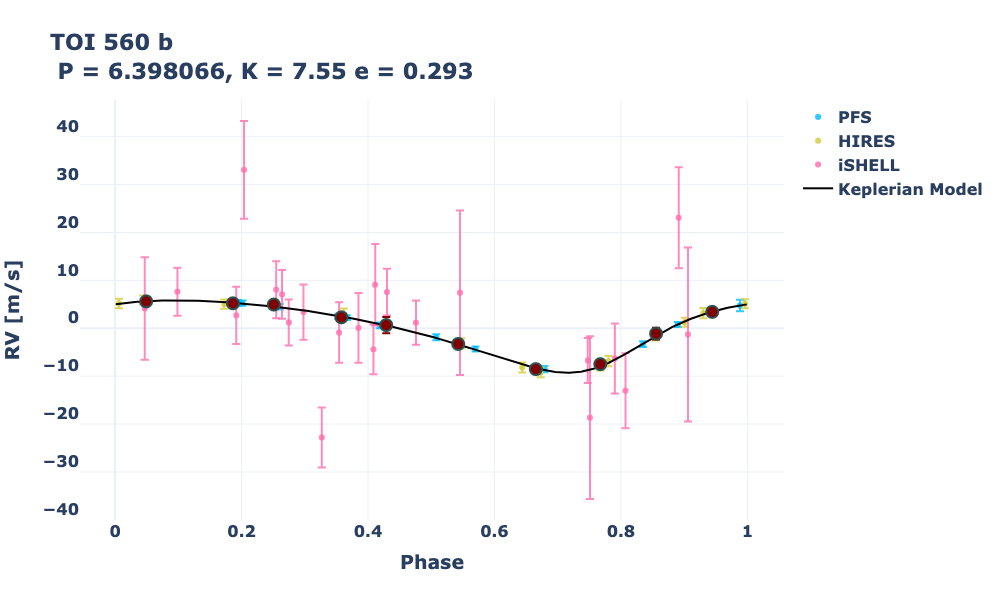}
    \includegraphics[width=.45\textwidth]{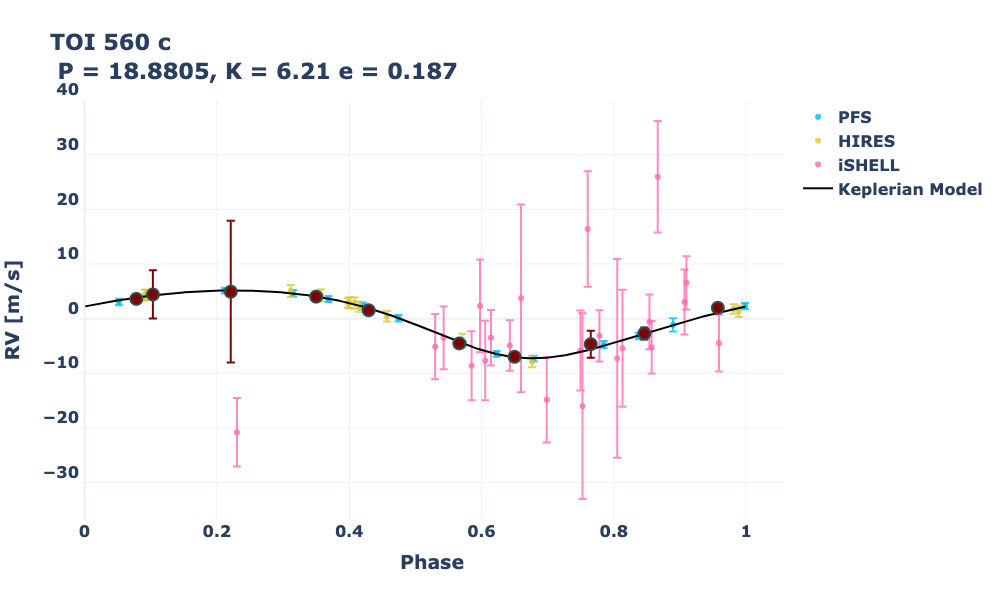}
    \caption{RV time-series plot for the disjoint GP model with the 12.2 day prior on $\eta_{P}$ , phased to the period of b and c respectively, after subtracting the best-fit stellar activity model and the other planet.}
    \label{fig:disjoint2}
\end{figure*}
\begin{figure*}[htp] 
    \centering
    \includegraphics[width=1.05\textwidth]{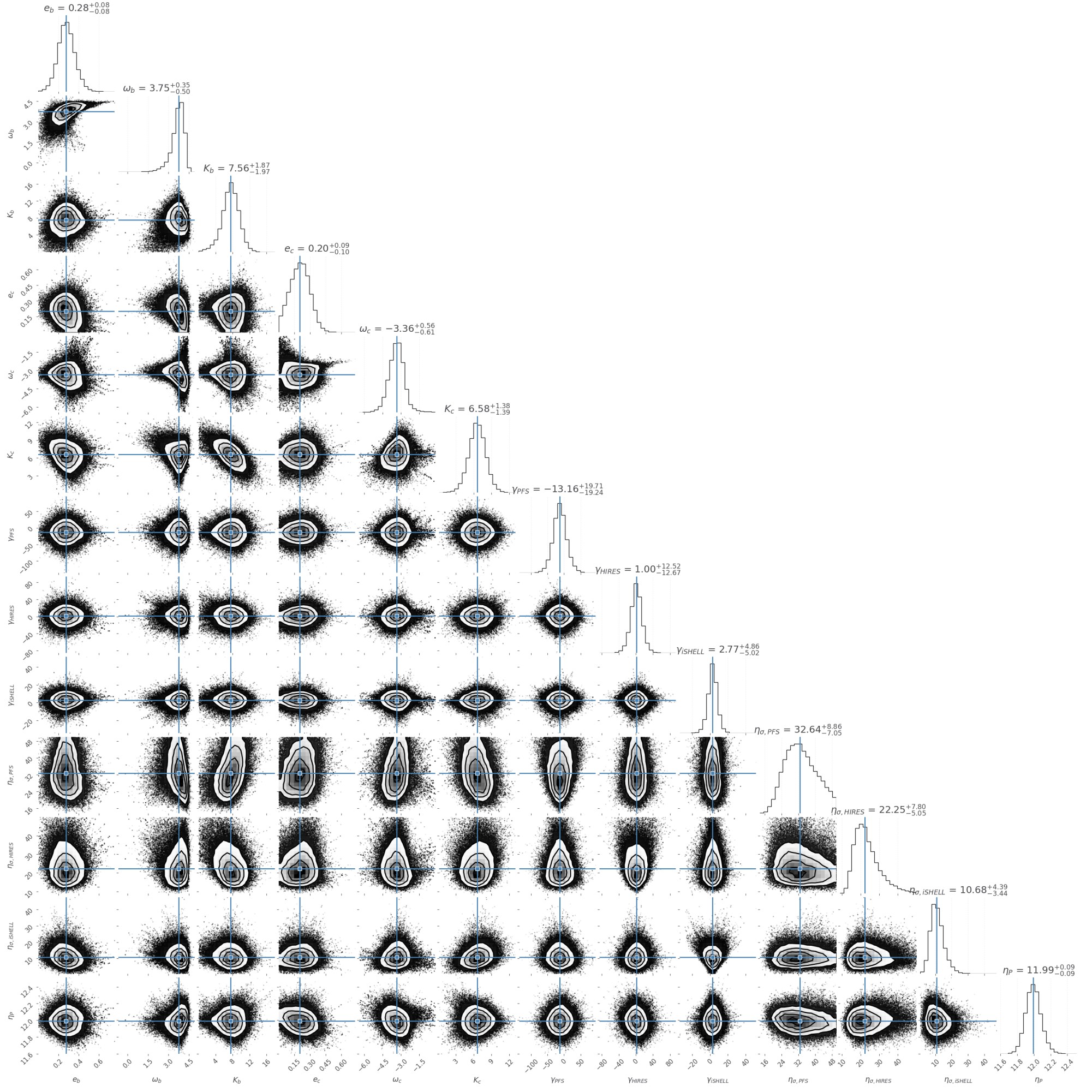}
    \caption{MCMC cornerplot of our disjoint GP model for all spectrographs with the 12.2 day prior on $\eta_{P}$, showing the posterior distributions and covariances of each model parameter that we allowed to vary.}   
    \label{fig:disjoint3}
\end{figure*}

\clearpage 

\section{Summary of the priors used in the $J_{1}$ and disjoint kernels}

\begin{table*}[http]
\centering
\begin{tabular}{lccccccc}
    \hline
    Parameter [units]& Initial & Priors &  MAP & MCMC & MAP & MCMC \\
    &Value&& Value& Posterior& Value& Posterior\\
    &($P_{0}$)& & ($J_{1}$) &($J_{1}$) & (Disjoint)& (Disjoint) \\ 
    \hline
    \hline
    $P_b$ [days] & 6.3980661 & \faLock &  -- & -- & -- & -- \\
    $TC_b$ [days] & 2458517.68971 & \faLock &  -- & -- & -- & -- \\
    $e_b$ & 0.294 & $\mathcal{U}(0,1)$; & 0.30 & $0.27^{+0.09}_{-0.08}$ & 0.29 & $0.28^{+0.08}_{-0.08}$\\
    & & $\mathcal{N}(P_{0},0.13)$& & & & \\
    $\omega_b$ & $130\pi/180$ & $\mathcal{U}(P_{0}-\pi,P_{0}+\pi)$; & 3.87 & $3.75^{+0.35}_{-0.59}$&3.81&$3.75^{+0.35}_{-0.50}$\\
    & &  $\mathcal{N}(P_{0},45\pi/180)$ & & & & \\
    $K_b$ [m\,s$^{-1}$] & 10 & $\mathcal{U}(0, \infty)$  & 7.36 & $7.43^{+1.10}_{-2.21}$ & 7.54 & $7.56^{+1.88}_{-1.97}$\\
    \hline
    $P_c$ [days] & 18.8805 & \faLock &  -- & -- & -- & --\\
    $TC_c$ [days] & 2458533.593 & \faLock & -- & -- & -- & --\\
    $e_c$ & 0.093 & $\mathcal{U}(0,1)$;  & 0.20 & $0.21^{+0.09}_{-0.09}$ & 0.19 & $0.20^{0.10}_{-0.10}$\\
    & & $\mathcal{N}(P_{0},0.13)$& & & & \\ 
    $\omega_c$ & $190\pi/180$ & $\mathcal{U}(P_{0}-\pi,P_{0}+\pi)$; & -3.61 & $-3.36^{+0.53}_{-0.62}$ & -3.57 & $-3.36^{+0.56}_{-0.62}$ \\
    & & $\mathcal{N}(P_{0},45\pi/180)$& & & & \\ 
    $K_c$ [m\,s$^{-1}$] & 10 & $\mathcal{U}(0, \infty)$ & 6.29 & $6.61^{+1.49}_{-1.49}$ & 6.21 & $6.58^{+1.39}_{-1.40}$\\ 
    \hline
    $\gamma_{iSHELL}$ [m\,s$^{-1}$] & $\mathrm{MEDIAN}(RV_{iSHELL})$+1$^{a}$ & $\mathcal{N}$($P_{0},100)$ & 3.73 & $3.33^{+5.21}_{-5.29}$ &3.15 & $2.77^{+4.88}_{-5.04}$\\ 
    $\gamma_{PFS}$ [m\,s$^{-1}$] & $\mathrm{MEDIAN}(RV_{PFS})+1^{a}$ & $\mathcal{N}(P_{0},100)$ & -13.85 & $-12.10^{+19.52}_{-20.47}$ &-14.20 & $-13.16^{+19.80}_{-19.31}$\\
    $\gamma_{HIRES}$ [m\,s$^{-1}$] & $\mathrm{MEDIAN}(RV_{HIRES})$+1$^{a}$ & $\mathcal{N}(P_{0}+1, 100)$ & 4.34 &  $5.27^{+13.12}_{-12.22}$ & 1.19 & $0.996^{+12.57}_{-12.73}$\\ 
    $\eta_{P}$ & 12.03 & $\mathcal{N}(P_{0},0.07)$ & 111.97 & $11.99^{+0.09}_{-0.08}$ & 11.97 & $11.99^{+0.09}_{-0.09}$ \\
    $\eta_{\ell}$ & 0.44 & \faLock &  -- & -- & -- & --\\
    $\eta_{\tau}$ & 57.96 & \faLock & -- & -- & -- & --\\
    $\eta_{\sigma,iSHELL}$ & $\mathrm{STDDEV}_{iSHELL}$ & $\mathcal{J}(0.67,50)$ & 9.54 & $12.43^{+5.34}_{-3.92}$& 8.43 & $10.68^{+4.41}_{-3.45}$ \\
    $\eta_{\sigma,PFS}$ & $\mathrm{STDDEV}_{PFS}$ & $\mathcal{J}(0.67,50)$ &  29.36 & $32.95^{+9.27}_{-6.86}$& 30.04 & $32.64^{+8.90}_{-7.07}$ \\
    $\eta_{\sigma,HIRES}$ & $\mathrm{STDDEV}_{HIRES}$ & $\mathcal{J}(0.67,50)$ &  13.63 & $18.77^{+7.49}_{-5.30}$& 17.34 & $22.25^{+7.84}_{-5.06}$\\
    \hline
\end{tabular}
\caption{The circular model parameters and prior distributions used in our model that considers the transiting b and c planets, as well as the recovered MAP fit and MCMC posteriors for the $J_{1}$ and disjoint Kernel models. \faLock\ indicates the parameter is fixed.  Gaussian priors are denoted by $\mathcal{N}(\mu, \sigma)$, uniform priors by $\mathcal{U}($lower bound, upper bound$)$, and Jeffrey's priors by $\mathcal{J}($lower bound, upper bound$)$. The initial values for $\eta_{\sigma}$ are set to the standard deviation of the respective datasets. \\
$^{a}$ We want the initial value to be the median of the RVs for that spectrograph; the $+1$ is used incase the median is already zero, as Nelder-Mead solvers cannot start at zero.}

\label{tab:bc_priors3}
\end{table*}

\clearpage 
\section{Summary of the priors used in the no GP RV analysis}
\begin{table*}[http]
\centering
\begin{tabular}{lccccc}
    \hline
    Parameter [units]& Initial Value ($P_{0}$)& Priors &  MAP Value & MCMC Value\\
    \hline
    \hline
    $P_b$ [days] & 6.3980661 & \faLock & --& --\\
    $TC_b$ [days] & 2458517.68971 & \faLock &  --&--\\
    $e_b$ & 0.294 & $\mathcal{U}(0,1)$; $\mathcal{N}(P_{0},0.13)$ &  0.26 & $0.13^{+0.1}_{_0.1}$\\
    $\omega_b$ & $130\pi/180$ & $\mathcal{U}(P_{0}-\pi,P_{0}+\pi)$, $\mathcal{N}(P_{0},45\pi/180)$ &2.88&$-3.2^{+1.4}_{-1.6}$\\
    $K_b$ [m s$^{-1}$] & 10 & $\mathcal{U}(0, \infty)$ & 2.3 &$2.97^{+2.49}_{-1.85}$\\
    \hline
    $P_c$ [days] & 18.8805 & \faLock & --&--\\
    $TC_c$ [days] & 2458533.593 & \faLock & --&--\\
    $e_c$ & 0.093 &  $\mathcal{U}(0,1)$; $\mathcal{N}(P_{0},0.13)$;& 0.093&$0.13^{+0.1}_{-0.1}$\\
    $\omega_c$ & $190 \pi/ 180$ &  $\mathcal{U}(P_{0}-\pi,P_{0}+\pi)$, $\mathcal{N}(P_{0},45\pi/180)$& -3.3 & $-3.2^{+1.4}_{_1.6}$\\
    $K_c$ [m s$^{-1}$] & 10 & $\mathcal{U}(0, \infty)$ &  $7.4 \times 10^{-5}$ &$1.15^{+1.58}_{-0.91}$\\
    \hline
    $\gamma_{iSHELL}$ [m\,s$^{-1}$] & $\mathrm{MEDIAN}(RV_{iSHELL})+\pi/100^{a}$ & $\mathcal{N}$($P_{0}, 100)$ & -0.32 & $0.18^{+2.78}_{-2.77}$\\
    $\gamma_{PFS}$ [m\,s$^{-1}$] & $\mathrm{MEDIAN}(RV_{PFS})+\pi/100^{a}$ & $\mathcal{N}(P_{0},100)$ &  -10.9 & $-10.83^{+1.95}_{-2.03}$\\
    $\gamma_{HIRES}$ [m\,s$^{-1}$] & $\mathrm{MEDIAN}(RV_{HIRES})+\pi/100^{a}$ & $\mathcal{N}(P_{0}, 100)$ &  0.15 & $-0.26^{+3.2}_{-3.1}$\\ 
    \hline
\end{tabular}
\caption{The circular model parameters and prior distributions used in our model that considers the transiting b and c planets, as well as the recovered MAP fit and MCMC posteriors for the no GP runs. \faLock\ indicates the parameter is fixed.  Gaussian priors are denoted by $\mathcal{N}(\mu, \sigma)$. \\
$^{a}$ We want the initial value to be the median of the RVs for that spectrograph; the $+1$ is used incase the median is already zero, as Nelder-Mead solvers cannot start at zero}.

\label{tab:bc_priors2}
\end{table*}

\clearpage
\section{Dave Results} 
Figures \ref{fig:modshift8} and \ref{fig:modshift34} below show the full DAVE vetting results (top table) for the Sectors 8 and 34 light curves respectively.  \tess\ transit data (top row), the binned data (2nd row), and different phased diagnostic plots to look for odd-even effects (third row) secondary, tertiary and ``negative'' eclipses such as would be produced by false positives (fourth row). Neither analysis identifies statistically significant evidence for a false-positive scenario for TOI 560 b.

Figure \ref{fig:photocenter} shows the photocenter difference images and PSFs for tbe \textit{TESS} light curves.  No significant photocenter motion in transit is observed, helping exclude blended eclipsing binary scenarios.

\begin{figure*}[htp]
    \centering
    \includegraphics[width=0.63\textwidth]{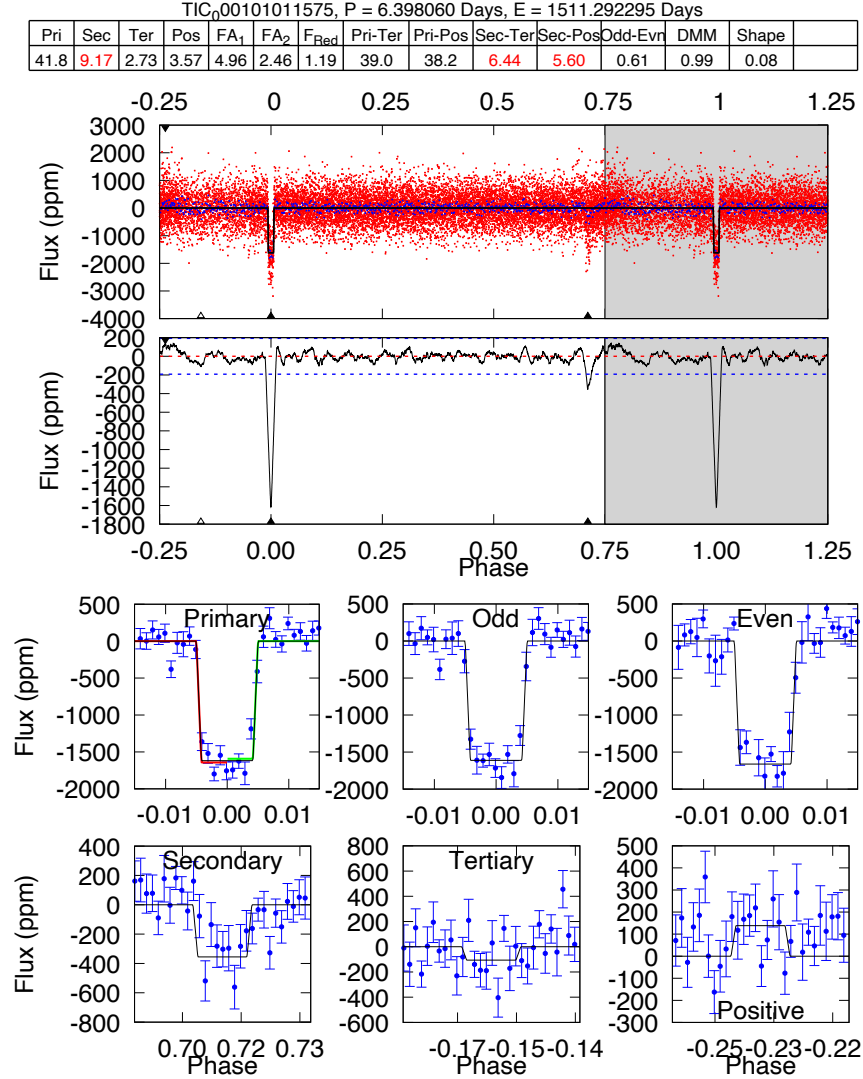}
    \caption{Sector 8: full transit data (top), convolved (middle), and different phased scenarios (bottom, labeled) showing primary, odd, even, secondary, tertiary, and positive transits. The table on top show the DV Model-Shift Uniqueness Test. The top line shows the TCE ID and associated orbital period and epoch. The table lists the values for the significances of each event (Pri=primary, Sec=secondary, Ter=tertiary , and Pos=positive), the false alarm detection thresholds (FA1 and FA2), and the ratio of the noise level on the timescale of the transit duration (red noise) divided by the Gaussian noise (Fred). The difference in significance between the primary and tertiary events (Pri-Ter), the primary and positive events (Pri-Pos), the secondary and tertiary events (SecTer), the secondary and positive events (Sec-Pos), and odd- and even-numbered events (Odd-Evn) are listed next. Finally the values for the Depth Mean-to-Median (DMM), Shape, and the Transit Asymmetry Test (TAT) tests are shown}.
    \label{fig:modshift8}
\end{figure*}

\begin{figure*}[htp]
    \centering
    \includegraphics[width=0.89\textwidth]{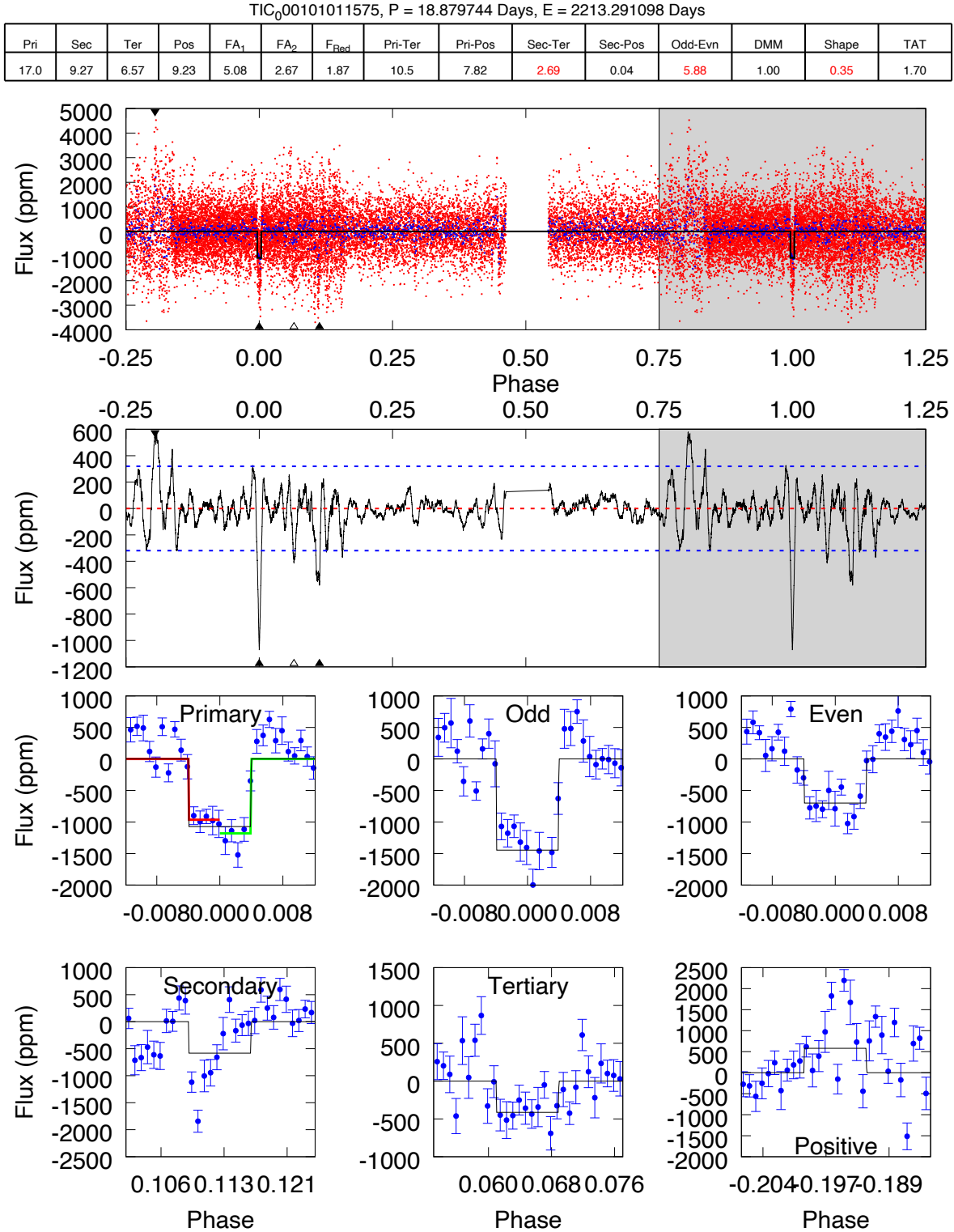}
    \caption{Sector 34: full transit data (top), convolved (middle), and different phased scenarios (bottom, labeled) showing primary, odd, even, secondary, tertiary, and positive transits. The table on top show the DV Model-Shift Uniqueness Test. The top line shows the TCE ID and associated orbital period and epoch. The value of Sec-Ter appears red because Sec-Ter $<$ FA2.
The value of Odd-Evn appears in red because Odd-Evn $>$ FA1.
The value of Shape appears in red if it $>$ 0.3
}.
    \label{fig:modshift34}
\end{figure*}

\begin{figure*}[htp]
    \centering
    \quad
    \includegraphics[width=.85\textwidth]{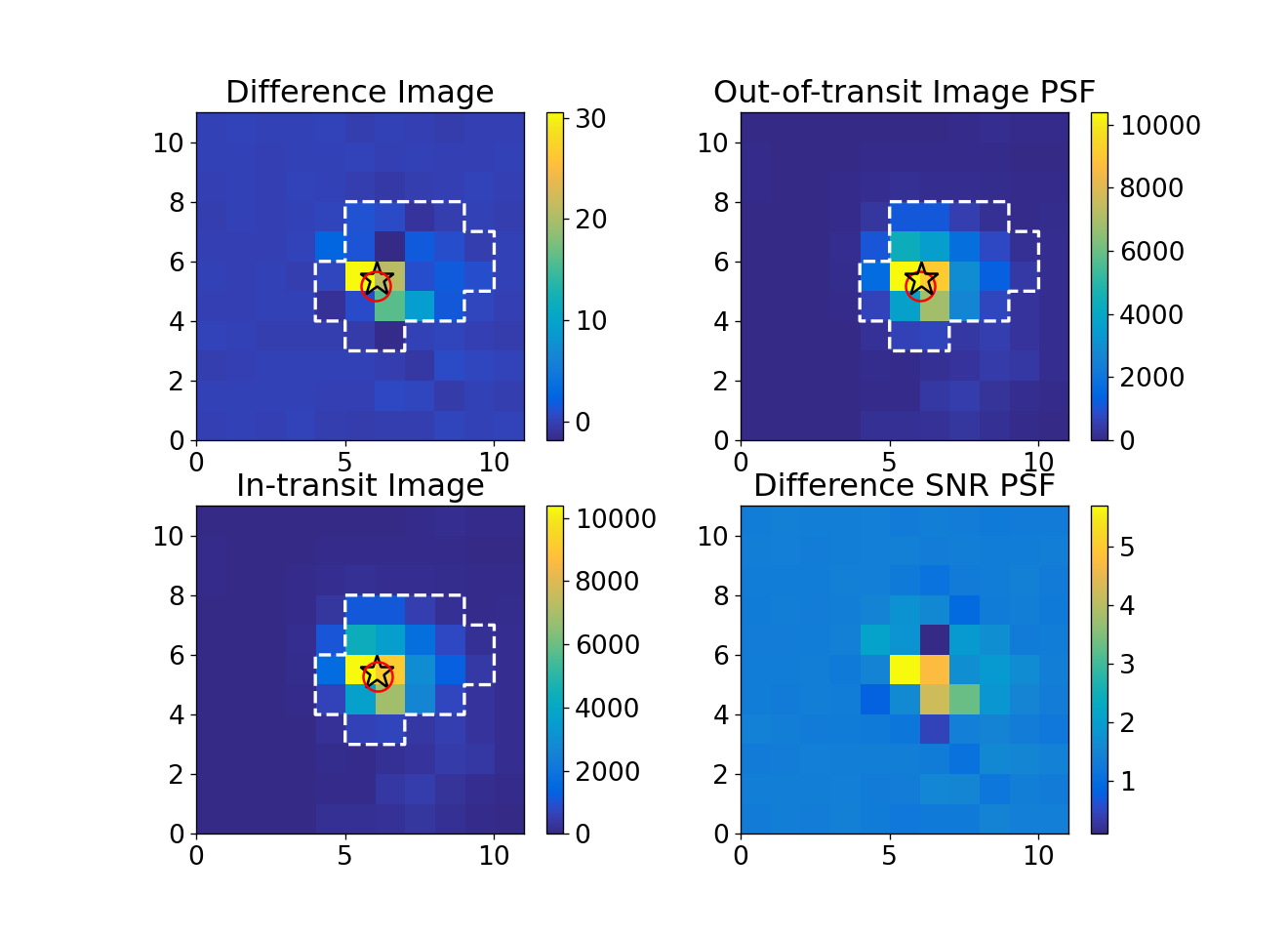}
    \includegraphics[width=.85\textwidth]{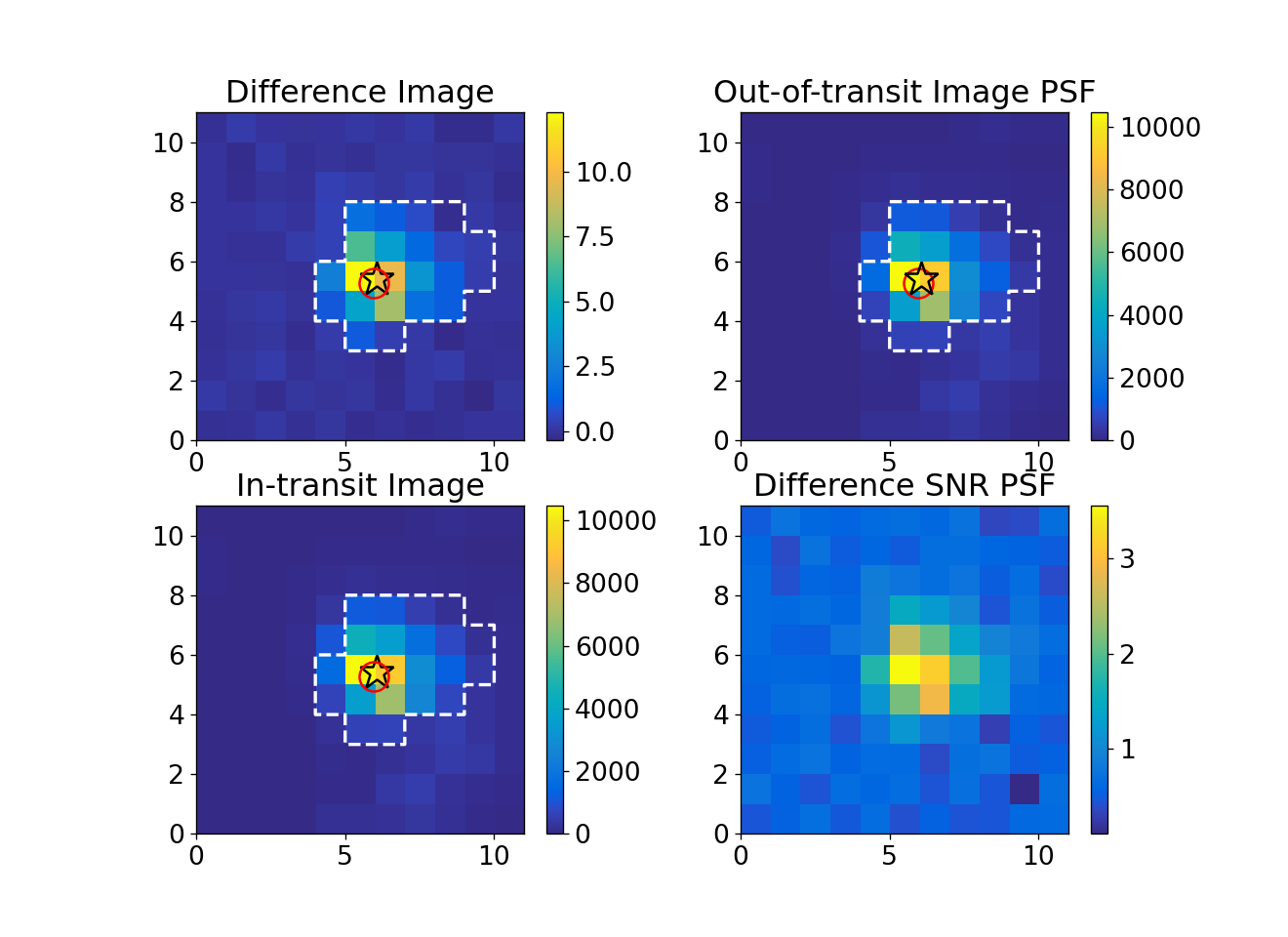}
    \caption{Photocenter difference images and PSFs for sector 8 (top two rows) and 34 (bottom two rows) .  The black star indicates the TIC position, while the red circle is the observed photocenter. The white dashed line indicates the \textit{TESS} target pixel aperture used to extract the light curve, just as the orange outlines shown in the TPF plot (Figure \ref{fig:tpfs1}).}
    \label{fig:photocenter}
\end{figure*}

\clearpage 
\subsection{RVs of our EXOASTv2 analysis}
\begin{figure*}[htp]
    \centering
    \includegraphics[width=.49\textwidth]{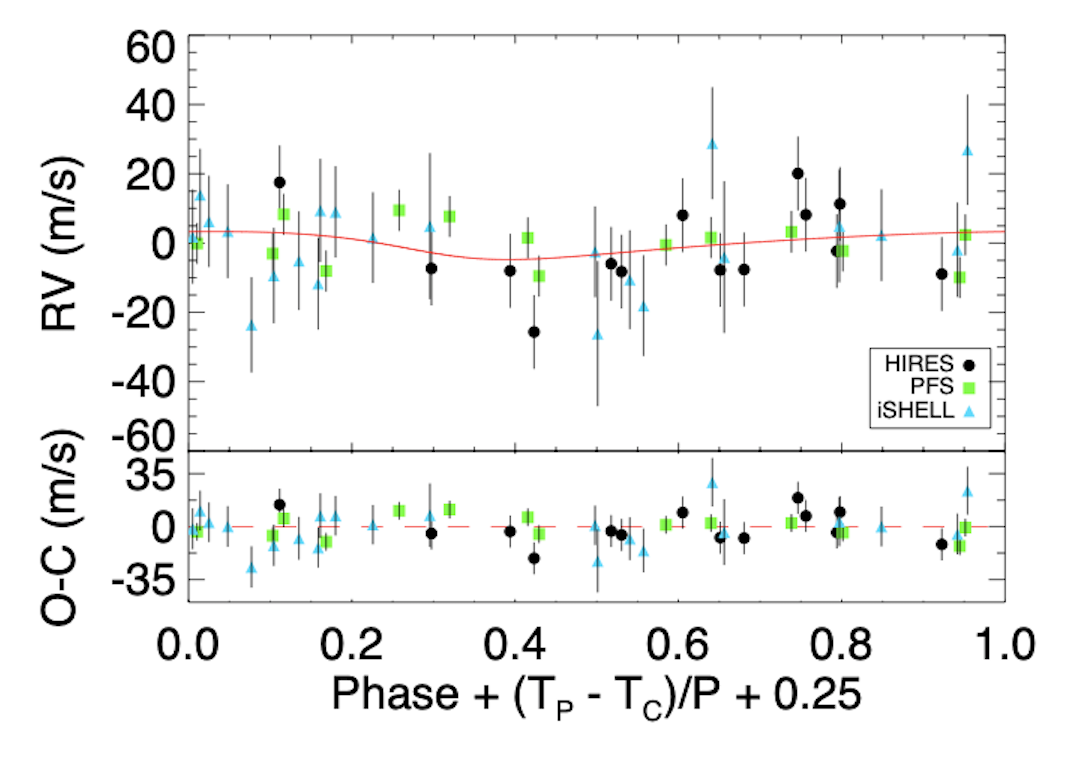}
    \includegraphics[width=.49\textwidth]{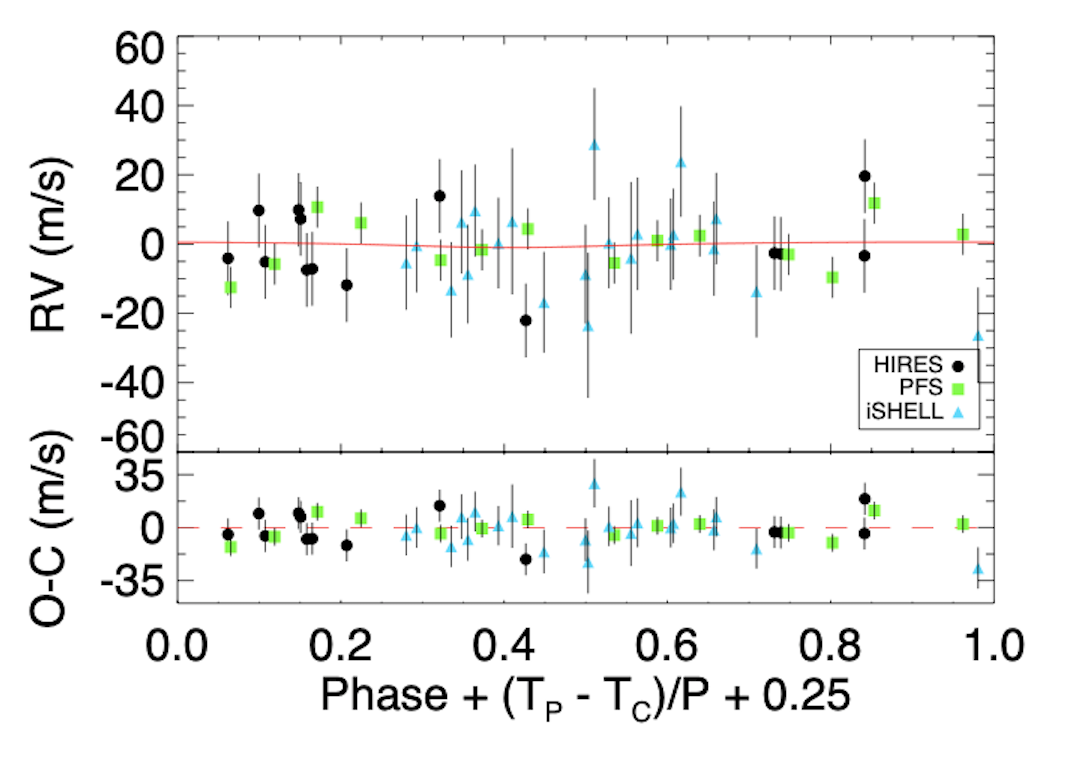}
    \includegraphics[width=.49\textwidth]{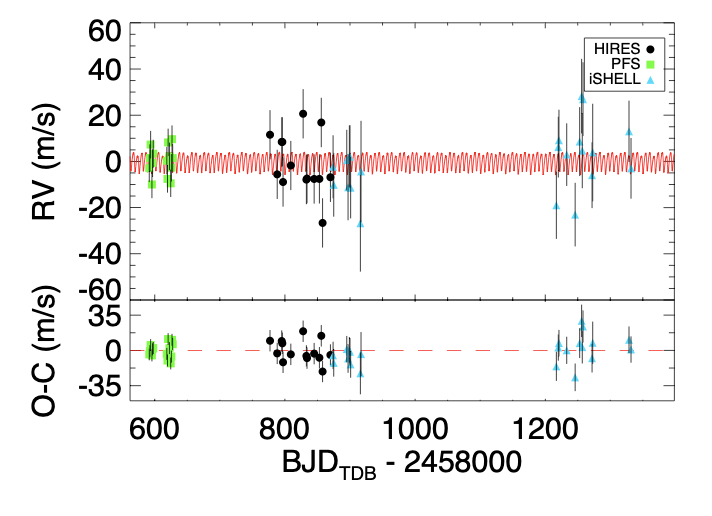}
    \caption{Radial velocities obtained from our \texttt{EXOFASTv2} analysis. Top row: Phased RVs and residuals, with a best-fit model in red.  Bottom Row: Unphased RVs.} 
    \label{fig:eastmanrvs} 
\end{figure*}

With \texttt{ExoFASTv2}, we carry out an independent RV analysis with no GP to account for stellar activity, as a means of cross-checking our RV analysis with \texttt{pychell}.  We recover similar upper-limits and posteriors with both approaches.  The phased and unphased RVs from the \texttt{ExoFASTv2} analysis  are shown in Figure \ref{fig:eastmanrvs}.

\clearpage 
\section{Radial Velocities}
Table \ref{tab:RVs} lists the RVs used in our analysis from three different spectrographs, iSHELL, PFS, and HIRES.

\startlongtable
\begin{deluxetable*}{|c|c|c|c|}\label{tab:RVs}
    \tablecaption{RVs from the different spectrographs we have used in our analysis.}
    \tablehead{time & mnvel & errvel & tel} 
    \startdata
2458874.008515  & -1.466217          & 5.969914          & iSHELL \\
2458875.0489    & -9.277676          & 6.318843          & iSHELL \\
2458895.020254  & 1.423469           & 4.620647          & iSHELL \\
2458897.033761  & -10.358407         & 7.313267          & iSHELL \\
2458899.005408  & 3.073978           & 5.014754          & iSHELL \\
2458900.0048987 & 2.879875           & 5.961691          & iSHELL \\
2458900.988993  & -10.525324         & 5.177452          & iSHELL \\
2458915.97533   & -25.92508          & 16.986916         & iSHELL \\
2458916.96779   & -3.498715          & 18.169736         & iSHELL \\
2459217.043971  & -18.129482         & 7.833479          & iSHELL \\
2459220.036259  & 7.051904           & 4.807669          & iSHELL \\
2459221.028598  & 10.012615          & 4.878999          & iSHELL \\
2459232.98009   & 3.793529           & 5.757477          & iSHELL \\
2459245.961775  & -22.181634         & 6.264058          & iSHELL \\
2459252.899543  & 9.439542           & 8.503822          & iSHELL \\
2459255.971086  & 29.092975          & 10.547535         & iSHELL \\
2459256.967084  & 5.499865           & 10.704609         & iSHELL \\
2459257.97079   & 27.775656          & 10.204183         & iSHELL \\
2459271.925394  & -5.116571          & 7.272103          & iSHELL \\
2459272.952189  & 4.726328           & 17.173103         & iSHELL \\
2459328.734044  & 13.861937          & 5.093228          & iSHELL \\
2459331.828947  & -2.145161          & 4.680086          & iSHELL \\
2458591.59721   & -13.67             & 0.44              & PFS    \\
2458592.55596   & -9.3               & 0.53              & PFS    \\
2458593.61288   & -3.46              & 0.51              & PFS    \\
2458595.61376   & -20.71             & 0.66              & PFS    \\
2458596.60841   & -11.42             & 0.6               & PFS    \\
2458597.591     & -7.39              & 1.2               & PFS    \\
2458618.52548   & -10.43             & 0.59              & PFS    \\
2458619.53717   & -18.31             & 0.6               & PFS    \\
2458620.50553   & -2.51              & 0.52              & PFS    \\
2458622.55362   & -8.55              & 0.54              & PFS    \\
2458624.50368   & -20.21             & 0.59              & PFS    \\
2458625.51752   & -13.44             & 0.55              & PFS    \\
2458626.50784   & -1.11              & 0.62              & PFS    \\
2458627.5164    & -9.31              & 0.59              & PFS    \\
2458777.117606  & 12.8465553361664   & 0.874164164066315 & HIRES  \\
2458788.120576  & -4.3158532184062   & 0.883318662643433 & HIRES  \\
2458795.078952  & 9.74058435517379   & 0.96750670671463  & HIRES  \\
2458796.044556  & 9.71581891692533   & 0.99626761674881  & HIRES  \\
2458797.109961  & -7.62476348436288  & 0.957593381404877 & HIRES  \\
2458809.083128  & -0.466031437909477 & 0.951945245265961 & HIRES  \\
2458827.972435  & 21.9254428753361   & 0.960972011089324 & HIRES  \\
2458832.988014  & -6.58019230860952  & 1.08839976787567  & HIRES  \\
2458833.948473  & -6.12234576086814  & 1.06621098518372  & HIRES  \\
2458844.911132  & -6.30641359100671  & 1.05241250991821  & HIRES  \\
2458852.955442  & -6.28103672598236  & 0.923207104206085 & HIRES  \\
2458855.901039  & 18.1420626828827   & 1.26564979553223  & HIRES  \\
2458857.895081  & -25.3380374118174  & 1.04208660125732  & HIRES  \\
2458869.88686   & -5.56625046683735  & 1.06461870670319  & HIRES
    \enddata
\end{deluxetable*}

\section*{}

\end{document}